\documentclass[11pt,a4paper,fleqn]{article}
\pdfoutput=1
\usepackage{graphicx}
\usepackage[usenames,dvipsnames]{xcolor} % colour
\usepackage{amssymb,amsmath,mathtools,mathrsfs} % maths
\usepackage{epsfig,subfigure} % plots
\usepackage{booktabs,longtable} % tables
\usepackage{exscale,relsize} % scale 
\usepackage[normalem]{ulem} % editing
\usepackage{enumerate} 
\usepackage{jcappub}
\usepackage{aastex}
\usepackage{cancel}
\usepackage{caption}
\usepackage[noabbrev]{cleveref}
\usepackage{hyperref}
\usepackage{float}
\usepackage{appendix}
\usepackage{tabularx}
\usepackage[utf8]{inputenc}
\usepackage{xspace}
\usepackage{bm,bbm}
\usepackage{multirow}
\usepackage{nicefrac,xfrac}

\definecolor{darkgreen}{RGB}{0,100,0}

\def\be{\begin{equation}}
\def\ee{\end{equation}}
\def\beq{\begin{equation}}
\def\eeq{\end{equation}}

\def\bea{\begin{eqnarray}}
\def\eea{\end{eqnarray}} 

\def\ze{z}
\newcommand{\dl}{d_L}
\allowdisplaybreaks

\title{
\vskip-2.1cm
Measuring the propagation speed of gravitational waves with LISA
}

\author[a]{Tessa Baker~\footnote{Project coordinators. Emails: t.baker@qmul.ac.uk;\,g.tasinato@swansea.ac.uk},}
\affiliation[a]{Department of Physics \& Astronomy, Queen Mary University of London, Mile End Road, London, E1 4NS, UK}
\author[b]{Gianluca Calcagni,}\affiliation[b]{Instituto de Estructura de la Materia, CSIC, Serrano 121, 28006 Madrid, Spain}
\author[a]{Anson Chen~\footnote{ Corresponding author. E-mail: a.chen@qmul.ac.uk },}
\author[c,d]{Matteo Fasiello,}
\affiliation[c]{Instituto de Fisica Téorica UAM-CSIC, C/ Nicolas Cabrera 13-15, 28049,
Madrid, Spain}
\affiliation[d]{Institute of Cosmology \& Gravitation, University of Portsmouth, PO1 3FX, UK}
\author[e]{Lucas Lombriser,}
\affiliation[e]{Département de Physique Théorique, Universit\'e de Gen\`eve, 24~quai Ernest Ansermet, 1211~Genève~4, Switzerland}
\author[f]{Katarina Martinovic,}
\affiliation[f]{
Physics  Department,  King's College London, %\, University \, of London, 
 Strand,  London,
WC2R  2LS,  UK}
\author[g]{Mauro Pieroni,}
\affiliation[g]{Blackett  Laboratory, Imperial College London, London, SW7 2AZ,
UK}
\author[f]{Mairi Sakellariadou,}
\author[h]{Gianmassimo Tasinato {\color{blue}$^1$},}\affiliation[h]{Physics Department, Swansea University, SA28PP, UK} \author{\hskip1cm} \author[i]{Daniele Bertacca,}\affiliation[i]{Dipartimento di Fisica e Astronomia ``Galileo Galilei'', Universit\`a degli Studi di Padova; INFN, Sezione di Padova, via F. Marzolo 8, I-35131; INAF - Osservatorio Astronomico di Padova, Vicolo dell'Osservatorio 5, I-35122, Padova, Italy.} \author[j]{Ippocratis D. Saltas}\affiliation[j]{CEICO, Institute of Physics of the Czech Academy of Sciences, Na Slovance 2, 182 21 Praha 8, Czechia}

\author[ ]{ \\ 

\vskip0.1cm

  \texttt{(For the LISA Cosmology Working Group)}}

\begin{document}
\begin{figure}
\begin{flushright}
\href{https://lisa.pages.in2p3.fr/consortium-userguide/wg_cosmo.html}{\includegraphics[width = 0.2 \textwidth]{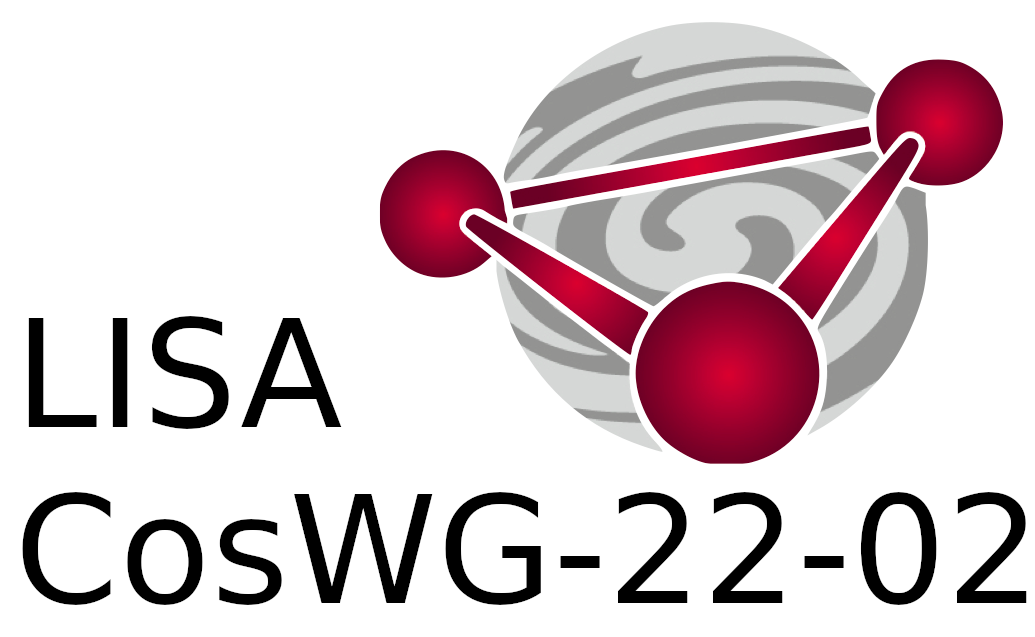}}\\[5mm]
\end{flushright}
\end{figure}

\abstract{
The propagation speed of gravitational waves, $c_T$, has been tightly constrained by the binary neutron star merger GW170817 and its electromagnetic counterpart, under the assumption of a frequency-independent $c_T$. Drawing upon arguments from Effective Field Theory and quantum gravity, we discuss the possibility that modifications of General Relativity allow for transient deviations of $c_T$ from the speed of light at frequencies well below the band of current ground-based detectors. We motivate two representative Ans\"atze for $c_T(f)$, and study their impact upon the gravitational waveforms of massive black hole binary mergers detectable by the LISA mission. We forecast the constraints on $c_T(f)$ obtainable from individual systems and a population of sources, from both inspiral and a full inspiral-merger-ringdown waveform. We show that LISA will enable us to place stringent independent bounds on  departures from General Relativity in  unexplored low-frequency regimes, even in the absence of an electromagnetic counterpart.}

%{\color{blue} propagation speed of GWs.%, luminosity distance}
%.}
% \arxivnumber{21XX.XXXXX}

\maketitle
\flushbottom

%--------------------------------------------------------------------------------------------------------------------------------------------------
%\begin{document}

\section{Introduction}
\label{sec:intro}

The detection of gravitational waves (GWs) by the LIGO-Virgo Collaboration has set stringent constraints on deviations from General Relativity (GR) \cite{LIGOScientific:2016aoc,LIGOScientific:2016sjg,LIGOScientific:2016lio,LIGOScientific:2016dsl} and offered additional probes of fundamental physics (e.g. \cite{Yunes:2016jcc}). 
The future spaceborne detector LISA \cite{LISA:2017pwj} will be able to further test properties of GW propagation over large cosmological distances, by measuring low-frequency GWs emitted by coalescing massive black hole binaries. 
The specific aim of this work is to investigate what LISA can teach us about the speed of gravitational waves, by means of analysis of GW waveforms \textit{only}. Our goal is part of a wider search for general, frequency-dependent modifications of GW propagation, which can be tested by the next generation of GW experiments (see e.g. \cite{Cornish:2011ys,Chamberlain:2017fjl}).
 
 The propagation speed of GWs, $c_T$, was most recently measured by the LIGO-Virgo collaboration using observations of the binary neutron star merger GW170817 \cite{LIGOScientific:2017vwq,LIGOScientific:2017zic,LIGOScientific:2017ync,2017PhRvL.119p1101A}. This impressively precise bound\footnote{The bound quoted here uses the minimum source distance of 26 Mpc, and allows up to 10 s delay before the emission of photons from the associated gamma ray burst \cite{2017PhRvL.119p1101A}.} of $-3\times 10^{-15}\leq c_T-1\leq 7\times 10^{-16}$ (in $c=1$ units) was translated into a constraint on the landscape of dark energy and extended  gravity models in \cite{Lombriser:2015sxa,Bettoni:2016mij,Creminelli:2017sry,Sakstein:2017xjx,Ezquiaga:2017ekz,Baker:2017hug}, where it proved fatal for a handful of theories.
\smallskip

Indeed, the constraint from GW170817 is widely considered a major challenge to extended gravity theories predicting a non-standard GW propagation speed. However, it can also inform discussions on properties required for these gravity models to possess a healthy ultraviolet (UV) completion. This is the viewpoint 
  of \cite{deRham:2018red}, %who
  which added a degree of subtlety to the interpretation of the data that has not yet been considered widely in the literature (though see e.g. \cite{Noller:2020afd} for further theoretical work on the topic). In \cite{deRham:2018red},  compelling arguments and examples are presented suggesting that the speed of propagation of GWs may vary as a function of the energy scale. 
  The starting point is the observation that at low energies, most  theories spontaneously break Lorentz invariance through a time-dependent vacuum expectation value of an additional field(s). Such a time-dependent vacuum expectation value is essential for driving cosmic acceleration, but it usually leads to  a tensor speed  $c_T<1$ due to non-minimal couplings between
  extra fields and gravity. Explicit examples of this phenomenon arise in the context of Horndeski theories and their extensions, Beyond Horndeski or DHOST \cite{Horndeski:1974wa,Deffayet:2011gz,Zumalacarregui:2013pma,Gleyzes:2014dya,Langlois:2015cwa,Crisostomi:2016czh,BenAchour:2016fzp}.

  On the other hand, if the UV completion of an extended gravity theory is required to be Lorentz invariant (as is usually the case), then {\it necessarily} the graviton speed becomes luminal 
  %TB: this originally. said superluminal -- this is incorrect
  at high energies. The transition between non-luminal and luminal speed is likely to occur well before (or at most, around) the strong-coupling scale 
 of the theory, which for Horndeski-like theories is typically $\Lambda\,=\,(M_{\rm Pl}\,H_0^2)^{1/3}\,\sim\,260$~Hz. This is within the frequency band of ground-based
 GW detectors:  
 as a consequence,
 ground measurements might correspond to the frequency range for which the Lorentz invariance %(of the UV completion) 
 of the theory has already enforced luminal propagation speed. At lower frequencies, for example in the LISA frequency band ($\sim 10^{-5} - 0.1$ Hz), the speed of GWs may instead be different from one. 
    
In a broader context, an intriguing picture about sub- and super-luminality of GWs is emerging from recent literature on so-called positivity bounds.  %in the cosmological context. 
Such a programme aims at using criteria of unitarity, causality, locality (and Lorentz invariance) to ascertain whether low-energy effective theories admit a standard UV completion. In the cosmological context or near black holes, it has often been assumed that the speed of GWs ought to be (sub- or at the most) luminal, leading to theoretical constraints on several models beyond Einstein gravity on a Friedmann--Robertson--Walker (FRW) background \cite{Armendariz-Picon:2005oog,Adams:2006sv,Dubovsky:2006vk,Babichev:2006vx,DeFelice:2006pg,Calcagni:2006ye}.

These criteria are an extension of, or rather an extrapolation from, seminal results on causality bounds derived for flat spacetime. The issue is subtler in curved spacetimes (FRW being the key example here), as the QED case studied in \cite{Drummond:1979pp} demonstrates. Whenever curvature becomes important, super-luminality of GWs does \textit{not} imply a lack of causality. In curved spacetime, the speed itself\footnote{In the sense that in a gravitational theory (a setup that spontaneously breaks Lorentz invariance) the notion of low-energy speed, that is any velocity inferred from a low-energy EFT, is not frame independent, see \cite{deRham:2020zyh} for a more in-depth analysis.} may be frame-dependent (see, e.g., \cite{Creminelli:2014wna} for an example in the cosmological context) and therefore not a reliable indicator of causality. Remarkably, in the standard EFT treatment of GR one finds that loop contributions from massive fields lead to a non-luminal speed of GWs on cosmological backgrounds; positivity arguments suggest a super-luminal speed of GWs at low energies \cite{deRham:2019ctd,deRham:2020zyh}. Such findings are not at all in conflict with causality, and have in several examples been shown to be necessary precisely to guarantee causality. In \cite{deRham:2020zyh}, a notion of causality\footnote{See also \cite{Chen:2021bvg} for a very recent work where the notion of ``infrared causality'' is introduced and studied in detail vis-\`{a}-vis asymptotic causality.} more reminiscent of the standard lore has been shown to be more than compatible with positivity bounds whenever a well-defined decoupling limit of the (helicity-2 modes of the) theory exists\footnote{ In this context, the allowed super-luminality is Planck-suppressed and one cannot resolve the deviation from luminality. This result, however, hinges on there being a well-defined decoupling limit. This is not the case in all frames and one must not therefore extrapolate it to EFTs of dark energy, modified gravity.}.

A frequency-dependent propagation speed can also arise in any scenario of gravity where the spectral dimension of spacetime changes with the probed scale. This scale-dependent behaviour of geometry is typical of a broad class of theories of quantum gravity \cite{tHooft:1993dmi,Carlip:2009kf,Calcagni:2009kc,Calcagni:2016xtk,Carlip:2017eud,Mielczarek:2017cdp} and is due to the presence of at least one fundamental scale in the texture of spacetime (see also \cite{Mirshekari:2011yq,Ellis:2016rrr,Arzano:2016twc,Yunes:2016jcc,Calcagni:2016ivi}; also we make some further comments in \S\ref{subsec:eom_amp}). The ensuing dispersion relation  features a non-trivial mixing between time and momentum and leads to a mixed redshift-frequency dependence of $c_T(z,f)$. Also, a frequency dependent GW speed  arises in brane-world models motivated by string theory \cite{Csaki:2000dm}.

Lastly, we should mention that a massive graviton
(or the related bigravity) scenario can lead to a  frequency-dependent GW velocity, with interesting and testable
 consequences for GW waveforms (as first pointed out in  \cite{Will:1997bb}). We refer the reader to the recent \cite{Ezquiaga:2021ler}, and references therein, for thorough  analysis of this case.
 
\smallskip
Our aim in this work is 
to  develop a general theoretical and numerical toolkit for quantifying  the perspective of LISA to measure a frequency-dependent $c_T$ {\it only through} its effects
on GW waveforms from merging massive black hole (MBH) binaries, without 
  relying on specific modified gravity scenarios\footnote{We refer the reader to \cite{Joyce:2014kja} for a review of modified gravity models, and
\cite{Gair:2012nm,Will:2014kxa,Yunes:2013dva,Yunes:2016jcc} for some studies on how to constrain modified gravity  with GW observations.}. 
We implement
two representative Ans\"atze for a frequency-dependent
GW propagation velocity. The 
first Ansatz is motivated from a perturbative
expansion in powers of  $(f/f_\star)$,
with $f_\star$ a fiducial frequency controlling the onset of deviations from GR. The second Ansatz describes
scenarios with rapid changes in $c_T$,  which  smoothly change from $c_T\neq 1$ at small frequencies to $c_T=1$ at larger frequencies. For both Ans\"atze we derive how the GW waveforms are modified with respect to GR. The tools we develop, although applied to two representative scenarios, are very flexible, and can be used in future for testing any new theoretical models predicting transitory variations of  $c_T$ as function of frequency.

We will show that LISA can obtain good constraints on both the GR and new parameters involved, even without electromagnetic (EM) counterparts. In fact,
a major advantage of our work is that it does not rely on detection of unique EM counterparts for LISA sources. Whilst LISA standard sirens can serve as a further tool to test gravity (see e.g. \cite{Cutler:2002ef,Finn:2013toa,Lombriser:2015sxa,Tamanini:2016zlh,Caprini:2016qxs,Belgacem:2019pkk,Baker:2020apq}), the rate of EM counterparts adds a further layer of uncertainty to that already coming from the massive black hole  population models. Furthermore, constraints from standard sirens can only be obtained very close to or after the merger, when the sky localisation is good enough to narrow down candidate host galaxies. In principle, one can imagine the analysis we present here being performed on-the-fly as a system inspirals, as done for regular GR parameters in \cite{Mangiagli:2020rwz}. 
\smallskip

This work is organized as follows. \S\ref{sec:theory} develops general theoretical considerations regarding the effects of a frequency-dependent $c_T(f)$ on GW propagation and corresponding observables, and presents the two Ans\"atze for $c_T(f)$ that will be used in our analysis. In \S\ref{sec:waveforms} we carefully derive the expressions for the GW waveforms in this context. We make use of a Post Newtonian (PN) expansion for describing the inspiral phase, and we adapt the PhenomA waveform \cite{Ajith:2007kx} and ppE approach of \cite{Yunes:2009ke} to describe the merger and ringdown epochs. In \S\ref{sec:dataanalysis} we discuss the GW data analysis tools we implement for our forecasts. We compare Fisher forecast techniques with Monte Carlo Markov chains, showing that a Fisher analysis is  adequate in this context. \S\ref{sec:forecast} presents
the Fisher forecasts: we derive the prospective constraints on GR parameters and our Ans\"atze parameters from GW detection of MBH binaries.  \S\ref{sec:conclude} contains our conclusions, and it is followed by five technical appendixes. Appendix \ref{sec:appendix_table} and \ref{sec:integrand} collect details on the Fisher forecast analysis; appendix \ref{theorymotEFT} contains some theoretical motivations on one of our Ans\"atze; appendix \ref{sec:luminalrecovery} is an analysis on the conditions to meet for recovering a luminal $c_T$ at high frequencies. Finally, appendix \ref{sec:fredep}  discusses future directions for further extending and developing our results; moreover, it makes more explicit the relation among our parametrizations and the ppE framework.

\section{Theoretical framework}
\label{sec:theory}

We assume that the dynamics of GW at emission and detection is described by GR -- possibly thanks
to  screening mechanisms, see. e.g. \cite{Burrage:2017qrf,Babichev:2013usa} for reviews (but see also \cite{Bezares:2021dma} for a different point of view). Deviations from GR can occur during the propagation
 of GW through cosmological space-time from source to observation. We focus  on exploring  consequences of a frequency-dependent speed of GW propagation
 $c_T\,=\,c_T (f)$. Except in appendix \ref{sec:fredep}, this is the only modification that we will allow with respect to the standard propagation equations of GR. In this paper, we will be agnostic with respect to the origin of these deformations and will collectively refer to them as \emph{modified gravity}. This term includes any model where the gravitational sector is altered with respect to GR, from purely \emph{ad hoc} phenomenological models and EFT results to models embedded into, or at least motivated by, a fundamental, self-consistent, predictive theory (e.g. UV completion of existing low-energy scenarios, quantum gravity, emergent gravity).
 
   We start in \S\ref{sec_pre} with  general kinematic
   considerations on the consequences
   of a $c_T(f)$ for GW observables. We then present in \S\ref{sec-theory-ansatze}
   two Ans\"atze for $c_T(f)$ that
   will constitute benchmark scenarios
   for our analysis \footnote{
 In  appendix \ref{sec:fredep}, we  will extend the  formulation of this \S\ref{sec:theory} to a more general case
 including GW friction,
 thus 
  linking the present discussion with scenarios 
   studied in \cite{Belgacem:2019pkk}.}.

 \subsection{Preliminary considerations}
 \label{sec_pre}
 
 We assume that GW are massless, and propagate
 freely through a cosmological background from their source -- an inspiralling binary -- to detection. 
 We consider the following quadratic action for the linearized transverse-traceless GW  modes
 %\cite{Grishchuk:1974ny,Grishchuk:1977zz}
\be
\label{act1}
S_T\,=\,\frac{ M_{\rm Pl}^2}{8} \,\int d t\,d^3 x\;a^3(t)\,\bar \alpha\,\left[   \dot h_{ij}^2- \frac{c_T^2 (f)}{a^2(t)}  (\vec \nabla h_{ij})^2
\right],
\ee
with $ M_{\rm Pl}$ the reduced Planck mass, and $\bar \alpha$ a dimensionless normalization constant that
we will fix with appropriate  physical considerations in what comes next.  It is straightforward to prove that the linearized evolution equation obtained from eq. \eqref{act1} describes a free GW, propagating through a cosmological space-time with arbitrary speed $c_T(f)$. 
The frequency dependence of $c_T(f)$ appearing in eq.  \eqref{act1} is physically
interpreted as the frequency of GW as emitted by an
inspiralling binary
process.  We can then make the 
hypothesis that $f\,=\,f(t)$ with $t$ related to the coalescence time (up to a constant shift). Hence  all quantities in 
eq. \eqref{act1} depend on time only. We do not need to make any further assumptions about the functional dependence of $c_T(f)$ in this subsection.

It is convenient to distinguish three notions of time for the system under consideration (see e.g. \cite{Maggiore:2007ulw}):
 \begin{itemize}
\item[-] Time $t_o$ as measured by ticks of a distant observer's clock
\item[-] Time $t_s$ as measured by clock ticks near the source region (local wave zone)
\item[-] Time $t_e$ when the signal is emitted (a cosmological time scale).
\end{itemize}
The frequency of GW at emission, $f_s$, can be different from the frequency
at detection, $f_o$, due to both the expansion of the universe and to modified gravity effects.
 Let us study this phenomenon in the system at hand. 
 
The action \eqref{act1} describes 
%
%
%
%Action \eqref{act1} describes 
 a free GW travelling through a geodesics in a Friedmann-Lemaitre-Robertson-Walker (FRW) metric, characterized by a line element %(see appendix of  arXiv:1906.01593)
\be\label{metle1}
d s^2\,=\, 
%a^2(\eta)\,e^{ 2 \int d \eta\,{\cal H}(\eta)\,\Gamma_\alpha (\eta,\,f)}\,
c_T(f)\,\bar \alpha\left[-c_T^2(f)\,d t^2+a^2(t)\, d \vec x^2\right] \,.
%-c_T^3(f)\,\bar \alpha\,d t^2+c_T(f)\,a^2(t)\, \bar \alpha\,d \vec x^2 \,.
\ee
This is an effective metric which we use for describing the propagation of the GW \cite{Belgacem:2019pkk}. In fact,  
denoting the associated metric tensor $\tilde g_{\mu\nu}$, the Lagrangian density for a free spin-2 field propagating through it reads
\begin{eqnarray}
{ L}_T&=&\sqrt{-\tilde g} \left[ \tilde g^{\mu\nu} \partial_\mu h_{ij}\partial_\nu h_{ij} \right]
 \\
 &=&
  a^3\,\bar \alpha\,
\left[  \dot h_{ij}^2-\frac{c_T^2 }{a^2} \left( \vec \nabla h_{ij} \right)^2
\right]\,,
\end{eqnarray}
corresponding to the Lagrangian density in the integrand of eq.~\eqref{act1}. 
With the help of eq.~\eqref{metle1}
 we  write comoving and physical distances as

\be \label{excodi}
r^{\rm GW}_{\rm com} (t)\,=\,\int_0^r d r'\,=\,\int_{t_e}^t\,\frac{c_T [f(t')]}{a (t')}\,d t'
\ee
and
% GW physical distance
\be
\label{exphydi}
r^{\rm GW}_{\rm phys} (t)\,=\, a (t)\,c^{1/2}_T(f)\,{\bar \alpha}^{1/2}\,%e^{\int dt \, H \, \Gamma_\alpha}\,
 r^{\rm GW}_{\rm com} (t)\,.
\ee
We  make the hypothesis that, in proximity of the source, modified gravity effects 
have no time to develop, i.e.
\be
\lim_{t\to t_s}\,\frac{r^{\rm GW}_{\rm phys} (t)}{r^{\rm GW}_{\rm com} (t)}\,=\,a(t_s)\,.
\ee
This fixes $\bar \alpha\,=\,c^{-1}_T(f_s)$ hence we conclude that
\be
\label{exphydi2}
r^{\rm GW}_{\rm phys} (t)\,=\, a (t)\,\left[\frac{c_T\left(f(t) \right)}{c_T\left(f_s\right)} \right]^{\frac12}\,%e^{\int dt \, H \, \Gamma_\alpha}\,
 r^{\rm GW}_{\rm com} (t)\,.
\ee

%\subsubsection*{How clocks tick}
We can use  relation \eqref{excodi} to find how 
the typical time scale related with the evolution 
of the GW phase differs between
 source and at detection. For two signals emitted at the same physical distance, one has 
\be
c_T(f_o)\,d t_o\,=\,(1+\ze)\,c_T(f_s)\, d t_s\,.
\ee
The relation between frequencies at source ($f_s$) and at detection ($f_o$), which scale as the inverse of time differences ($f\sim1/\Delta  t$), reads
\be\label{mairel1}
\frac{f_o}{c_T (f_o)}\,=\,\frac{f_s}{(1+\ze)\,c_T( f_s)}\,,
\ee
where $\ze=z_e$ is the redshift of the source. Notice that, in the frequency regimes where $c_T(f)$ is frequency-independent, we %can cancel $c_T$ from both sides and 
find
%have the usual relation
\be
\label{usrel1}
f_s\,=\,(1+\ze)\,f_o %\hskip1cm {\text{if $c_T$ is a constant, flat function of frequency}}\,,
\ee
which is the standard relation connecting frequencies at emission and at detection. In general, however,  a frequency-dependent GW velocity requires to generalize eq.~\eqref{usrel1} to eq.~\eqref{mairel1}.

\smallskip
It is convenient to define a dimensionless quantity $\Delta$ that measures the deviation from the standard
relation \eqref{usrel1} for GWs propagating through cosmological distances:
\bea
\label{defDEL}
\Delta&=&\frac{f_s- (1+\ze)\, f_o(f_s, \ze)}{f_s}
\\
&=&1-\frac{c_T(f_o)}{c_T ( f_s)}\,.
\label{eq:Delta_def}
\eea
$\Delta$ can be expressed as function of $f_s$, or of $f_o$, depending on which is more convenient. A value $\Delta\neq0$ indicates that $c_T$ is a non-constant
function of frequency.
Using the parameter $\Delta$, the clock ticks at source and observer are related by
\bea\label{refatim}
d t_o&=&\frac{f_s}{f_o}\,d t_s
\\
&=&\frac{(1+\ze)\,d t_s}{1-\Delta}\,,\label{dtodts}
\eea
then integrating
\be
t_o\,=\,(1+\ze)\,\int_0^{t_s} \frac{d t'_s}{1-\Delta(t'_s)}\,.
\ee
Simple manipulations lead to the equality
\bea
\label{frder1}
\frac{d f_o}{d t_o}&=&\frac{d f_s}{d t_s}\,\frac{(1-\Delta(f_s))^2}{(1+z)^2}\,\left[ 1+\frac{d\,\ln (1-\Delta(f_s))}{d\,\ln f_s}\right]\,,
\eea
or equivalently
\bea
\label{frder2}
\frac{d f_s}{d t_s}
&=&\frac{d f_o}{d t_o}\,
\frac{(1+z)^2}{(1-\Delta(f_o))^2}
\left[ 1-\frac{d\,\ln (1-\Delta(f_o))}{d\,\ln f_o}\right]\,.
%\frac{d f_o}{d t_o}
%\nonumber\\
%&=&\frac{d f_s}{d t_s}
%\,\frac{(1-\Delta)^2}{(1+z)^2}\,.
\eea

As an immediate, general 
application of the formulas we derived, we conclude this subsection by %, we apply our results to 
deriving an expression for the GW luminosity
distance in scenarios with $\Delta\neq0$, following the arguments of  %obtained
 %in 
 \cite{Belgacem:2019pkk}. 
We call ${\cal F}$
the energy flux %(energy per unit area per unit time) 
at observer position:
\be
{\cal F}\,=\,\frac{d E_o/d t_o}{\rm Area}
\ee
where Area$= 4 \pi ( r^{\rm GW}_{\rm phys}  )^2$. 
Then we introduce the luminosity at the source position, ${\cal L}$:
 \be
 {\cal L}\,=\,\frac{d E_s}{d t_s}\,=\,\frac{(1+z_e)^2}{(1-\Delta)^2}\,\frac{d E_o}{d t_o}
 \,,\ee
where \eqref{dtodts} has been used. The luminosity distance  $d_L^{\rm GW}$ is defined
 in terms of  
 the  following relation
\be
{\cal F}\,\equiv\,\frac{\cal L}{4 \pi \,(d_L^{\rm GW})^2}
\,.
\ee
Using these formulas, as well as relation
\eqref{exphydi2} to connect comoving and physical distance, we obtain % we obtain
\bea
d_L^{\rm GW}&=&{(1+z_e)}\,\left(1-\Delta \right)^{-\frac12}\, r^{\rm GW}_{\rm com}\,,
\label{expLD}
\eea
so the effects of a $c_T$ varying with frequency are contained in the dependence on $\Delta$ as defined in \eqref{defDEL}. 
As we will learn in \S\ref{sec:waveforms}, the luminosity distance $d_L^{\rm GW}$  and other relations we derived here play an important role  for characterizing the properties  of the GW waveforms.  %$c_T$.  

 \subsection{Two Ans\"atze for \texorpdfstring{$c_T(f)$}{cT(f)}}
% for a frequency-dependent speed of GW}
 \label{sec-theory-ansatze}
 
 After the previous  considerations,
 in this subsection we discuss two   representative 
 Ans\"atze for $c_T$.
 They will represent our benchmark scenarios for the LISA forecasts developed 
 in the next sections. In fact, after discussing
 the Ansatz functional forms, we briefly anticipate
 the level of constraints we will be able to obtain with LISA on the parameters
 characterizing them. Importantly,
 these Ans\"atze aim to discuss possible ways to parametrize deviations from $c_T=1$ around LISA frequencies, and are not built for automatically satisfying at the same time constraints on $c_T$ within ground-based frequency ranges. To do so, further corrections
 to their frequency dependence  might be needed in the intermediate frequency band
 between LISA and ground-based experiments. We will comment on this point through the text, and above all in Appendix~\ref{sec:luminalrecovery}.

\subsubsection*{Polynomial {Ansatz}}

Inspired by the scale-dependent choice originally put forward in \cite{Nishizawa:2017nef}, our first model parameterizes $c_T(f)$ as a polynomial in frequency:
\begin{equation}
c_T(f) = %1+\Gamma_\beta(f) = 
 c_0 + \sum_{n} \beta_n \left(\frac{f}{f_*}\right)^n .
\label{eq:cT_power}
\end{equation}
Here $n\neq 0$ can be a positive or negative integer,  $\beta_n$ is a set of parameters controlling deviations from the limiting speed $c_0$, and $f_*$ is a fixed frequency scale controlling the onset of the deviations. In what follows we  study both positive and negative values of $n$ as separate cases. Note that, for simplicity, we do not allow $\beta_n$ to be function of time; this possibility will nevertheless be explored in appendix \ref{sec:fredep}. Notice that our Ansatz \eqref{eq:cT_power} includes  more than one free parameter, hence it goes beyond the one-parameter parametrization proposed in \cite{Mirshekari:2011yq} \footnote{In fact,
it is interesting to compare our Ansatz \eqref{eq:cT_power}  to the parameterisations used in \cite{Yunes:2016jcc} and 
\cite{LIGOScientific:2019fpa,LIGOScientific:2020tif,LIGOScientific:2021sio} to set constraints on the GW dispersion relations, and their effects on the GW speed.  In Appendix \ref{app_rough} we briefly  review the approach of \cite{Yunes:2016jcc} to obtain rough bounds on parameters controlling the GW  speed,  and apply their methods to the case of LISA.}.

In the negative power ($n<0$) case, it is natural to set $c_0=1$ such that consistency with GR is enforced at high frequencies. In the positive power ($n>0$) case $c_0$ is the (unknown) low-frequency GW propagation speed. If $c_0<1$ ($c_0>1$) then the GW speed rises (falls) back towards the GR value of $c$ for $\beta_n>0$ ($\beta_n<0$). This case turns out to be the mathematically simplest model we study; however, it implicitly requires that some additional physics terminates the power-law growth of $c_T$ between the LISA band and the band of ground-based detectors, again to maintain consistency with current bounds (see Appendix~\ref{sec:luminalrecovery}). In the next subsection we will see an example of such additional physics in the context of an EFT-inspired model, for which eq.(\ref{eq:cT_power}) is a low-frequency Taylor expansion.

In practical terms, we will see in \S\ref{subsub:polynomial_phase} that values of $c_0 \neq 1$ only results in a rescaling of $\beta_n$ coefficients in the phase of the waveform. The only other change it brings is a scaling of the waveform amplitude by a factor of $c_0^{1/6}$. As such, constraints on $c_0$ are largely non-degenerate with other parameters in our forecasts, and including it will only weaken them by a small fraction (see Appendix A for exact results). Therefore, without loss of generality, we will evaluate the constraining power of LISA using $c_0=1$ in both polynomial cases for simplicity. Our results remain valid for values of $c_0\neq 1$ with negligible changes.

In both the positive-power and negative-power cases alike, we assume $(f/f_*)^{{\rm sgn}(n)}$ to be a small quantity, allowing us to truncate our polynomial series for $c_T(f)$ (assuming that the $\beta_n$ are not large enough to violate the validity of the  expansion). We will see later that expanding $c_T(f)$ up to quadratic order will prove sufficient to study the dominant corrections to the waveform that may be detectable with LISA. In our forecast in \S\ref{sec:forecast}, we  mainly consider MBH binaries with total masses between $10^4$ and $10^7M_\odot$, as these generally give signal-to-noise ratio (SNR) $>10$ in LISA (see Figure~\ref{fig:waterfall}). The frequency range for these waveforms is between $\sim 10^{-5}$ and $\sim 10^{-1}$ Hz, so $f_*$ is required to stay outside this range in order for $(f/f_*)^{{\rm sgn}(n)}$ to remain stmall. In addition, $f_*$ should be lower than the LIGO lower sensitivity bound of $\sim 10$ Hz. Therefore the typical `safe' values of $f_*$ we use 
%for systems with different masses 
in the positive- and negative-power cases are $2$ Hz and $2\times 10^{-7}$ Hz, respectively; in this context, safe means that the deviations from GR will remain small for any astrophysical system detectable by LISA. 

Values of $f_*$ within the LISA band can be considered, and will result in tighter parameter constraints, but also imply that \textit{some} LISA systems could show non-perturbative departures from GR. Such non-perturbative effects lie beyond the scope of the current work.
It is worth noting that constraints on eq.~\eqref{eq:cT_power} are degenerate in $\beta_n/f_*^n$ and so constraints on $\beta_n$ can be translated from one $f_*$ to another (Appendix.~\ref{sec:luminalrecovery}). 

The negative-power case is arguably the most natural prescription of deviations from GR here, because at high frequencies $c_T/c\rightarrow 1$ without the need to invoke additional physics. However, the bounds on $|c_T/c-1|$ from GW170817 are so impressively tight that they are hard to satisfy even in this model. Using the values of $f_*$ we discussed above, and assuming no finely-tuned cancellations between the $n=-1$ and $n=-2$ terms, formally we need $|\beta_1|\lesssim 10^{-4}$ to satisfy the existing bounds ($\beta_2$ remains virtually unconstrained).  However, recognising that our power-law models would at best be only crude representations of the underlying physics, we do not apply the latter prior on $\beta_1$ in most of this work. In \S\ref{subsec:results_IMR} we present results with only $\beta_2$ allowed to vary, which require no further assumptions to be consistent with GW170817.

\subsubsection*{An EFT-inspired {Ansatz}}\label{secgmnotes}

\begin{figure}
\centering
 \includegraphics[width = 0.47 \textwidth]{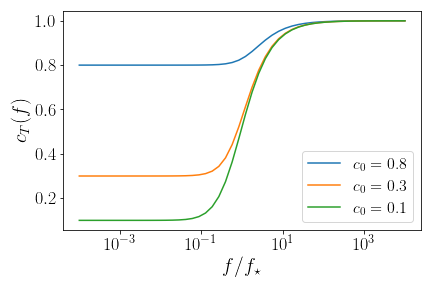}
 \caption{\small Plot of the EFT ansatz for $c_T$ as a function of frequency, as given by eq. \eqref{cTexpA}.} 
 \label{fig:plotCS1}
\end{figure}

The second parametrization we consider has the
property of rapidly changing from a value
of $c_T$ smaller than one at small frequencies
to  $c_T\,=\,1$ at high frequencies, as broadly motivated by the scenarios discussed in the Introduction (see Figure \ref{fig:plotCS1}):

\be\label{cTexpA}
c_T(f)\,=\,\left[ 
1+\frac{f_\star^2}{f^2}-\frac{f_\star^2}{f^2}\,\sqrt{1+2 \left(1-c_0^2 \right) \,\frac{f^2}{f_\star^2} }
\right]^{1/2}.
\ee 
The parametrization \eqref{cTexpA} is controlled
by two free parameters: a fiducial frequency $f_\star$ around which $c_T$ changes rapidly, 
and a low-frequency speed 
$c_0$ with $0\,<\,c_0\,\le\,1$. Ansatz \eqref{cTexpA}
 is motivated by the analysis in \cite{deRham:2018red}
of an UV completion of a scalar field theory, where the scalar velocity depends 
on the energy, and smoothly (but rapidly) connects
from $c_0$ to $1$ as the energy increases. 
 The transition from $c_0$ to unity occurs within a relatively small interval as the
 frequency increases; the width of the transition is not a free parameter and depends entirely
 on $c_0$. See Appendix~\ref{theorymotEFT} for more details on theoretical characterization of this Ansatz and Appendix~\ref{sec:luminalrecovery} for a discussion of its compatibility with the GW170817 bound. Further model-dependent choices of $c_T$ with similar properties might be considered, and their consequences for LISA can be analyzed with the tools we develop in this work.
 
The convenient, 2-parameter parametrization of \eqref{cTexpA} is in some sense a UV-complete version of the positive polynomial model described above. It already contains the high-frequency transition back to $c_T=1$ without further intervention. In fact, one can show that the positive-power polynomial case in equation \eqref{eq:cT_power} represents a low-energy Taylor expansion of equation \eqref{cTexpA} in the limit $\beta_1=0$ and $\beta_2=(1-c_0^2)^2/4c_0$.

\begin{figure}%[h!]
\centering
 \includegraphics[width = 0.49 \textwidth]{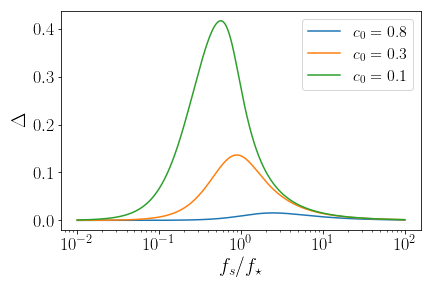}
  \includegraphics[width = 0.49 \textwidth]{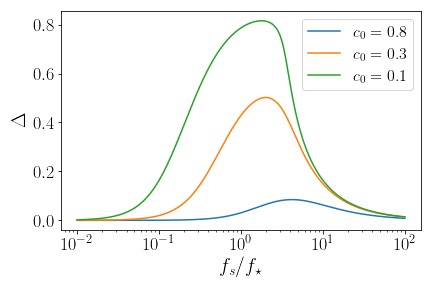}
 \caption{\small Plot of $\Delta$ for the EFT-inspired Ansatz, as defined in \eqref{defDEL}. Left panel: $z=0.2$; Right panel: $z=2$. } %We choose $c_0\,=\,0.1$ (dot dashed green lines), $c_0\,=\,0.3$ (dashed blue lines),  $c_0\,=\,0.8$ (black lines).}
 \label{fig:plot2A}
\end{figure}

\begin{figure}%[h!]
\centering
 \includegraphics[width = 0.49 \textwidth]{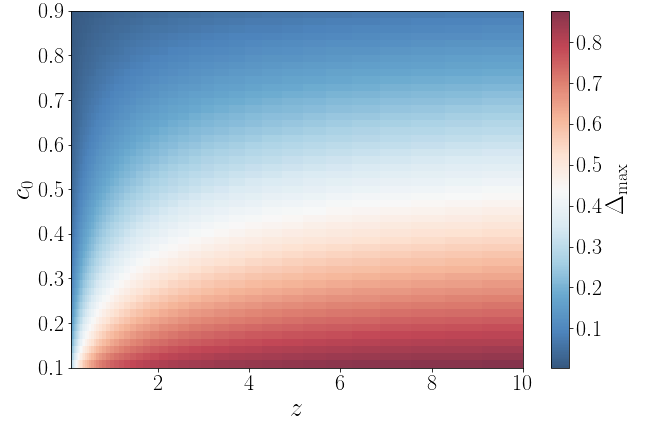}
  \includegraphics[width = 0.49 \textwidth]{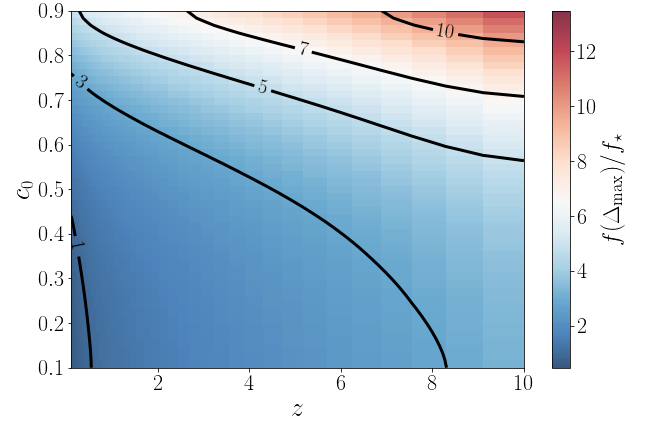}
 \caption{\small Variation of maximum value of $\Delta$ (left panel) and the position of the maximum (right panel) with redshift and $c_0$, for the EFT-inspired Ansatz of \S\ref{secgmnotes}.}
 \label{fig:colormap}
\end{figure}

\smallskip
A frequency profile for $c_T(f)$ as \eqref{cTexpA}
implies that all the frequency-dependent effects studied
in \S\ref{sec_pre} occur in a relatively small frequency band centered around $f_\star$. One can easily compute numerically 
the function $\Delta(f)$, introduced in  \eqref{defDEL}, which is the important quantity that controls the deviations from GR. We plot  $\Delta(f)$
 in 
Figure \ref{fig:plot2A} for representative choices
of parameters. We notice that this function has a pronounced peak, whose maximal value $\Delta_{max}$ depends
 on $c_0$, but also on the redshift $z$ at which the GW source event occurs. To understand better how $\Delta(f)$ evolves over the $z-c_0$ parameter space, we evaluate the amplitude and the position of the maximum of the function  for redshifts log-uniformly distributed from 0.1 to 10, and values of $c_0$ uniformly distributed between 0.1 and 0.9, see Figure \ref{fig:colormap}.  We see that maximum deviation from GR occurs at frequencies of the order $f_{\star}$ and for small $c_0$ and large $z$, as expected. We numerically found a simple phenomenological fit relating
  $\Delta_{\rm max}$ to $c_0$ and $z$ that is valid up to large redshifts ($z=15$):
\begin{equation}
\Delta_{\rm max}(c_0, \,z)\,=\,\left(1.07-1.04\,c_0\right)\,\left[1-\frac{1}{
(1+z)^{(1.07-0.84\,c_0)}}
\right]\,.
\label{eq:deltamax}
\end{equation}
For more details on the expression above we refer the reader to Appendix~\ref{theorymotEFT}. This relation suggests that if we were able
to measure with good precision deviations from
GR induced by Ansatz~\eqref{cTexpA}, we might then be able to extract independent information on the {\it redshift} of the source, which might be helpful to build a Hubble diagram with GW sirens. We leave the exploration of this idea to future work.

The two parameters $f_\star$ and $c_0$ controlling  the location and height of the transition (with $c_0=1$ corresponding to the GR case) can indeed be constrained very well with LISA.  
In \S\ref{sec:forecast} we forecast LISA capabilities
to measure these quantities, and find that both parameters influence considerably GW waveforms. We conclude that 
for MBH  binaries
in specific mass ranges (around $M_{\rm tot}\sim 10^5 M_\odot$), the parameters $f_\star$ and $c_0$ characterizing Ansatz~\eqref{cTexpA}, can be constrained
to a fractional error  of order percent level or better,  with respect
to their fiducial values.

%%%%%%%%%%%%%%%%%%%%%%%%%%%%%%%%%%%%%%%%%%%%%%%%%%%%%%%
\section{Waveform computation}
\label{sec:waveforms}

In this section we compute how gravitational waveforms are modified in models where $c_T$ is a function of frequency, 
making use of the two Ans\"atze discussed in the previous section. Both the waveform amplitude and the phase are affected. We combine methods first introduced in \cite{Will:1997bb} in the context of a massive
graviton with tools motivated by the standard post-Newtonian approach to GW observables. We start
in \S\ref{sub_lDGWa} and \S\ref{subsec:waveform_phase} by discussing how the waveform amplitude and phase are sensitive to a frequency-dependent $c_T$, focussing on the inspiral epoch only. Then in \S\ref{sec:IMR_extension} we take an additional
step and consider extended gravitational waveforms that include also the merger and ringdown epochs. We adopt the frequency-domain PhenomA waveforms 
of \cite{Ajith:2007kx}, and follow similar lines to the ppE approach of \cite{Yunes:2009ke} for the phase of a system. 

\subsection{GW luminosity distance and GW amplitude}
\label{sub_lDGWa}

As we learned in \S\ref{sec_pre}, eq.~\eqref{expLD}, when $c_T$
is function of frequency the
GW luminosity distance is given by 
\begin{equation}
\dl^{\rm GW} = (1+\ze)\,r^{\rm GW}_{\rm com}\, \sqrt{\frac{c_T(f_s)}{c_T(f_o)}}.
 \label{eq:distance}
\end{equation}
while 
the relation between frequencies at source and detection is

\begin{equation}
f_o\,=\,\frac{f_s}{(1+\ze)} \frac{c_T( f_o )}{c_T( f_s )} = f_z \frac{c_T( f_o )}{c_T( f_s )},
\label{eq:fo_fs}
\end{equation}
where in the second equality we define $f_z = f_s/(1+z)$ as the redshifted frequency as in GR. 

\begin{figure}
    \centering
    \includegraphics[width=0.6\textwidth]{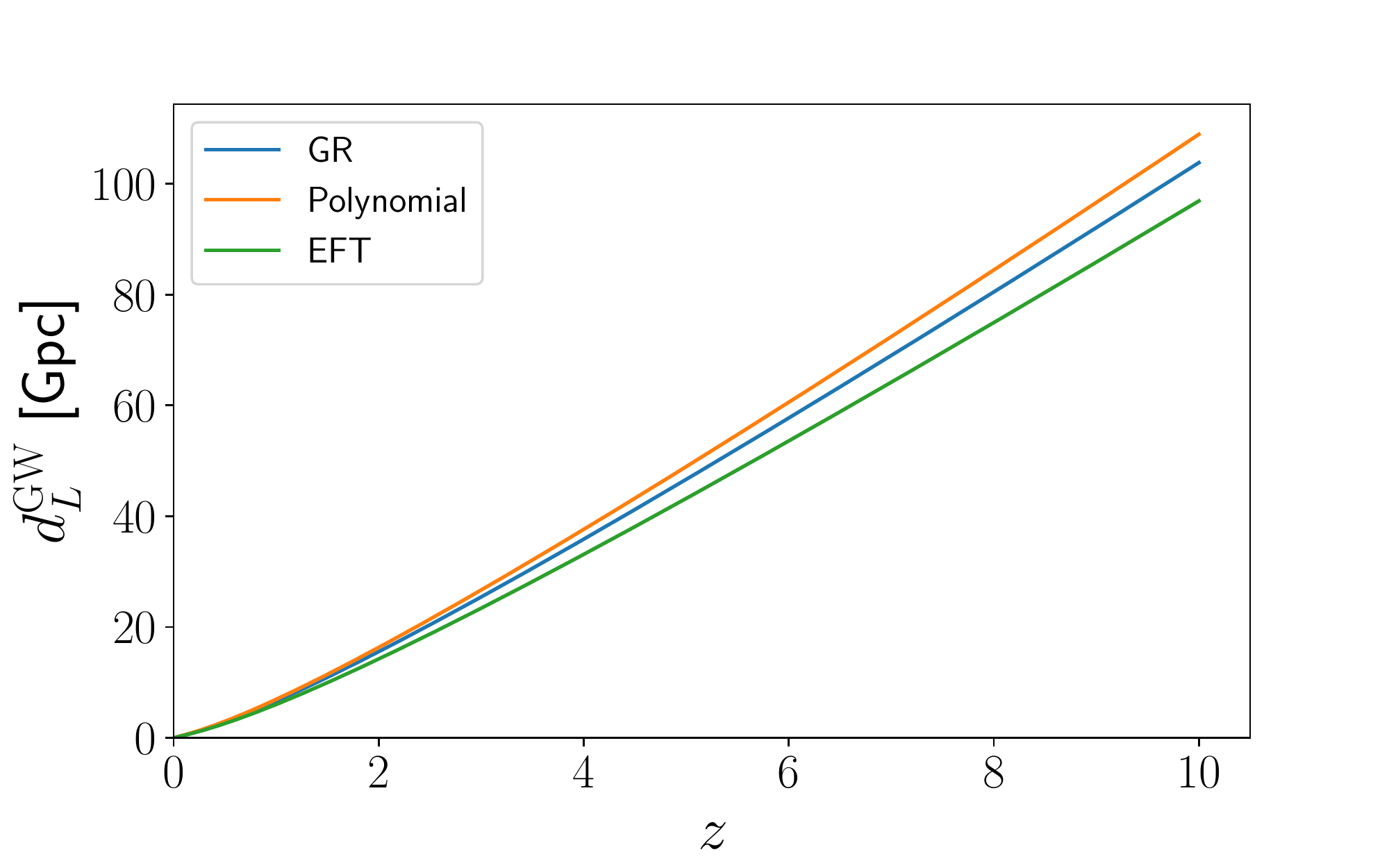}
    \caption{\small $\dl^{\rm GW}$ vs $z$ in GR, the polynomial and the EFT Ans\"atze. $\dl^{\rm GW}$ in the polynomial case is computed in the positive-power case with exaggerated values of $\beta_1=\beta_2=10$, at the frequency of $(f_o/f_*)=10^{-2}$. $\dl^{\rm GW}$ in the EFT case is computed choosing $c_0=0.9$ at the frequency of $(f_o/f_*)=1$. }
    \label{fig:DL_z}
\end{figure}

We plot in Figure \ref{fig:DL_z} the GW luminosity distance versus $z$ in GR, the polynomial Ansatz and the EFT-inspired Ansatz respectively. The values of $\dl^{\rm GW}$ in the polynomial case are larger than in GR for positive values of parameters $\beta_1$ and $\beta_2$ (and vice-versa for negative $\beta_1$ and $\beta_2$). For the EFT-inspired case $\dl^{\rm GW}$ is suppressed with respect to its GR behaviour.

\begin{figure}
\centering
    \includegraphics[width=0.7\textwidth]{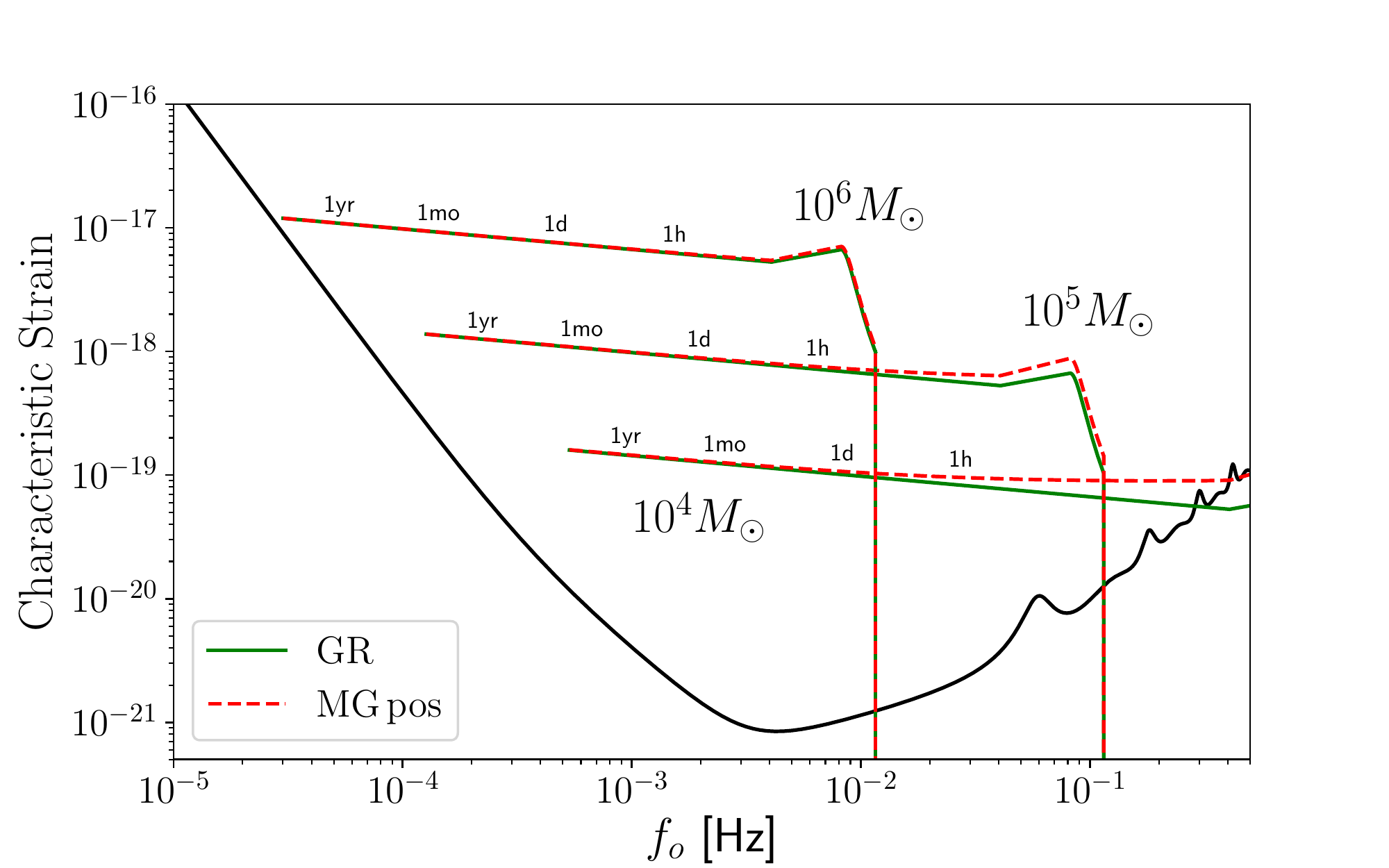}\\
    \includegraphics[width=0.7\textwidth]{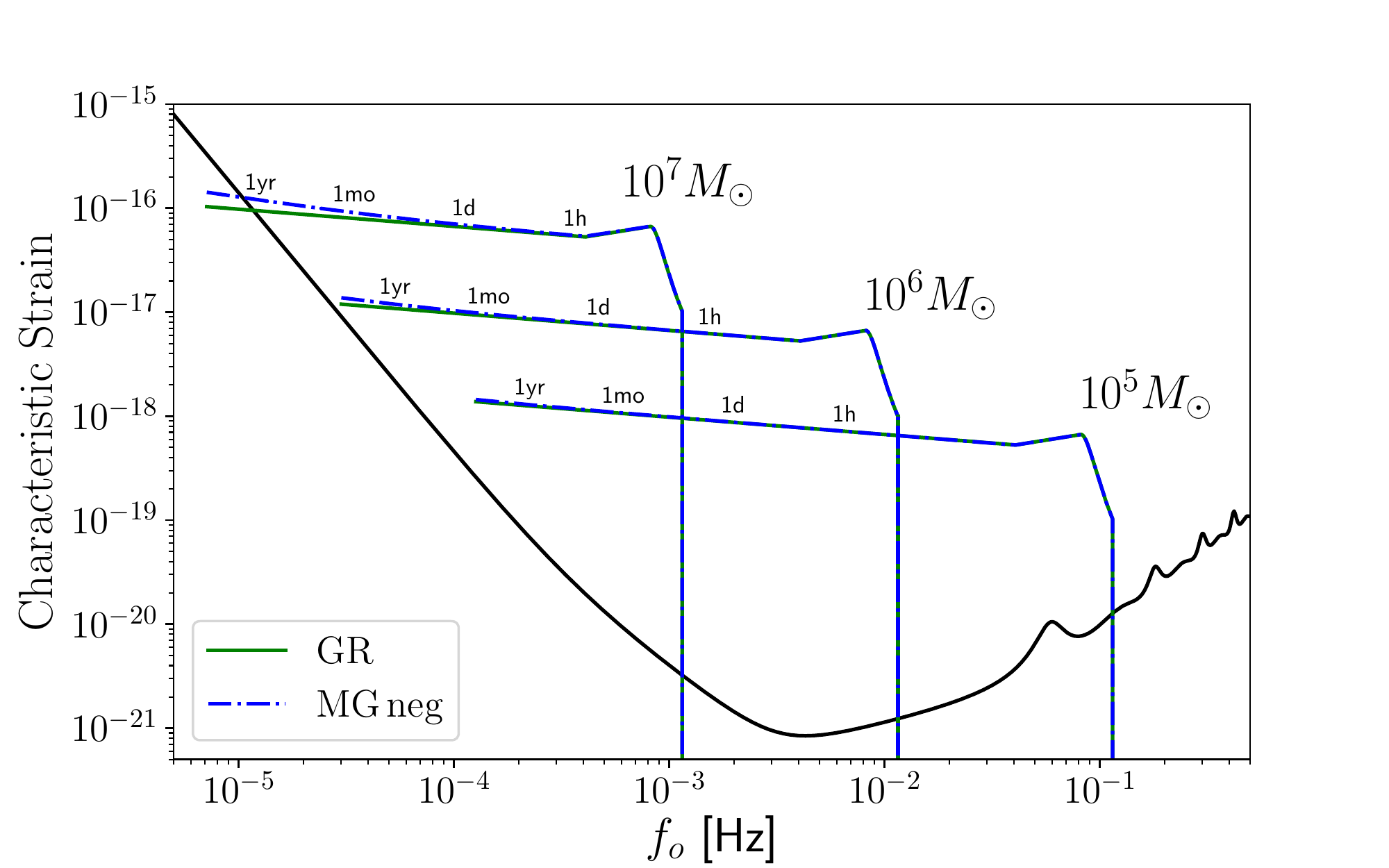}
    \caption{\small The characteristic strains in GR (green), the positive-power polynomial Ansatz (red) and the negative-power polynomial Ansatz (blue) for binaries with different total masses at $\ze=1$. The characteristic strain is the modulus of the strain scaled by the frequency, i.e.~$2f_o |h(f_o)|$ \cite{Moore:2014lga}. The sensitivity curve is plotted as $\sqrt{f_o Sn(f_o)}$. We use exaggerated values $\beta_1=\beta_2=100$ for the positive-power case and $\beta_1=\beta_2=200$ for the negative-power case to visualize the modified gravity effects. We use $f_*=2$ Hz and $f_*=2\times10^{-7}$ Hz for the positive and negative-power cases respectively. The timeline shown on the amplitude is the time before merger computed at Newtonian order.}
    \label{fig:amp_pds}
\end{figure}

The two helicities of the GW waveform for the binary compact object inspiral in Fourier space are given by (see e.g. \cite{Robson:2018ifk})
\begin{align}
    & h_{+}(f) = A(f) \frac{1+\cos^2\iota}{2} e^{i\Psi(f)}, \\
    & h_{\times}(f) = iA(f) \cos \iota ~e^{i\Psi(f)},
\end{align}
where $A(f)$ is the amplitude of the waveform and $\Psi(f)$  the phase (to be discussed   in the next section).  $\iota$ is the inclination angle of the orbit relative to the line of sight. 
The  GW amplitude in GR, without accounting for the redshift, is given by 
\begin{equation} A^{\rm GR}(f_s) = \sqrt{\frac{5\pi}{24}} \frac{{\cal M}_s^2}{a(t_s)r_{\rm com}} (\pi{\cal M}_sf_s)^{-7/6} \,. \end{equation}
It is derived from the time-dependent GW amplitude using the stationary phase approximation  in the Fourier transform of the waveform \cite{Maggiore:2007ulw}. ${\cal M}_s$ is the chirp mass of the binary system at the source, defined by ${\cal M}_s=M_{\rm tot} \eta^{3/5}$, with $M_{\rm tot}$ the binary total mass, $\eta=m_1m_2/M_{\rm tot}$ the reduced mass parameter, and $m_1$, $m_2$ the two component masses. Since the signal observed by the detector is redshifted, we rewrite the waveform using the redshifted chirp mass ${\cal M}_z = (1+\ze){\cal M}_s$, redshifted frequency $f_z=f_s/(1+\ze)$, and using $1/a(t_s) = (1+\ze)$. The redshifted GW waveform amplitude is then given by  \begin{equation} A^{\rm GR}(f_z) = \sqrt{\frac{5\pi}{24}} \frac{{\cal M}_z^2}{(1+\ze)r_{\rm com}} (\pi{\cal M}_zf_z)^{-7/6}\,. \end{equation}
 In modified gravity, the quantities involved in GW propagation are not only scaled by redshift, but also scaled by $c_T(f_o)/c_T(f_s)$. Hence we define the observed chirp mass as
 \begin{equation} {\cal M}_o = {\cal M}_z\frac{c_T(f_s)}{c_T(f_o)}\,.%\label{eq:Mrel} 
 \label{eq:app_Mo_Mz} 
 \end{equation}
We can replace the physical distance ${(1+\ze)r_{\rm com}}$ by $\dl^{\rm GW}$ using eq.~(\ref{eq:distance}), and replace ${\cal M}_z$ by ${\cal M}_o$, so to  finally obtain the modified GW amplitude as 
  \begin{equation}
 A^{\rm MG}(f_o) = \sqrt{\frac{5\pi}{24}} \frac{{\cal M}_o^2}{\dl^{\rm GW}} (\pi{\cal M}_of_o)^{-\frac76} \left[\frac{c_T(f_o)}{c_T(f_s)}\right]^{\frac32} .\label{eq:AMG}
 \end{equation}
The amplitudes of the characteristic strains (defined by $2f_o|h(f_o)|$ \cite{Moore:2014lga}) in GR as well as the positive- and the negative-power polynomial cases are plotted in Figure~\ref{fig:amp_pds}, with exaggerated values of $\beta_1$ and $\beta_2$. Also plotted is the effective sensitivity curve of LISA with angular averaging over the sky 
%{\color{red} [What does it mean "sky angles"?]} 
 and the polarisation angle adopted from reference~\cite{Flauger:2020qyi}. It shows that the modified amplitudes deviate from their GR equivalents as $f_o$ approaches $f_*$. Note that the amplitudes in the figure extend to the merger and the ringdown phases using the PhenomA waveform, which we  discuss in \S\ref{sec:IMR_extension}. Since $f_*$ for the positive and the negative-power polynomial cases are in opposite extrema of the LISA band, the modification effects are more manifest in systems with different total masses in the two cases. Lighter systems are preferred for detecting beyond Einstein models described by the positive-power polynimal Ansatz, and heavier systems for the negative-power polynomial Ansatz.

\subsection{Phase}
\label{subsec:waveform_phase}

The phase of the GW during inspiral can be computed analytically using methods based on the Post-Newtonian (PN) expansion. We first set up the calculation using a general $c_T(f)$, and then we specialise our results to the polynomial and EFT-inspired Ans\"atze described in \S\ref{sec:theory}. As the focus of our work is on GW propagation effects, we do not consider modifications to the  physics of the merging process \textit{at the source position}. As such, we expect the rate of change of GW frequency in the source frame to match that of GR. This is the starting point of our calculation, and is given by (we expand up to 2.5 PN order):
 \begin{equation}
          \frac{d f_s}{d t_s}\,=\,\frac{96}{5 \pi\,{\cal M}_s^2}\,{u}^{\frac{11}{3}}\Bigg[1 + \psi_{ 1}u^{\frac{2}{3}} + \psi_{1.5}u + \psi_{2}u^{\frac{4}{3}} + \psi_{2.5}u^{\frac{5}{3}} \Bigg],
    \label{eq:dfs_dts}
 \end{equation}
 where $u$ is defined
 as
 \begin{equation} \label{MuMz}
 u = \pi {\cal M}_s f_s = \pi {\cal M}_z f_z = \pi {\cal M}_o f_o\,,
\end{equation} 
 and is frame-independent, while $\psi_k~ (k=1,1.5,2,2.5)$ are the PN phase parameters. In this work we specialise to non-spinning binary systems on circular orbits, recognising that  parameters associated with spin/non-circular orbits will be included in a later analysis, for example following the methods of \cite{Berti:2004bd,Berti:2005qd}. In this case, the PN coefficients read \cite{Buonanno:2002fy}: 
 \begin{align}
     &\psi_{1} = - \Bigg(\frac{743}{336} + \frac{11}{4}\eta \Bigg)\eta^{-\frac{2}{5}}\\
     &\psi_{1.5} = 4\pi \eta^{-\frac{3}{5}}\\
     &\psi_{2} = \Bigg(\frac{34103}{18144} + \frac{13661}{2016}\eta + \frac{59}{18}\eta^2 \Bigg) \eta^{-\frac{4}{5}}\\
     &\psi_{2.5} = -\frac{\pi}{\eta} \left(\frac{4159}{672}+\frac{189}{8}\eta \right),
 \end{align}
where $\eta=m_1m_2/(m_1+m_2)^2$ is the symmetric mass ratio. %, $m_1$ and $m_2$ being the two component masses of the binary. %Already defined
We include up to the 2.5 PN term, as this is dominant for the latest stage of inspiral phase we consider -- Figure~\ref{fig:PNterm_neg} shows this for GR and the negative polynomial Ansatz (we do not show the positive polynomial Ansatz as its deviations from GR are less pronounced). 
We  verified that the 3 PN term remains subdominant in all our calculations.
 
 We express our results in terms
 of the quantity $\Delta(f)$ introduced in \eqref{defDEL}  for parameterizing  deviations from GR. 
 Making use of formulas \eqref{frder1} and \eqref{frder2}, we find
\begin{align}  
%\frac{ d f_o}{d t_o}&= %\frac{(1-\Delta)^2}{ (1+ \ze)^2}\left[
%\frac{1}{ 1 + f_o(1-\Delta) \frac{ \partial}{\partial f_o}\left(\frac{1}{1-\Delta}\right) }\right]
 % \frac{ d f_s}{d t_s}\\
 \frac{ d f_o}{d t_o}  %& = \frac{(1-\Delta)^2}{ (1+ \ze)^2}
%\left(\frac{1} {1 + \frac{f_o}{1-\Delta} \frac{ \partial \Delta}{\partial f_o}} \right) \frac{ d f_s}{d t_s}\\
  & = (1-\Delta)^2 \left(\frac{1} {1 + \frac{f_o}{1-\Delta} \frac{ \partial \Delta}{\partial f_o}} \right) \frac{96}{5 \pi\,{\cal M}_z^2}\,u^{\frac{11}{3}}\Bigg[1 + \psi_1 u^{\frac{2}{3}} + \psi_{ 1.5}u + \psi_{2}u^{\frac{4}{3}} + \psi_{2.5}u^{\frac{5}{3}} \Bigg].
\label{eq:df_dt_delta}
\end{align}
Note that the mass appearing in the line above is now the redshifted chirp mass, and all references to source-frame quantities have been eliminated. The next step of the calculation is to integrate this expression twice, to find the time to coalescence and then the GW phase. At this point, we separate the discussion % calculations 
for the polynomial and EFT Ans\"atze.

\subsubsection{Polynomial parametrization models}
\label{subsub:polynomial_phase}

For the polynomial case only, we  make an additional simplification by setting $c_T(f_s)=1$. LISA binaries are located inside galaxies, a region where existing observations \cite{Will:2014kxa,Williams:2004qba,BeltranJimenez:2015sgd,Capozziello:2014rva} constrain gravity to be very close to GR. We will coarsely model this behaviour by fixing $c_T$ to unity at the starting point of the GW \textit{extragalactic} path as well. %The GW then accumulates deviations from GR as it travels across intergalactic space, obeying eq.(\ref{gwprop}) until it reaches the Milky Way. 
We do not attempt to model what happens when the GW exits or enters a galaxy, as our simple Ansatz in eq.~(\ref{eq:cT_power}) contains no environmental dependence. However, we assume that entrance to a screened region does not completely erase the accumulated beyond-GR changes to the signal\footnote{We note that such erasure \textit{does} happen to the amplitude changes induced by modified GW damping in some scalar-tensor theories \cite{Dalang:2019fma}. In these models the GW amplitude depends only on the start and end points of the GW trajectory. We are not aware of any reason similar behaviour should happen when $c_T$ is modified, or in models that lie outside the standard Horndeski canon (such as those represented here).}. Then eq. (\ref{eq:df_dt_delta}) becomes:
\begin{equation}
\frac{d f_o}{d t_o} = \frac{c_T(f_o)^2}{1-\frac{\partial \ln c_T(f_o)}{ \partial \ln f_o}} \frac{96}{5 \pi\,{\cal M}_z^2}{u}^{\frac{11}{3}}\Bigg[1 + \psi_{1}u^{\frac{2}{3}} + \psi_{1.5}u + \psi_{\rm 2}u^{\frac{4}{3}} + \psi_{2.5}u^{\frac{5}{3}} \Bigg].
\end{equation}
Rearranging for $t_o$ and integrating over the observed frequency we have
\begin{equation}
t_o(f_o) - t_c  = \int_{f_c}^{f_o} \frac{1}{c_T(\tilde{f}_o)^2}\left[ 1-\frac{\partial \ln c_T( \tilde{f}_o)}{ \partial \ln \tilde{f}_o} \right] \frac{5 \pi {\cal M}_z^2}{96} \tilde{u}^{-\frac{11}{3}} \Bigg[1 + \psi_{ 1}\tilde{u}^{\frac{2}{3}} + \psi_{1.5}\tilde{u} + \psi_{ 2}\tilde{u}^{\frac{4}{3}} + \psi_{2.5}\tilde{u}^{\frac{5}{3}} \Bigg]^{-1} d\tilde{f}_o \,,
\label{eq:to_power}
\end{equation}
where $t_c$ and $f_c$ are the cutoff time and frequency that mark the end of the first  inspiral phase. In this paper we take $f_c$ to be twice the frequency of the inner-most stable circular orbit (ISCO) \cite{Maggiore:2007ulw}. To simplify the computation, we convert the integral eq. (\ref{eq:to_power}) to be with respect to $u$ instead of $f_o$. Differentiating $u = \pi \left({\cal M}_z/c_T(f_o)\right) f_o$ we find
\begin{equation}
\frac{du}{df_o} = %\pi {\cal M}_o + \pi f_o \frac{d{\cal M}_o}{df_o} = \pi {\cal M}_o + \pi f_o {\cal M}_z \frac{d}{df_o} \Bigg( \frac{1}{c_T(f_o)} \Bigg) \\
%& = \pi {\cal M}_o - \pi f_o {\cal M}_z  \frac{1}{c_T(f_o)^2}  \frac{dc_T(f_o)}{df_o} \\
%& = \pi {\cal M}_o \Bigg[ 1 -  \frac{f_o}{c_T(f_o)} \frac{dc_T(f_o)}{df_o} \Bigg] =
\pi \frac{{\cal M}_z}{c_T(f_o)}  \Bigg[1 - \frac{d \ln c_T(f_o)}{d \ln f_o} \Bigg],
\label{defuf0}
\end{equation}
so that the integration of the time interval, when expressed with the help of the function $\Delta(f)$,  reduces to
\begin{align}
\hspace{-1cm}t_o(u) - t_c %& = \int_{u_c}^{u} \frac{1}{c_T(\tilde{u})} \frac{5 {\cal M}_z}{96} \tilde{u}^{-\frac{11}{3}} \Bigg[1 + \psi_{ 1}\tilde{u}^{\frac{2}{3}} + \psi_{1.5}\tilde{u} + \psi_{ 2}\tilde{u}^{\frac{4}{3}} + \psi_{2.5}\tilde{u}^{\frac{5}{3}} \Bigg]^{-1} d\tilde{u} \nonumber \\
  & \simeq \frac{5{\cal M}_z}{96}\! \int_{u_c}^{u} \frac{1}{c_T(\tilde{u})} \tilde{u}^{-\frac{11}{3}} \Bigg[1 - \psi_{1}\tilde{u}^{\frac{2}{3}} - \psi_{1.5}\tilde{u} + (\psi_{1}^2 - \psi_{2})\tilde{u}^{\frac{4}{3}} + (2 \psi_{ 1}\psi_{1.5} - \psi_{2.5})\tilde{u}^{\frac{5}{3}}\Bigg] d\tilde{u}\,,
\label{to_MG}
\end{align}
and we Taylor expanded the square bracket in eq.~(\ref{eq:to_power}) to second order, since $u\ll1$. Under the stationary phase approximation \cite{Will:1997bb}, the phase of the gravitational waveform is then computed as
\begin{align}
\Psi &=2\pi\int_{f_c}^{f_o}[t_o(\tilde{f}_o)-t_{c}]\,d\tilde{f}_o + 2 \pi f_o t_c - \Psi_c - \frac{\pi}{4} \label{fasemod}\\
 & = 2\pi\int_{u_c}^{u}[t_o(\tilde{u})-t_{c}]\,\frac{c_T(\tilde{u})}{\pi {\cal M}_z  \left[1 - \frac{d \ln c_T(\tilde{f}_o)}{d \ln \tilde{f}_o} \right]}\, d\tilde{u} + 2 \pi f_o t_c - \Psi_c - \frac{\pi}{4},
 \label{Psi_MG}
\end{align}
where $t_c$ and $\Psi_c$ are nuisance parameters that mark the time and the phase at the end of the inspiral phase. 

Note that the integration in the phase above depends on $c_T(u)$. At first glance, the expansion of $c_T(f)$ in eq.~(\ref{eq:cT_power}) is not easily converted to $u$, due to the appearance of $c_T$ itself in eq.~(\ref{MuMz}). However, we can justify that the following form for $c_T(u)$ is equivalent to $c_T(f)$ up to $n=2$:
\begin{equation}
    c_T(u) = 1 + \beta_1 \left( \frac{u}{u_*} \right) + \beta_2 \left( \frac{u}{u_*} \right)^2,
\label{eq:cT_u}
\end{equation}
 where we define a new fixed scale $u_*\equiv\pi {\cal M}_z f_*$. Consider the following quantity
\begin{equation}
    \frac{u}{u_*} = \frac{\pi \frac{{\cal M}_z}{c_T(f)} f_o}{\pi {\cal M}_z f_*} \simeq \frac{f_o}{f_*} \left[ 1-\beta_1 \left(\frac{f_o}{f_*} \right) + {\cal O}\left(\frac{f_o}{f_*}\right)^2 \right], %= \frac{f_o}{f_*} - \beta_1 \left(\frac{f_o}{f_*}\right)^2,
\end{equation}
where we have expanded $1/c_T(f)$ since $|c_T(f)-1|\ll1$. Then eq. (\ref{eq:cT_u}) can be expanded to be
\begin{align}
    c_T(u) &= 1 + \beta_1 \left[\frac{f_o}{f_*} - \beta_1 \left(\frac{f_o}{f_*}\right)^2 \right] + \beta_2 \left(\frac{f_o}{f_*}\right)^2 + {\cal O}\left(\frac{f_o}{f_*}\right)^3  \\
    &= 1 + \beta_1 \left( \frac{f_o}{f_*} \right) + \tilde{\beta}_2 \left( \frac{f_o}{f_*} \right)^2 + {\cal O}\left(\frac{f_o}{f_*}\right)^3 ,
    \label{eq:ctu}
    %
    %&= c_T(f_o) + {\cal O}\left(\frac{f_o}{f_*}\right)^3.
\end{align}
where $\tilde{\beta}_2=\beta_2-\beta_1^2$. Notice that  eq. (\ref{eq:ctu}) has precisely the same form as eq. (\ref{eq:cT_power}) up to $n=2$, simply with a redefinition of one of the coefficients. Hence we will use $c_T(u)$ from here on; we will drop the tilde on $\tilde{\beta}_2$ since it is operationally equivalent to the non-tilde quantity.

Following similar arguments, we can replace the denominator in eq. (\ref{Psi_MG}) as:
\begin{align}
\left[1 - \frac{d \ln c_T}{d \ln f_o} \right]^{-1} %&=\left[1 - \frac{d \ln c_T}{d \ln u} \frac{d \ln u}{d \ln f_o} \right]^{-1} =\left[1 - \frac{d \ln c_T}{d \ln u} \left(1 + \frac{d \ln {\cal M}_o}{d \ln u}\right) \right]^{-1} \nonumber\\
%&=\left[1 - \frac{d \ln c_T}{d \ln u} \left(1 - \frac{d \ln c_T}{d \ln u}\right) \right]^{-1} %\approx \left[1 - \frac{d \ln c_T}{d \ln u} \left(1 + \frac{d \ln c_T}{d \ln u}\right)^{-1} \right]^{-1} \nonumber\\
&\simeq 1+\frac{d \ln c_T}{d \ln u} = 1+\frac{1}{c_T}\left[\beta_1\left(\frac{u}{u_*}\right)+2\beta_2\left(\frac{u}{u_*}\right)^2\right].\label{cTappr}
\end{align}
We expand eq. (\ref{Psi_MG}) a final time, and carry out the integration. We arrive at the modified inspiral phase of the GW waveform, for the positive-power case:
\begin{align}
    \Psi^{\rm pos}(u)&= \frac{5}{48} u^{-\frac{5}{3}} \Bigg\{ \frac{9}{40} - \frac{1}{2} \psi_{1}u^{\frac{2}{3}} - \frac{3}{2} \left[ \frac{3}{5} \psi_{1.5} -\frac{3}{20}\frac{1}{u_*} \beta_1 \right]u + \frac{9}{4} (\psi_{1}^2 - \psi_{2}) u^{\frac{4}{3}}\nonumber \\
    & - (2 \psi_{1}\psi_{1.5} - \psi_{2.5}) u^{\frac{5}{3}} \ln u \Bigg\}  + 2 \pi f_o t_c - \Psi_c - \frac{\pi}{4}.\label{posPsi}
\end{align}
We have dropped the terms where the order of $u$ is higher than 2.5 PN order in the curly bracket. This is since we only expand up to 2.5 PN, and higher-order terms will be integrated into the 3 PN term. We have verified that the new 3 PN term is still insignificant. Note that only the 1.5 PN term is modified from GR by the appearance of $\beta_1$.

If we use the setup of $c_T(f)$ with the additional parameter $c_0$ as in eq. \eqref{eq:cT_power}, we can pull the factor $c_0$ out by
\begin{equation}
    c_T(f) = c_0 \left[1 + \frac{\beta_1}{c_0}\frac{f}{f_*} + \frac{\beta_2}{c_0}\left(\frac{f}{f_*}\right)^2 \right],
    \label{eq:cT_power_c0}
\end{equation}
and then rename $\beta_1/c_0$ and $\beta_2/c_0$ as the new $\beta_1$ and $\beta_2$, so that $c_0$ just works as a scaling of $\beta_1$ and $\beta_2$. The factor $c_0$ outside the square bracket in eq. \eqref{eq:cT_power_c0} is then cancelled by $1/c_T(u)$ in eq. \eqref{to_MG}, and following the same derivation in eq. \eqref{eq:cT_u}-\eqref{cTappr}, $c_T(u)/[1-d\ln c_T/d\ln f_o]$ in eq.\eqref{Psi_MG} remains unchanged as well. Hence the phase is actually unchanged up to a redefinition of $\beta_1$ and $\beta_2$.

Repeating our steps in the negative-power polynomial case, we find a significantly lengthier expression: 
\begin{align}
    & \Psi^{\rm neg}(u) = \frac{5}{48} u^{-\frac{5}{3}} \Bigg\{  \frac{3}{11} \Bigg[ \Bigg(\frac{3}{14} \beta_1^2 -\frac{33}{56} \beta_2 \Bigg) u_*^2 - \frac{3}{8} \beta_1 \beta_2 u_*^3 \psi_{\rm 1.5} \Bigg] u^{-2} \nonumber\\
&+ \frac{9}{70} \beta_1\beta_2 u_*^3 (\psi_{\rm 1}^2 - \psi_{\rm 2})u^{-\frac{5}{3}} \nonumber\\
& + \frac{1}{3} \Bigg[ \Bigg(\frac{3}{4} \beta_2 - \frac{1}{4} \beta_1^2 \Bigg) u_*^2 \psi_{\rm 1} + \frac{1}{2} \beta_1 \beta_2 u_*^3 (2 \psi_{\rm 1}\psi_{\rm 1.5} - \psi_{\rm 2.5}) \Bigg] u^{-\frac{4}{3}} \nonumber\\
& + \frac{3}{8} \Bigg[ - \frac{3}{11} \beta_1 u_* + \Bigg( \frac{48}{55} \beta_2 - \frac{3}{11} \beta_1^2 \Bigg)u_*^2 \psi_{\rm 1.5} \Bigg] u^{-1} \nonumber\\
& + \frac{3}{7} \Bigg( \frac{3}{10}\beta_1^2 - \frac{21}{20} \beta_2 \Bigg)u_*^2 (\psi_{\rm 1}^2 - \psi_{\rm 2}) u^{-\frac{2}{3}}\nonumber\\
& + \frac{1}{2} \Bigg[ \frac{1}{3} \beta_1 u_* \psi_{\rm 1} + \Bigg( \frac{1}{3} \beta_1^2- \frac{4}{3} \beta_2 \Bigg) u_*^2 (2 \psi_{\rm 1}\psi_{\rm 1.5} - \psi_{\rm 2.5}) \Bigg]u^{-\frac{1}{3}}\nonumber\\
& + \frac{3}{5} \Bigg( \frac{3}{8} + \frac{3}{8} \beta_1 u_* \psi_{\rm 1.5} \Bigg) - \frac{9}{28} \beta_1 u_* (\psi_{\rm 1}^2 - \psi_{\rm 2}) u^{\frac{1}{3}} \nonumber\\
& + \Bigg[ -\frac{1}{2} \psi_{\rm 1} - \frac{1}{2} \beta_1 u_* (2 \psi_{\rm 1}\psi_{\rm 1.5} - \psi_{\rm 2.5})  \Bigg] u^{\frac{2}{3}} - \frac{9}{10} \psi_{\rm 1.5} u \nonumber\\
& +\frac{9}{4} (\psi_{\rm 1}^2 - \psi_{\rm 2})u^{\frac{4}{3}} - (2 \psi_{\rm 1}\psi_{\rm 1.5} - \psi_{\rm 2.5})u_o^{\frac{5}{3}} \ln u \Bigg\}+ 2 \pi f_o t_c - \Psi_c - \frac{\pi}{4} \,.\label{negPsi}
\end{align}
In this case, negative PN terms are generated, but no terms higher than 2.5 PN appear. We drop terms lower than $-3$ PN order, to keep the lowest order the same as in $c_T$. Recall that for this model, $u_*$ is chosen below the LISA band, i.e., it is a small number. Terms with order of $u$ below $-3$ PN also come with higher powers of $u_*$, so that they are suppressed. 

\begin{figure}
    \centering
    \includegraphics[width=0.48\textwidth]{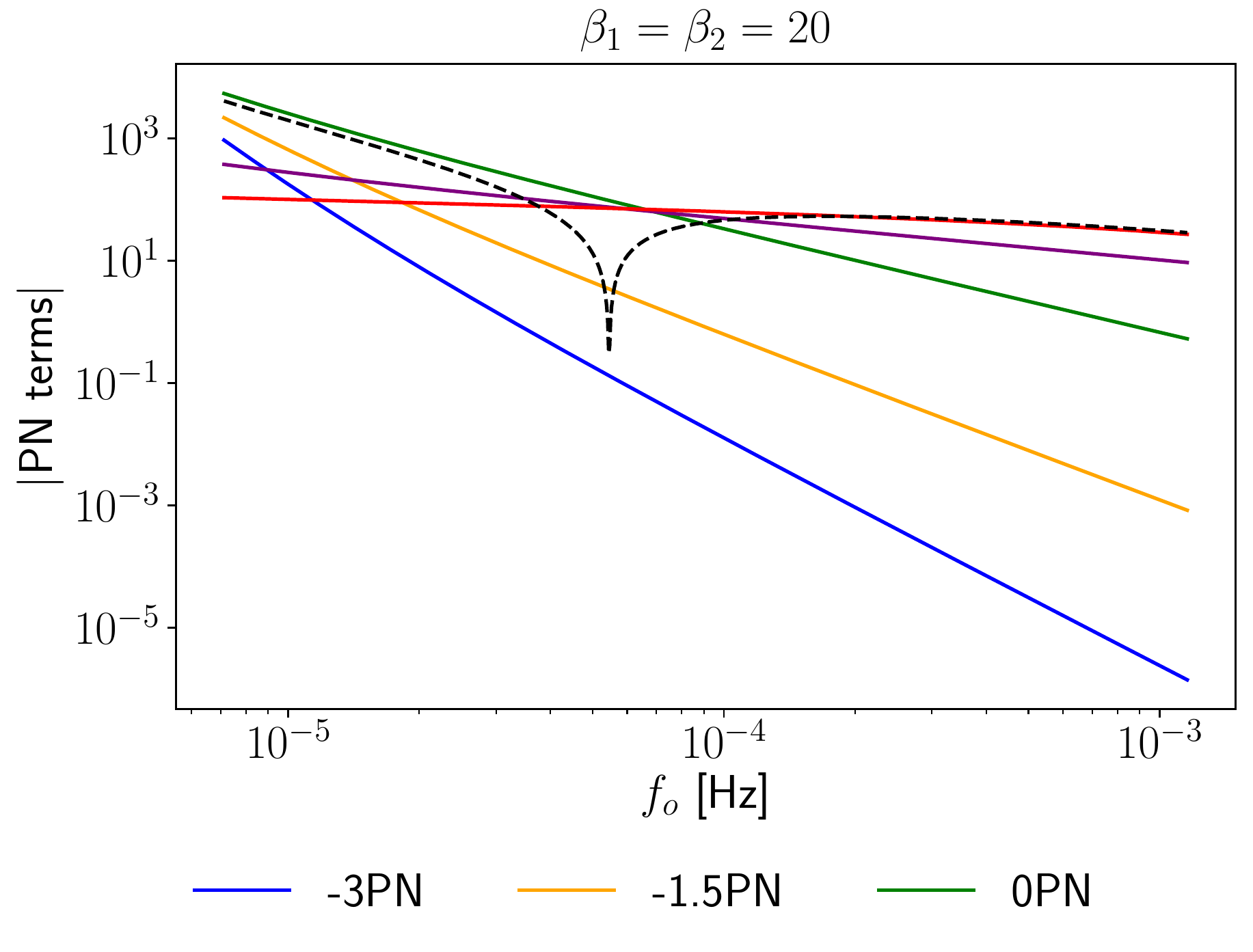}
    \includegraphics[width=0.48\textwidth]{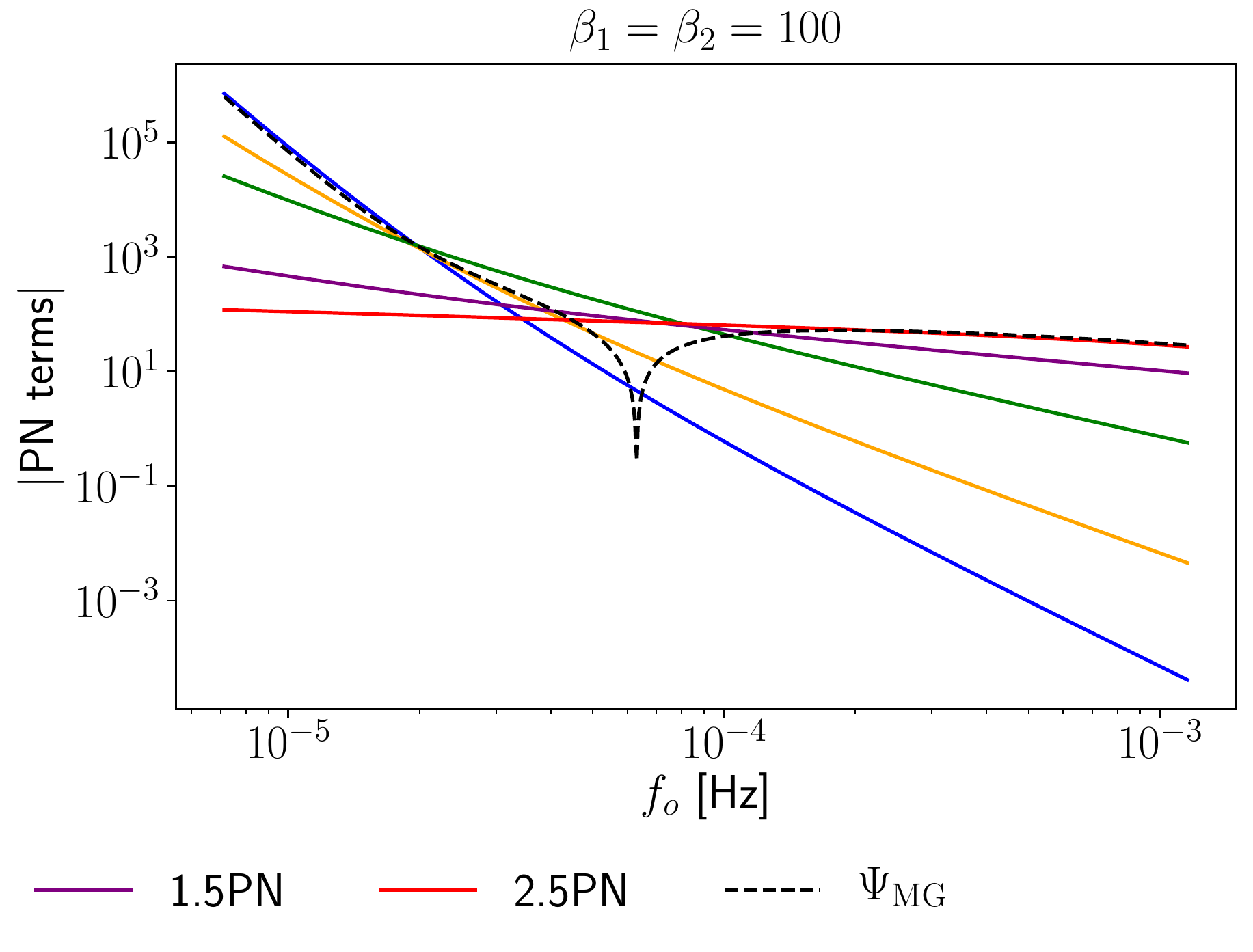}
    \caption{\small Absolute values of PN terms in the phase of the negative-power case for $M_{\rm tot}=10^7 M_\odot$, $z=1$ and $f_*=2\times10^{-7}$ Hz. The left panel shows the case with small values for $\beta_1$ and $\beta_2$, and the right panel shows the case with large values.}
    \label{fig:PNterm_neg}
\end{figure}

\begin{figure}[t]
\centering
    \includegraphics[width=0.49\textwidth]{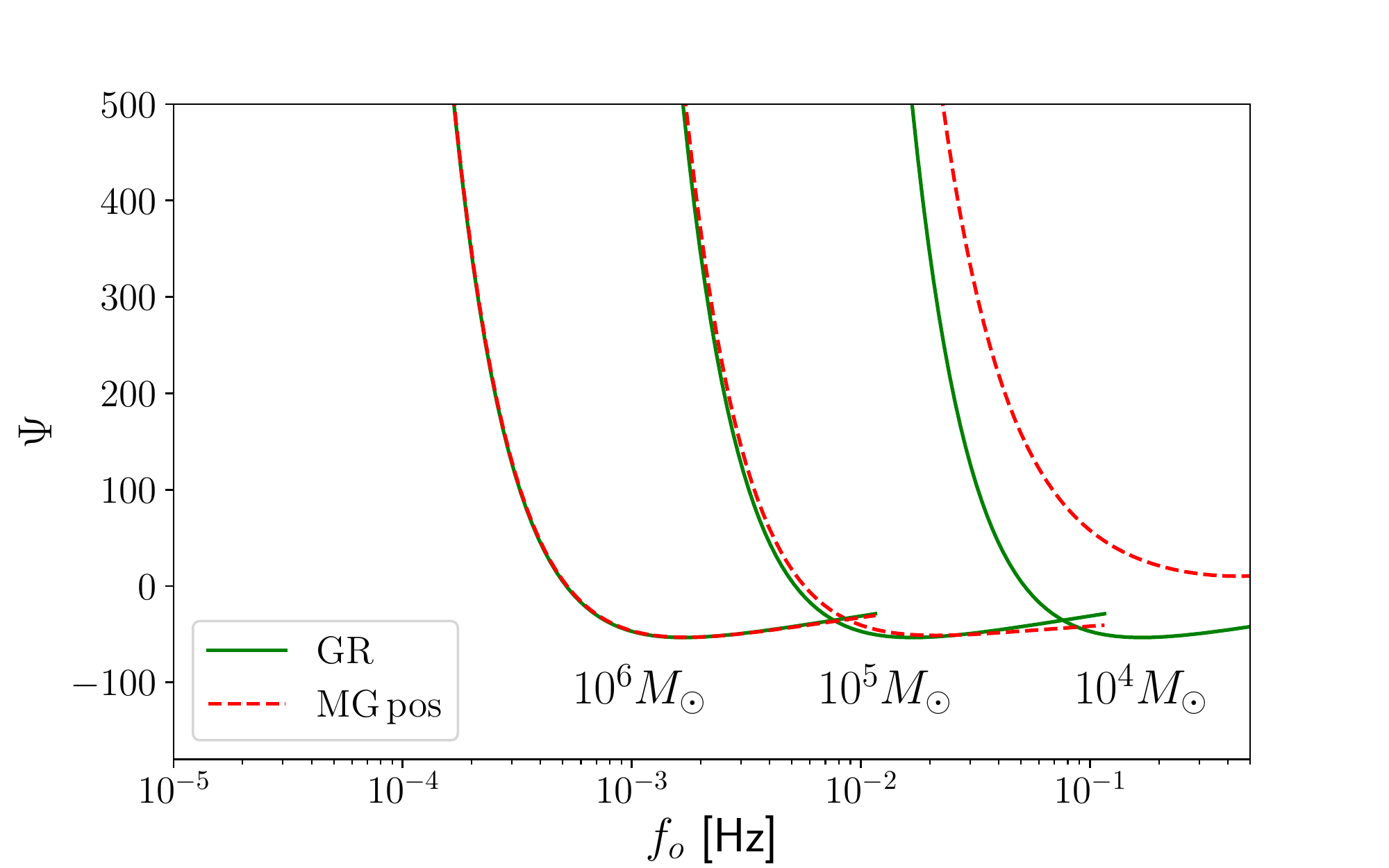}
    \includegraphics[width=0.49\textwidth]{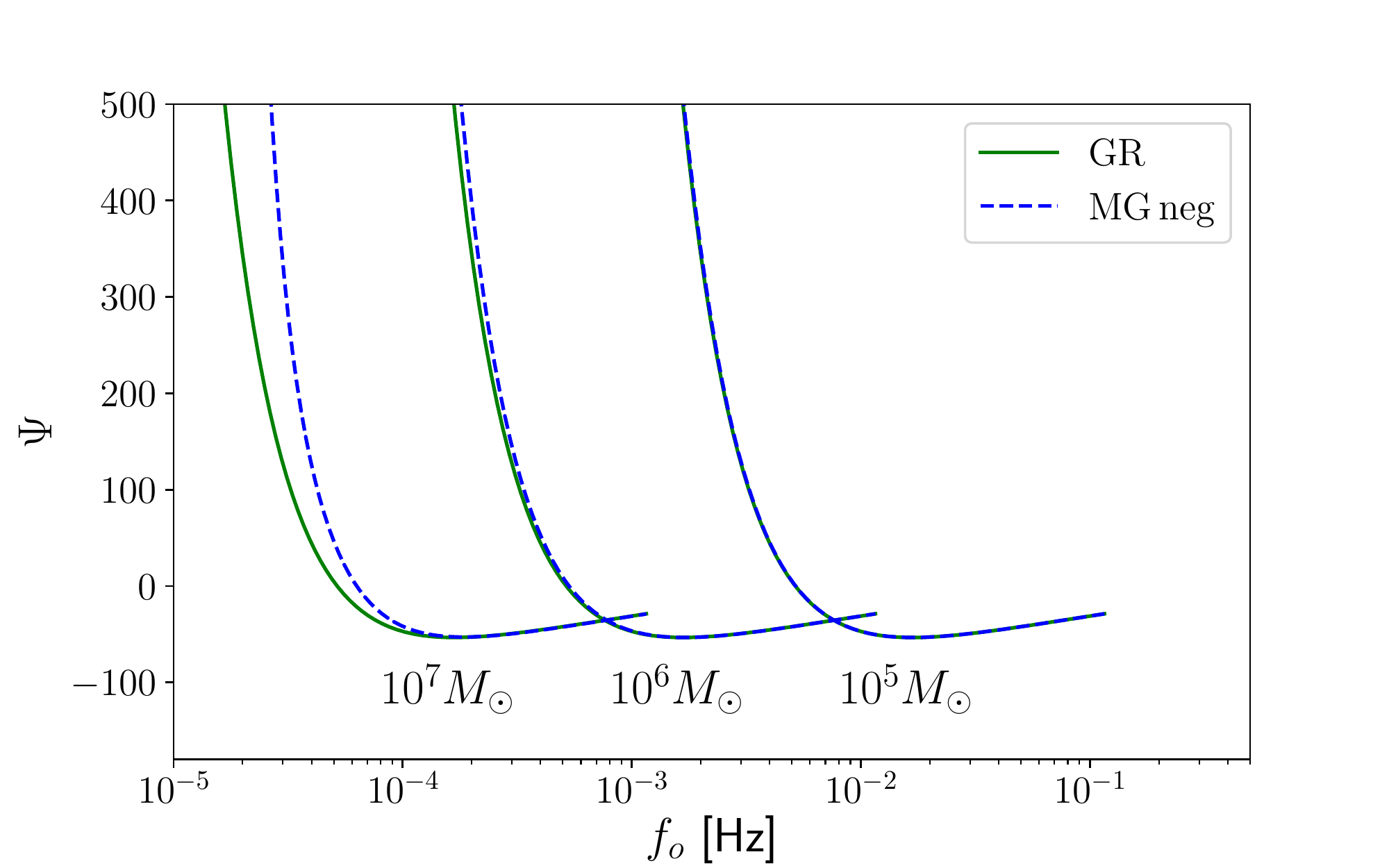}
    \caption{\small The phases in GR (green), the positive-power case (red) and the negative-power case (blue) for binaries with different total masses at $\ze=1$. We use exaggerated values $\beta_1=\beta_2=20$ for the positive-power case and $\beta_1=\beta_2=100$ for the negative-power case to visualize the modified gravity effect. We use $f_*=2$ Hz for the positive-power case, and $f_*=2\times10^{-7}$ Hz for the negative-power one. }
    \label{fig:phase}
\end{figure}

The modified early inspiral phase is dominated by the 0 PN term, and the late inspiral phase is dominated by the 2.5 PN terms, as in GR in the positive-power polynomial case. The scenario is the same in the negative-power case when $\beta_1$ and $\beta_2$ are small. However, the -1.5 PN and the -3 PN terms take over the early inspiral stage when $\beta_1$ and $\beta_2$ are large, as shown in Figure \ref{fig:PNterm_neg}. We plot the total phases of the inspiral waveform for different total masses in both polynomial cases in Figure~\ref{fig:phase}, with exaggerated $\beta_1$ and $\beta_2$, and appropriate $f_*$ as in Figure~\ref{fig:amp_pds}. Like Figure~\ref{fig:amp_pds}, Figure~\ref{fig:phase} shows deviations of the modified phases from their GR correspondences as $f_o$ approaches $f_*$. The deviations are larger for lighter binary systems in the positive-power case, and heavier binary systems in the negative-power case.

A feature this calculation highlights is that a power-law departure from $c_T=c$ leads to  a strong modification of gravity. The PN phase expansion operates naturally in powers of $f^{1/3}$, so corrections proportional to $f$ effectively `jump' three PN orders at once. It is for this reason that $\Psi^{\rm pos}$ is only minimally modified, and the corrections are rapidly pushed beyond $f_c$.

However, as we will see in \S\ref{sec:forecast}, detectability of these effects is determined by trade-off in number of binary orbits and SNR. High PN order corrections only dominate the GW phase for a very short period of time (few orbits), but they are also the regime in which the greatest SNR is accrued. For the negative power-law case, one  needs to measure hundred (thousands) of orbits over a timescale of months (years) to detect modifications to a noisy signal.

\subsubsection{EFT-inspired model}

\begin{figure}
\centering
    \includegraphics[width=0.49\textwidth]{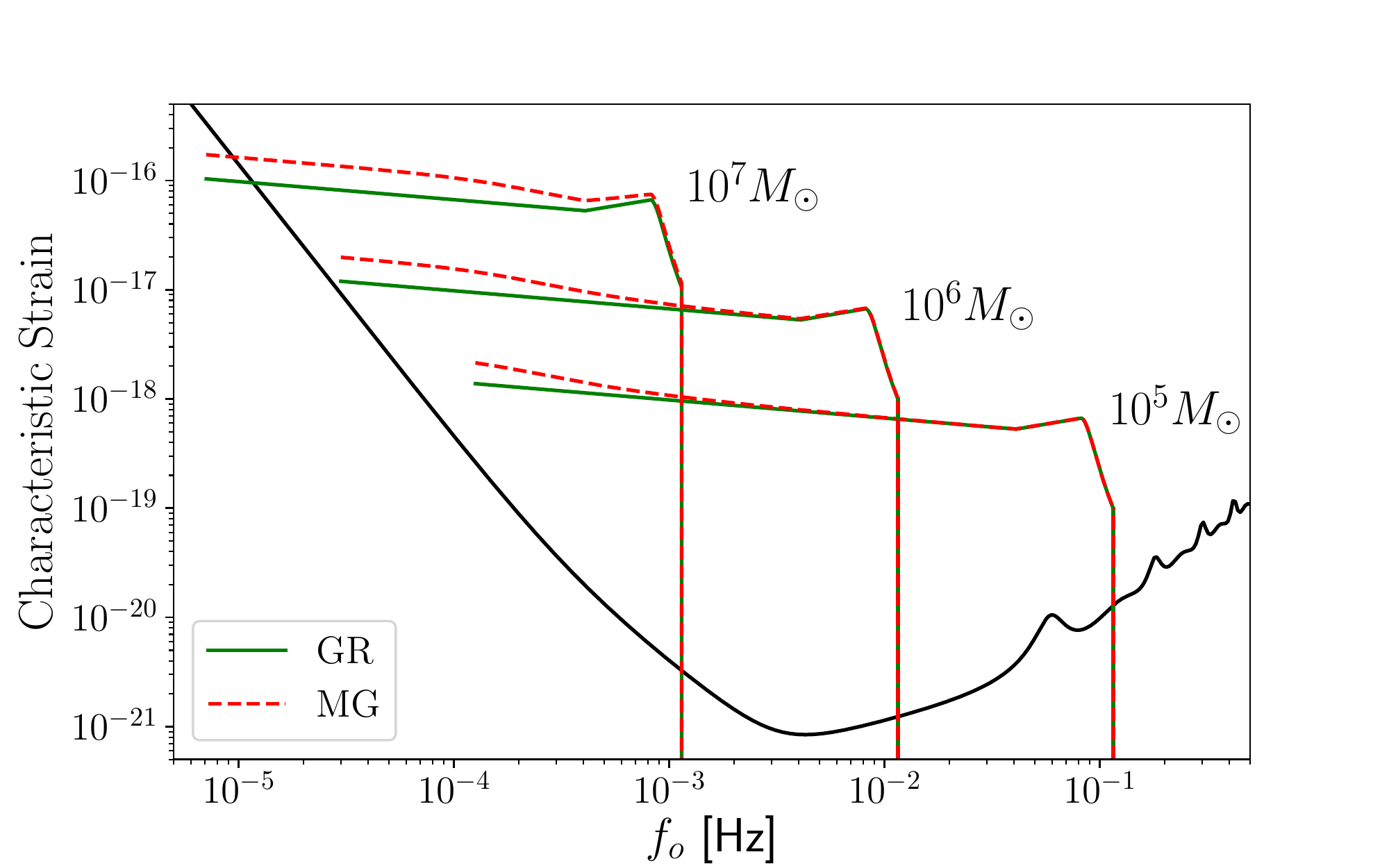}
    \includegraphics[width=0.49\textwidth]{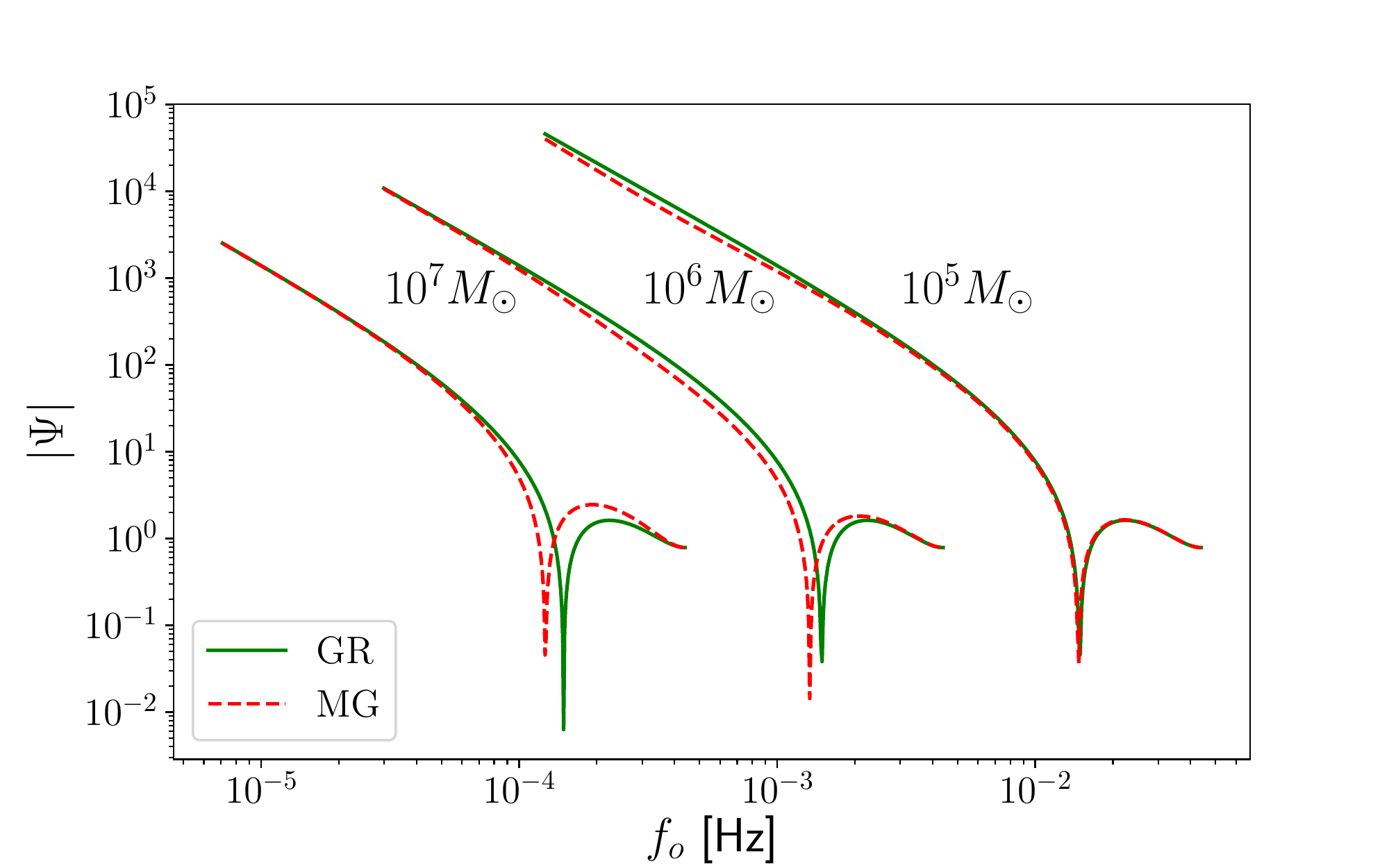}
    \caption{\small The characteristic strains and phases of the EFT Ansatz case (red) compared with GR (green) for binaries with different total masses at $\ze=1$. We use $c_0=0.6$ and $f_*=5\times10^{-4}$ Hz. }
    \label{fig:Delta_amp_phase}
\end{figure}

In the EFT Ansatz, $c_T(f_s)$ can be less than one  at low  frequency (i.e. we do not impose GR propagation near the source), so we need to work on the more general expression in eq.~(\ref{eq:df_dt_delta}). The integration of the time interval becomes
% Since $\Delta$ is computed in $f_s$ in the EFT case, we convert the expression with $f_o$ in eq. (\ref{eq:df_dt_delta}) to the one with $f_s$ as
% \begin{align}
%     \frac{df_o}{dt_o} = (1-\Delta)^2 \left(1 - \frac{f_s}{1-\Delta}\frac{\partial \Delta}{\partial f_s} \right) \frac{96}{5 \pi\,{\cal M}_z^2}\,u^{\frac{11}{3}}\Bigg[1 + \psi_1 u^{\frac{2}{3}} + \psi_{ 1.5}u + \psi_{2}u^{\frac{4}{3}} + \psi_{2.5}u^{\frac{5}{3}} \Bigg].
% \end{align}
\begin{eqnarray}
    t_o - t_c &=& \frac{5 \pi\,{\cal M}_z^2}{96} \int_{f_c}^{f_o} \frac{1}{(1-\Delta)^2} \left(1 + \frac{\tilde{f}_o}{1-\Delta} \frac{ \partial \Delta}{\partial \tilde{f}_o} \right) \tilde{u}^{-\frac{11}{3}}\nonumber\\
    && \times \Bigg[1 + \psi_1 \tilde{u}^{\frac{2}{3}} + \psi_{ 1.5}\tilde{u} + \psi_{2}\tilde{u}^{\frac{4}{3}} + \psi_{2.5}\tilde{u}^{\frac{5}{3}} \Bigg]^{-1} d\tilde{f}_o.
\label{eq:to_Delta}
\end{eqnarray}
The phase is then computed as :
\begin{align}
\Psi(f_o)&=2\pi\int_{f_{c}}^{f_o}[t_o(\tilde{f}_o)-t_{c}]\,d\tilde{f}_o - \frac{\pi}{4}\,. %\nonumber\\
% & = 2\pi\int_{u_{c}}^{u_o} \frac{1-\Delta}{\pi {\cal M}_z}\left(\frac{1}{1+\frac{f_o}{1-\Delta}\frac{\partial \Delta}{\partial f_o}}\right) [t_o(u)-t_{c}]\,du - \frac{\pi}{4}
\label{eq:Psi_Delta}
\end{align}
We evaluate the derivatives of $\Delta$ with respect to $f_o$ numerically, and compute the phase by numerically integrating eqs. (\ref{eq:to_Delta}) and (\ref{eq:Psi_Delta}). The amplitudes and the phases for systems with different total masses in the EFT Ansatz with $f_*=5\times10^{-4}$ Hz are plotted in Figure \ref{fig:Delta_amp_phase}. Recall that in this model, $f_*$ sets the position of the rapid growth of $c_T(f)$. We notice that at frequencies much higher than $f_*$, both amplitudes and phases are the same as in GR. The modified gravity effects start to become manifest when the observed frequency approaches $f_*$, resulting in a different $c_T$ at the source and observer.  
%\
 The modified amplitudes show constant offsets from their GR equivalences at low frequencies much smaller than $f_*$. This is because the comoving distance is modified by a factor of $c_T(f)$, and $c_T(f) \simeq c_0$ when $f\ll f_*$, so that the amplitude is suppressed by a factor of $1/c_0$ at low frequencies. The modified phase for the high mass system seems equivalent to GR values, but actually the deviation from GR of the phase does not vanish at low frequencies. The weakening of the deviation in the figure is caused by the fact that the deviation becomes less significant compared to the large values of the phase at low frequencies.

\subsection{IMR extension}
\label{sec:IMR_extension}

The inspiral waveform we have discussed above starts to become invalid above $f_{c}\sim2f_{\rm ISCO}$. In GR, extended template waveforms that include the complete Inspiral-Merger-Ringdown (IMR) phases can be obtained from numerical studies of binary black hole mergers. Here we lay out an approximate analytic prescription for adding the merger and ringdown to the inspiral waveforms obtained in the previous section. Note that this is possible because we are \textit{not} modifying the intrinsic strong-field dynamics of the source, but only modulations that affect its propagation.

However, our treatment involves some degree of approximation in the merger phase, the full extent of which can only be tested with dedicated with numerical relativity simulations. For this reason, in \S\ref{sec:forecast} we will present results using both the inspiral-only and full IMR waveform described below; these can be considered conservative and optimistic versions of our analysis, respectively.

We adopt a modified version of the frequency-domain PhenomA waveform from Ajith et al. \cite{Ajith:2007kx}. As our work neglects component spins, we do not require a more sophisticated waveform such as PhenomD \cite{Khan:2015jqa}. By fitting to a suite of numerical relativity simulations, the amplitude of the PhenomA waveform in GR is constructed piecewise as:
\begin{align}
&A_{\rm ins}(f) = C \left( \frac{f}{f_{\rm merg}} \right)^{-\frac76}, \label{Ains}\\
&A_{\rm merg}(f) = C \left( \frac{f}{f_{\rm merg}} \right)^{-\frac23}, \label{Amerg}\\
&A_{\rm ring}(f) = C \omega {\cal L}(f, f_{\rm ring}, \sigma),\label{Aring}
\end{align}
where the prefactor is
\begin{equation}
C = \sqrt{\frac{5}{24}} \pi^{-\frac23} \frac{{\cal M}_z^{\frac56}}{\dl}  f_{\rm merg}^{-\frac76},
\end{equation}
and
\begin{equation}
\omega = \frac{\pi \sigma}{2} \left(\frac{f_{\rm ring}}{f_{\rm merg}} \right)^{-2/3}.
\end{equation}
The Lorentzian function in the ringdown phase reads
\begin{equation}
{\cal L}(f, f_{\rm ring}, \sigma) = \left(\frac{1}{2\pi} \right) \frac{\sigma}{(f-f_{\rm ring})^2 + \sigma^2/4}
\end{equation}
The ends of inspiral, merger and ringdown phase are marked by $f_{\rm merg}$, $f_{\rm ring}$ and $f_{\rm cut}$ respectively. Their values, along with the value of $\sigma$, are computed via expressions of the form $f_k=(a_k \eta^2 + b_k \eta + c_k)/\pi M_{\rm tot}$, with the values of $a_k$, $b_k$ and $c_k$ are read from Table I in \cite{Ajith:2007kx}. One can verify the continuity of the above amplitudes across the IMR phase boundaries. We have verified that our computations in \S\ref{sub_lDGWa} \& \S\ref{subsec:waveform_phase} match the inspiral section of the PhenomA template when $c_T=1$, up to 2.5 PN order.

Since we are studying the effects of modified GW propagation, we assume that the PhenomA waveform in GR is valid \textit{at} the source. This implies the modification of the inspiral amplitude we derived in \S\ref{sub_lDGWa} -- which accounts for effects of varying $c_T$ during the propagation to the observer -- can be applied to all three pieces of the PhenomA amplitude. This allows us to replace
%\begin{align}
%&A_{\rm ins}^{\rm MG}(f_o) = C^{\rm MG} \left( \frac{f_o}{f_{\rm merg}} \right)^{-7/6}, \\
%&A_{\rm merg}^{\rm MG}(f_o) = C^{\rm MG} \left( \frac{f_o}{f_{\rm merg}} \right)^{-2/3}, \\
%&A_{\rm ring}^{\rm MG}(f_o) = C^{\rm MG} \omega {\cal L}(f_o, f_{\rm ring}, \sigma),
%\end{align}
%where
\begin{equation}
C\rightarrow C^{\rm MG} = \sqrt{\frac{5}{24}} \pi^{-\frac23} \frac{{\cal M}_o^{\frac56}}{\dl^{\rm GM}} \left(\frac{c_T(f_o)}{c_T(f_s)}\right)^{\frac32} f_{\rm merg}^{-\frac76},
\end{equation}
in eqs.~(\ref{Ains})--(\ref{Aring}). This produces the amplitude of the modified PhenomA template we use in our analysis. 

For the phase: we already have the modified phase for the inspiral from our previous computations in \S\ref{subsec:waveform_phase}. For the merger part of the signal, there is no straightforward prescription for how to adapt the GR PhenomA phase to our modified scenario. Hence, we will discard the merger phase of the PhenomA and replace it as follows.  

We will expand the phase as a power series in the frequency variable $u = \pi {\cal M}_o f_o$. This step is analogous to what is done in the Parameterized Post-Einsteinian (ppE) framework \cite{Yunes:2009ke,Yunes:2016jcc}, which describes deviations from GR in the waveform (see \S\ref{app_ppEmap} for further details on ppE). For prompt mergers, it is sufficient to truncate the series at linear order \cite{Yunes:2009ke}:
\begin{equation}
\Psi_{\rm merg}(f) = \bar{\Psi}_c + 2\pi \bar{t}_c f\,.
\label{eq:merger_psi}
\end{equation}
where $\bar{t}_c$ and $\bar{\Psi}_c$ are new constants. These are determined by enforcing the continuity of the the phase and its derivative at $f_{\rm merg}$, that is, $\Psi_{\rm ins}(f_{\rm merg})=\Psi_{\rm merg}(f_{\rm merg})$.
Explicitly this fixes the barred quantities to 
\begin{align}
  & \bar{\Psi}_c = \Psi_{\rm ins}(f_{\rm merg}) - 2\pi \bar{t}_c f_{\rm merg} \\
    & \bar{t}_c = \frac{1}{2\pi}\frac{d\Psi_{\rm ins}}{df} \Bigg|_{f=f_{\rm merg}}\,.
\end{align}
Hence the merger phase is fully determined by consistency with our inspiral calcuations. We will set the phase during the ringdown epoch to zero, as is done in ppE \cite{Yunes:2009ke} (modelling of quasinormal modes during ringdown lies beyond the scope of this work). Since the modified PhenomA waveform allows us to use the full GW signal, the SNR of our detections will increase. In addition, more cycles in the late inspiral epoch, and a merger era in the phase are included. These help tighten the constraints on our modified gravity parameters. We expect the additional amplitude information during the ringdown to be only weakly constraining.

%%%%%%%%%%%%%%%%%%%%%%%%%%%%%%%%%%%%%%%%%%%%%%%%%%%%%%%

\section{GW data analysis}
\label{sec:dataanalysis}

We employ data analysis techniques  
in order to find constraints both on GR quantities and on modified gravity parameters impacting the amplitude and phase of the waveform. In this section, we give a brief review of the process of measuring GWs and how to constrain parameters controlling sources and GW propagation. We start  the discussion in   general terms (i.e., we will not focus on any particular detector).  Then, in order to specialize our analysis to the case of LISA, we   choose the noise model and a particular set of sources. Since in the following sections we are interested in producing Fisher forecasts on a given set of parameters, we will conclude this section by comparing this approach to a Monte Carlo Markov Chain (MCMC) exploration of the parameter space. By construction, Fisher forecasts give the correct constraints on the model parameters assuming that the likelihood is Gaussian around the best fit, and that injected values are exactly recovered by the parameter estimation procedure. Since these assumptions are not trivially satisfied (for example one necessary condition for these to be true is for the signal to be sufficiently loud) and the Fisher approach can be misleading~\cite{Vallisneri:2007ev}, we perform this test to validate results of our Fisher analysis. In particular, we are able to define approximate thresholds  (corresponding to identifying regions of the parameter space) where we can assume our Fisher forecasts to be trustworthy. Readers eager to see the constraints themselves could skip ahead to \S\ref{sec:forecast}.  

\bigskip

%\subsection{Modeling the data}
Let us consider a single detector data stream which has already been cleaned from glitches and from other non-stationary effects beyond the GW signal we are interested in measuring. Under these assumptions, the data stream $d(t)$ in time domain can be expressed as a combination of signal $s(t)$ and noise $n(t)$ as $d(t) = s(t) + n(t)$. By considering a data segment of length $T$ (the observation time can be chosen appropriately for each source we will try to resolve), we can then perform a Fourier transform\footnote{In reality the data are sampled with some finite sampling rate, so that rather than a continuous function of $t$ they would be a discrete series of points and integrals would have to be replaced with sums. In order to keep the notation simpler, we proceed by ignoring this matter.} to express the data in frequency domain as
\begin{equation}
\tilde{d}\left(f\right) =  \int_{-T/2}^{T/2} d t  \;\textrm{e}^{ 2\pi i f t} d \left(t\right)\,.
\end{equation}
Let us proceed by assuming that the noise is Gaussian and with zero mean. If $n(t)$ is also stationary, then the ensemble average of its Fourier modes $\tilde{n}(f)$  obeys
\begin{equation}
\label{eq:singlesidedspn}
	\langle \tilde{n}(f) \tilde{n}^*(f') \rangle \equiv \frac{1}{2} \delta \left(f-f'\right) N\left(f\right) \; , 
\end{equation}
where $ N\left(f\right)$ is the single-sided noise power spectrum\footnote{The factor $1/2$ is introduced by convention in order to keep track of the fact that $N(f)$ is defined only for positive (physical) frequencies.} and, like $\tilde{n}(f)$, it has dimension Hz${}^{-1}$. $N\left(f\right)$ is real and positive and, since $d(t)$ is real, it obeys $N(f)=N(-f)$. While the noise is a stochastic variable with zero mean and some variance, if a resolvable GW signal is present we have $\langle \tilde{d}(f)  \rangle = \langle \tilde{s}(f) \rangle$. This means that we can write a likelihood defined using the standard matched filtering techniques, to describe the noise residuals as
\begin{equation}
    -2 \ln \mathcal{L} = \left(  \tilde{d} - \tilde{s}^{\rm th} (f, \vec{\theta}) |  \tilde{d} - \tilde{s}^{\rm th} (f, \vec{\theta})\right) \simeq  \left( s^{\rm th} (f, \vec{\theta}) | \tilde{s}^{\rm th} (f, \vec{\theta}) \right) - 2 \left(  \tilde{d} | \tilde{s}^{\rm th} (f, \vec{\theta}) \right)\; ,
  \end{equation} 
  where $\tilde{s}^{\rm th} (f, \vec{\theta})$ denotes the theoretical model for the signal (which at this level still contains both the waveform and the detector response function) which depends on a set of parameters $\vec{\theta}$, and $\left( a | b \right)$ denotes the noise weighted inner product:
  \begin{equation}
    (a|b) = 2 \int_{f_1}^{f_2} \frac{a(f)b^*(f)+a^*(f)b(f)}{N(f)} \; \textrm{d} f \; .
\end{equation}
Since in the following we will not be interested in estimating the source sky localization parameters, we can use a sky-averaged detector response function\footnote{This is equivalent to averaging over many realisations of the data with random source positions.}; for LISA, this contributes an overall factor of $\sim10/3$ to the strain sensitivity curve, see \cite{Robson:2018ifk}. After factoring this averaged detector response function out of the data (which are redefined as $\tilde{\bar{d}}_i$), we can define our likelihood as
\begin{equation}
\label{eq:our_likelihood}
    -2 \ln \mathcal{L} \simeq  \left( \tilde{h}^{\rm th} (f, \vec{\theta}) | \tilde{h}^{\rm th} (f, \vec{\theta}) \right) - 2 \left(  \tilde{d} | \tilde{h}^{\rm th} (f, \vec{\theta}) \right)\; ,
  \end{equation} 
where $\tilde{h}^{\rm th} (f, \vec{\theta})$ is the theoretical model for the GW waveform, and the inner product is now weighted with  the detector strain sensitivity $S_{n}(f)$. It is straightforward to show that the parameters $\vec{\theta}_0$ that maximize this likelihood are defined by solving:
\begin{equation}
    \left( \tilde{h}^{\rm th} (f, \vec{\theta})   \left|   \frac{ \partial \tilde{h}^{\rm th} (f, \vec{\theta}) }{ \partial \theta_l
    } \right. \right) =   \left(  \tilde{d}  \left|   \frac{ \partial \tilde{h}^{\rm th} (f, \vec{\theta}) }{ \partial \theta_l
    } \right.   \right)\; .
  \end{equation} 
We proceed by introducing the Fisher matrix, which provides a Gaussian approximation of the likelihood around its maximum, which is defined as:
\begin{equation}
   F_{lk} \equiv \left.  - \frac{ \partial^2 \ln \mathcal{L} }{\partial \theta_l \, \partial \theta_k} \right|_{\vec{\theta} = \vec{\theta}_0 } = \left.  \left( \left. \frac{ \partial \tilde{h}^{\rm th} (f, \vec{\theta}) }{ \partial \theta_l } \right| \frac{ \partial \tilde{h}^{\rm th} (f, \vec{\theta}) }{ \partial \theta_l } \right)  \right|_{\vec{\theta} = \vec{\theta}_0 } \; .
\end{equation}
As customary, the confidence intervals on the model parameters $\vec{\theta}$ can be drawn from the covariance matrix $C_{lk}$ which is obtained by inverting $F_{lk}$. \\

We only consider a subset of the parameters which are expected to significantly impact the GW waveform. In particular, we focus on five parameters common with  GR: $\ln \eta$, $\ln {\cal M}_z$, $\ln z$, $t_c$ and $\Psi_c$ defined as in~\S\ref{sec:waveforms}, meaning we will not consider the sky localization, the inclination nor the orientation of the binary as well as the spins of the two objects. Beyond these five parameters, we have the a set of modified gravity parameters $\vec{\theta}_{\rm MG}$, which can either represent the $\beta_1$ and $\beta_2$ defined as in eq.~\eqref{eq:cT_u} or alternatively the $f_*$, $c_0$ of eq.~\eqref{cTexpA}, so that our full parameter vector is:
\begin{equation}
\vec{ \theta} \equiv \{\ln \eta, \ln {\cal M}_z, \ln z, t_c, \Psi_c, \vec{\theta}_{\rm MG} \} \; .
\end{equation}
In the following, we will restrict our analysis to the case of LISA, so that the noise power spectra and the strain sensitivity can be taken for example from~\cite{Caprini:2019pxz, Flauger:2020qyi, Babak:2021mhe}. Since we are using a sky-averaged response function, we can also consider static arms lengths only and and do not require orbital information. We will adopt a procedure similar to the one employed in Sec.\ 2.4 of~\cite{Flauger:2020qyi}; we will use an effective combination of the three TDI channels~\footnote{For stochastic gravitational wave backgrounds the signal is quadratic in the data, so that the variance is quadratic in the noise PSD. On the other hand, for resolvable sources the signal is linear in the data and variance is thus linear in the noise PSD. This motivates the slightly different form of eq.~\ref{eq:effective_noise} from \cite{Flauger:2020qyi}.} in the AET basis (see for example~\cite{Hogan:2001jn}), where the noise is diagonal, defined as:
\begin{equation}
    \label{eq:effective_noise}
    \frac{1}{ S_{\rm eff} } \equiv \frac{1}{S_A} + \frac{1}{S_E} + \frac{1}{S_T} \; .
\end{equation}
Notice that this is formally equivalent to combining three likelihoods with the form given in eq.~\eqref{eq:our_likelihood}. We are assuming here that measurements in the three independent channels do not break any degeneracies among the signal or noise parameters. This is not strictly true, as in general the three detectors have different angular responses, and this could help to break degeneracies in, for example, sky localization. As a consequence, our assumption can be seen as a pessimistic one, and the constraints could be improved in a fully consistent, though more complicated, analysis which is postponed to future works on this topic.

We conclude this section by presenting a direct comparison between the constraints on $\vec{\theta}$ obtained from $F_{lk}$ and the ones obtained by directly evaluating eq.~\eqref{eq:our_likelihood}. This comparison is carried out for two sources: a high SNR case, where the two approaches are expected to match exactly and a low SNR one, where the real constraints obtained from the MCMC sampling of eq.~\eqref{eq:our_likelihood} are expected to start diverging from the Fisher ones. In both cases, we start by estimating the in-band time using~\cite{Maggiore:2007ulw} 
\begin{equation}
    T  = \int_{f_i}^{f_c} \left[ \frac{96 \, \pi^{8/3} }{5} \left( \frac{G \mathcal{M}_z}{c^3} \right)^{\frac{5}{3}} f^{\frac{11}{3}} \right]^{-1} {\rm d} f \; ,
\end{equation}
where $f_i$ is the initial frequency (i.e., the smallest frequency in the LISA band at which the source emits) and $f_c$ is the cutoff frequency. In practice, since most of the SNR comes from the high-frequency part of the signal, we cut the low-frequency part of the GW spectrum by choosing $f_i$ so that $T\simeq 10$ days\footnote{The main reason for this choice is that reducing the total observation time corresponds to reducing the frequency resolution. In practice this leads to a faster evaluation of the waveform and thus of the likelihood.}. We then generate a Gaussian realization of the noise on top of which we inject the signal, given by $h^{\rm th} (f, \vec{\theta})$, to get the $\tilde{d}$ to be used to evaluate eq.~\eqref{eq:our_likelihood}. The sampling of the parameter space is performed using~\texttt{Polychord}~\cite{Handley:2015fda, Handley:2015vkr} via its interface with~\texttt{Cobaya}~\cite{Torrado:2020dgo}. \\

\begin{figure}
    \centering
    \includegraphics[width=1.0\textwidth]{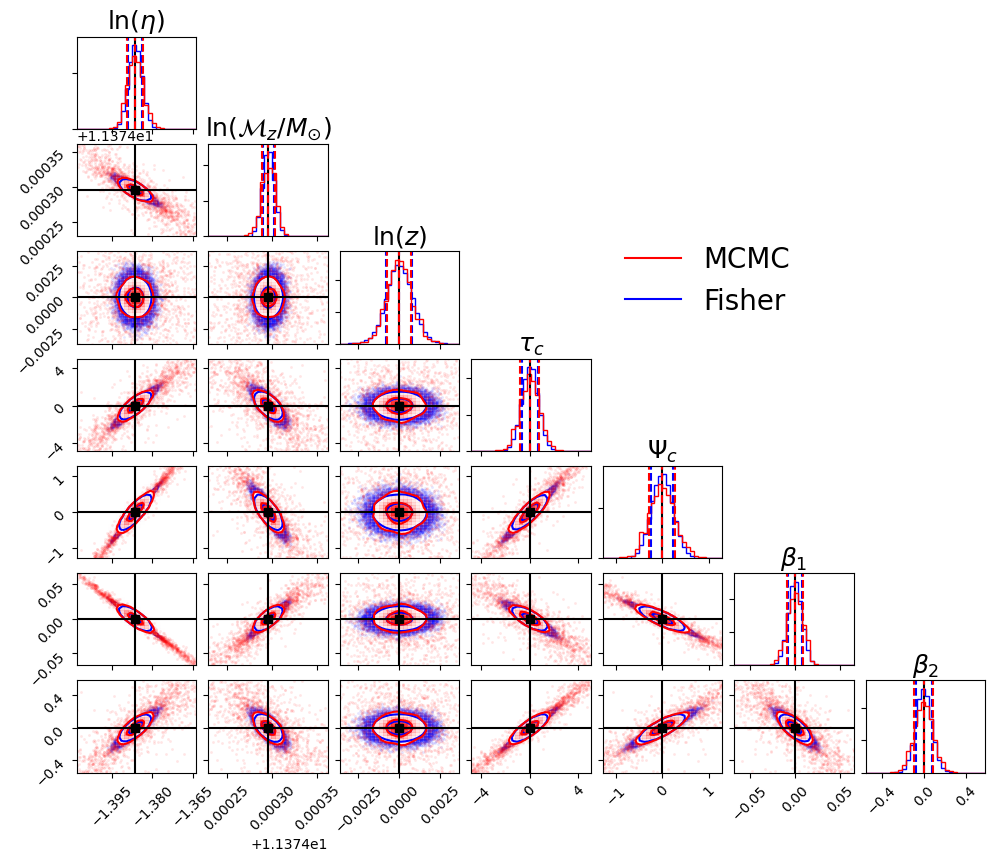}
    \caption{\small Comparison of MCMC and Fisher constraints for large ($\simeq 1020$) SNR. The values of injected parameters (not including redshift effects) are  $M_{\rm tot} = 10^5 M_{\odot}$, $\eta = 0.25$, $z=1$, $\tau_c=0$, $\Psi_c = 0$, $\beta_1 =0$, $\beta_2 = 0$. The positive-power polynomial Ansatz for $c_T(f)$ is used, with $f_*=0.1$ Hz. (Note that we have sampled the Fisher matrix here simply for convenience, they should be considered as perfect ellipses.)}
    \label{fig:SNR_high}
\end{figure}

\begin{figure}
    \centering
    \includegraphics[width=1.0\textwidth]{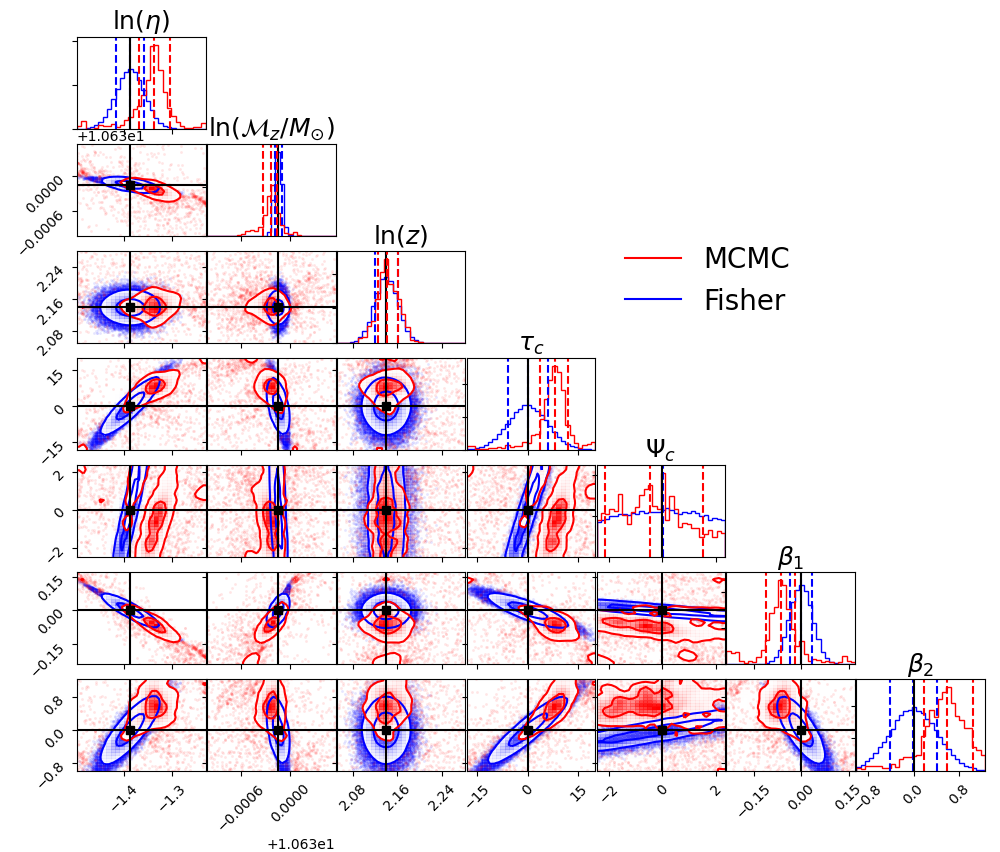}
    \caption{\small Comparison of MCMC and Fisher constraints for small ($\simeq 42$) SNR. The values of injected parameters (not including redshift effects) are $M_{\rm tot} = 10^4 M_{\odot}$, $\eta = 0.25$, $z=8.5$, $\tau_c=0$, $\Psi_c = 0$, $\beta_1 =0$, $\beta_2 = 0$. The positive-power polynomial Ansatz for $c_T(f)$ is used, with $f_*=0.1$ Hz. (Note that we have sampled the Fisher matrix here simply for convenience, they should be considered as perfect ellipses.)}
    \label{fig:SNR_low}
\end{figure}

In Figure~\ref{fig:SNR_high} we show the comparison between our Fisher forecast and a full MCMC sampling of the parameter space for a loud source with SNR $\simeq 1020$. It is manifest that for this event the two approaches lead to very similar results, confirming the validity of the Fisher approximations for similarly loud signals. On the other hand, in Figure~\ref{fig:SNR_low}, we compare the Fisher forecasts with the MCMC constraints for an event with SNR $\simeq 42$. Since in this case the event is much fainter, the validity of the Fisher approximation starts to break. As is clearly visible from the 1-dimensional marginalized constraints, the order of magnitude of the forecasted error bars are still accurate for all the parameters. However, we notice some deviations between the Fisher analysis and the real structure of the parameter space. This can be appreciated for example from displacement between the injected parameters (\emph{i.e.} the centers of the Fisher ellipses) and the best fit values recovered through the MCMC procedure. Nevertheless, since the recovered values are always $2\sigma$-compatible with the injected values, these displacements should not be interpreted as problematic. Events with similar SNR should be considered as thresholds to assume the Fisher forecasts are sufficiently robust.

%%%%%%%%%%%%%%%%%%%%%%%%%%%%%%%%%%%%%%%%%%%%%%%%%%%%%%%
\section{Forecasts}
\label{sec:forecast}
%Having validated
After confirming that Fisher forecasts give satisfactory results  for the parameters of interest, we focus on a Fisher analysis in what follows.   We will present forecasts for how well a four-year LISA mission can constrain both the polynomial and EFT-inspired Ans\"atze for $c_T(f)$ described in \S\ref{sec-theory-ansatze}, both with a single MBH merger (\S\ref{subsec:inspiral}) and a population of MBH mergers (\S\ref{subsec:results_IMR}).

\begin{figure}
    \centering
    \includegraphics[width=0.48\textwidth]{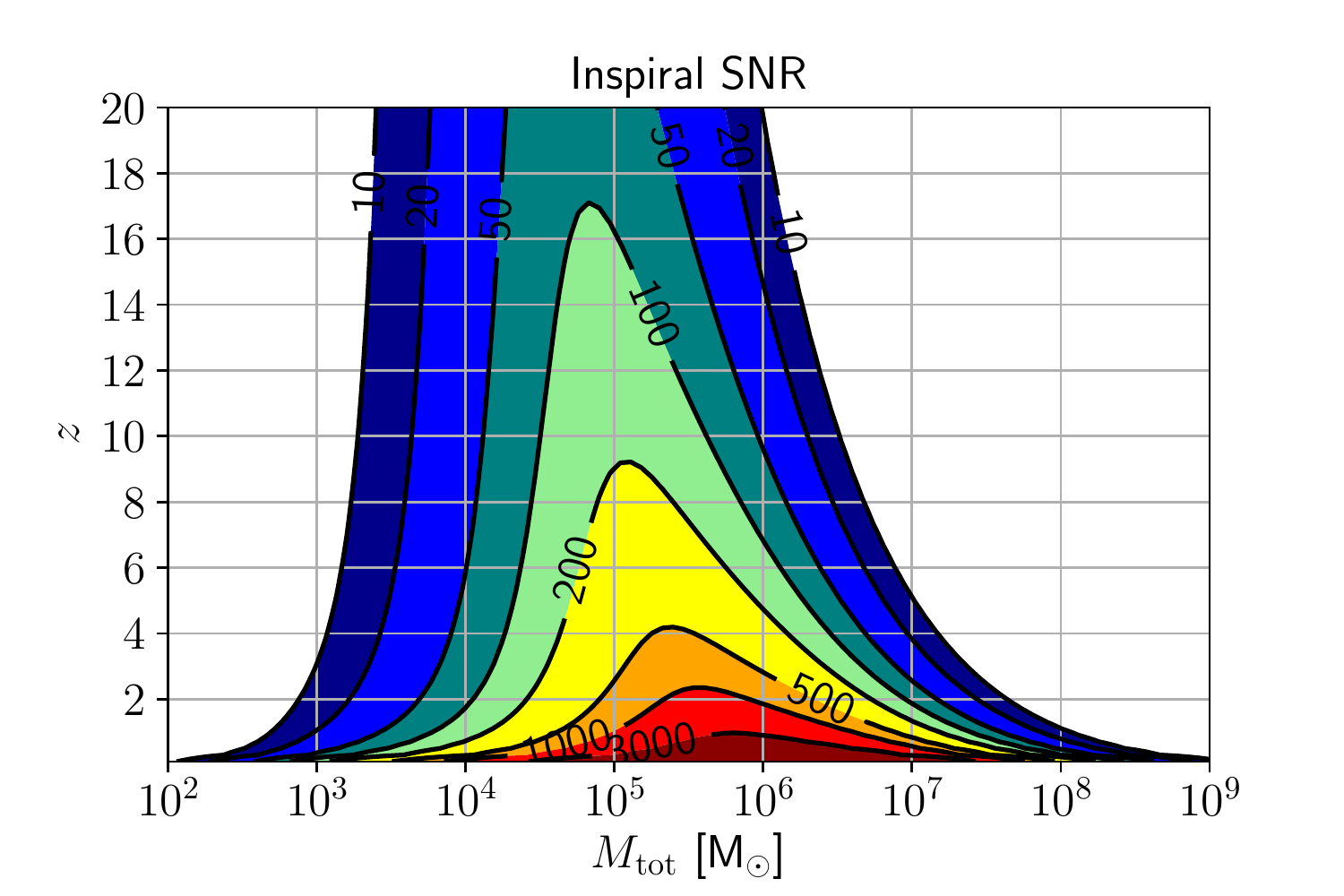}
    \includegraphics[width=0.48\textwidth]{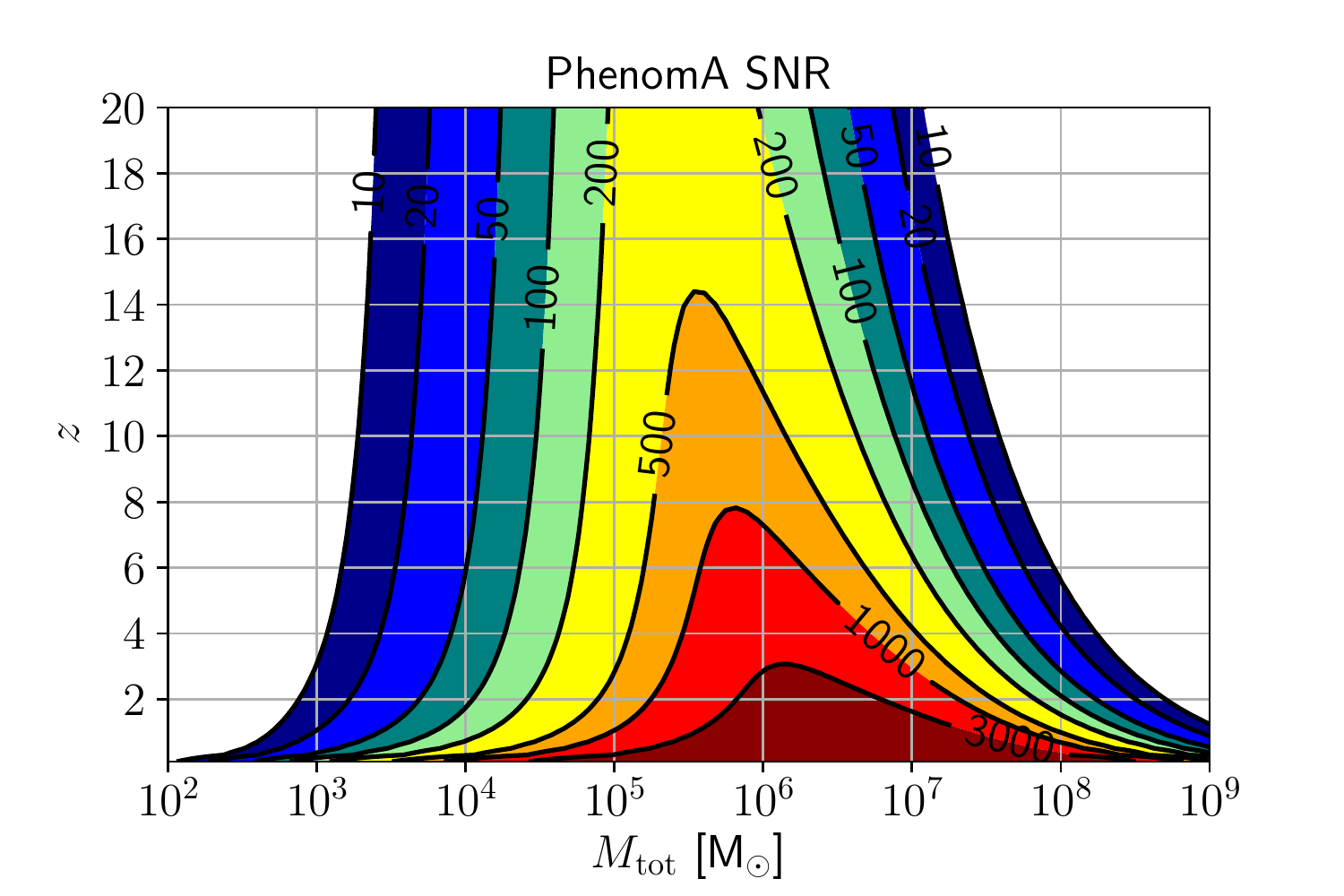}
    \caption{\small The SNR contours in the space of total mass and redshift for the inspiral-only waveform (left panel) and the entire PhenomA waveform (right panel).}
    \label{fig:waterfall}
\end{figure}

Naively one might expect that the best constraints on our modified gravity parameters will be obtained from systems with the highest total SNR. Figure~\ref{fig:waterfall} displays SNR contours for LISA detections within GR in terms of MBH total mass and redshift, using only the inspiral portion of the signal (left panel) and full inspiral-merger-ringdown signal (right panel). We notice that systems between $10^5$ and $10^7~M_\odot$ provide the highest SNR detections in both cases. We will show that these are \textit{not} necessarily the optimal systems for bounding $c_T(f)$, due to the frequency-dependent nature of our corrections.

\subsection{Inspiral-only results}
\label{subsec:inspiral}
%\\

\begin{figure}[h!]
    %\centering
    \includegraphics[width=1.1\textwidth]{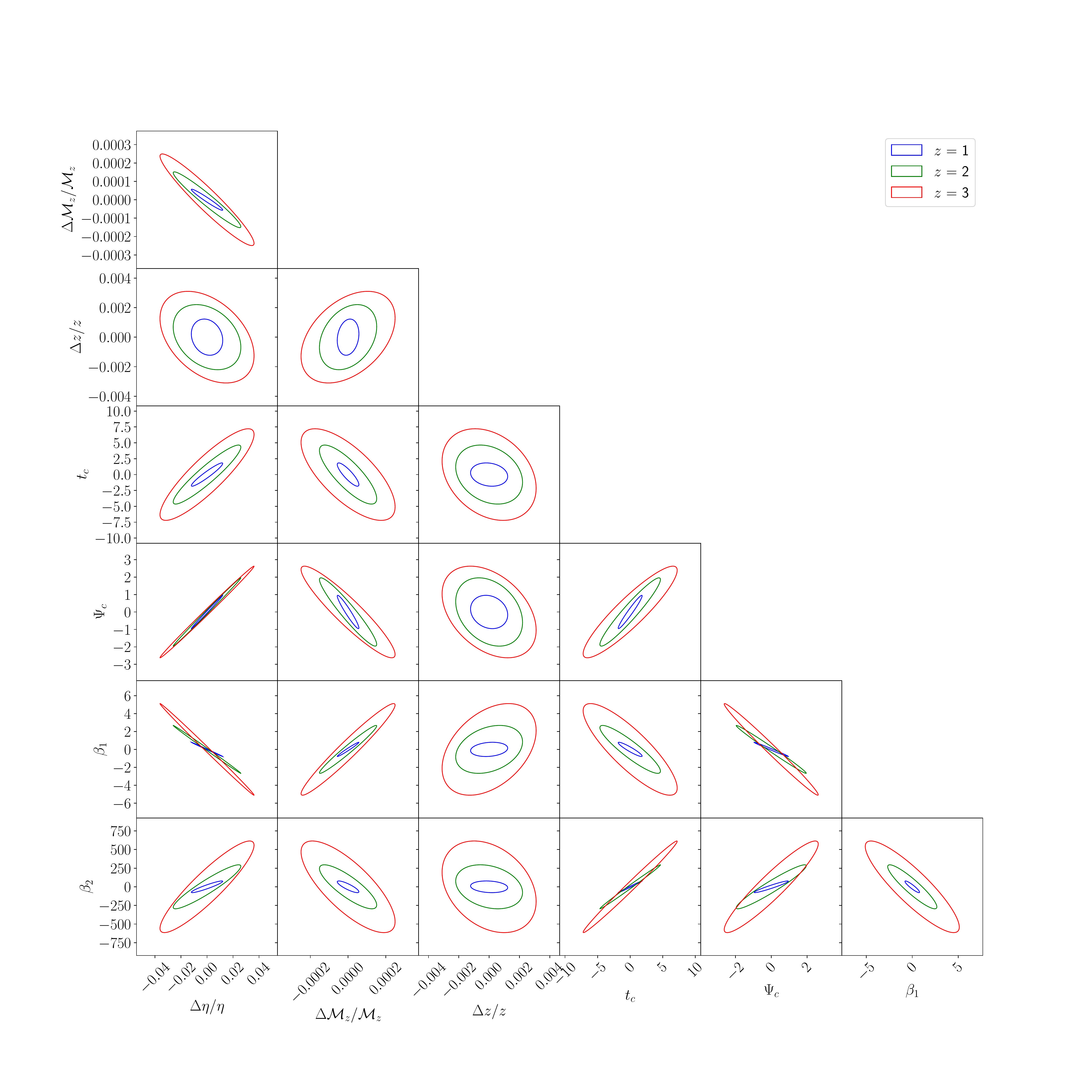}
    \caption{\small Forecast constraints for the positive-power polynomial case with $M_{\rm tot}=10^5~M_\odot$, $\eta=0.25$, $t_c=0$, $\Psi_c=0$, $f_*=2$ Hz, $\beta_1=0$ and $\beta_2=0$ at different redshifts using the inspiral waveform. We show the fractional constraints on $\eta$, ${\cal M}_z$, $z$ centered at 0, and the constraints on the rest of the parameters centered at their fiducial values. 
    }
    \label{fig:ellipse}
\end{figure}

\subsubsection*{Polynomial model}
We consider MBH binary mergers with total masses between $10^4$ and $10^7~M_\odot$, at redshifts of $z\,=\,1,~2,~3$ for each mass.
We aim to understand
  the effect of total mass and SNR on the parameter constraints. The component masses in these binary systems are equal, so $\eta=0.25$. We perform Fisher matrix analysis with the modified inspiral waveform, using 30 days of observation time. Although some sources will be detectable earlier than this (particularly when re-processing the data post-merger), the SNR of typical sources starts to rise steadily above $\sim 10$ at $\sim 30$ days before merger, and accumulates most of its final value after this \cite{Mangiagli:2020rwz}.
 
 As with the phase computation of \S\ref{subsub:polynomial_phase}, an advantage of our polynomial Ansatz is that it allows us fully analytical calculations. For this model we compute analytic derivatives of the waveform with respect to the parameters; we then verified our results with numerical derivatives. When computing the Fisher matrix, we put flat priors on $t_c$ in $(-50,50)$, $\Psi_c$ in $(-\pi,\pi)$, $\beta_1$ in $(-20,20)$, and $\beta_2$ in $(-1000,1000)$. As per the discussion of \S\ref{sec-theory-ansatze}, we fix $f_*=2$~Hz for the positive-power case, and $f_*=2\times10^{-7}$ Hz for the negative-power case. However, we note again that the forecasted constraints on $\beta_n$ can be translated to other values of $f_*$ (Appendix~\ref{sec:luminalrecovery}). Given the prior on $\beta_2$ and the values of $f_*$, we find that the frequency range of the waveform needs to be lower than $6\times10^{-2}$ Hz for the positive-power case, and higher than $6\times10^{-6}$ Hz for the negative-power case, so that the perturbation $\beta_2(f/f_*)^2<1$. This will cut off a part of the late-inspiral waveform for the cases with $M_{\rm tot}=10^4 M_\odot$ and smaller, which weakens the constraints in low-mass cases. The fiducial values we use for the modified gravity parameters are the GR values of $\beta_1=\beta_2=0$. The fiducial nuisance parameters are set to be $t_c=\Psi_c=0$.  Instead of constraining the value of $\eta$, ${\cal M}_z$ and $z$, we constrain their fractional errors in order to avoid incomparable magnitudes among the waveform derivatives that cause large off-diagonality in the Fisher matrix. 
 
 For the positive power model, we show the forecast constraint contours for the cases with total mass of $10^5~M_\odot$ at different redshifts in Figure~\ref{fig:ellipse}. As expected, the constraints are weaker for higher redshift due to an overall lowering of the SNR. Our results reflect a common difficulty of this analysis \cite{Will:1997bb}, namely that some of the parameters in the modified waveform are highly degenerate, e.g. the parameter pairs $(\ln \eta, \Psi_c)$, $(\ln \eta, \beta_1)$ and $(\Psi_c, \beta_1$). A possible reason for this result is that, apart from $z$, all other parameters are contained in the phase, so they are highly correlated with oneanother. 
 
\begin{figure}[h]
    \centering
    \includegraphics[width=0.6\textwidth]{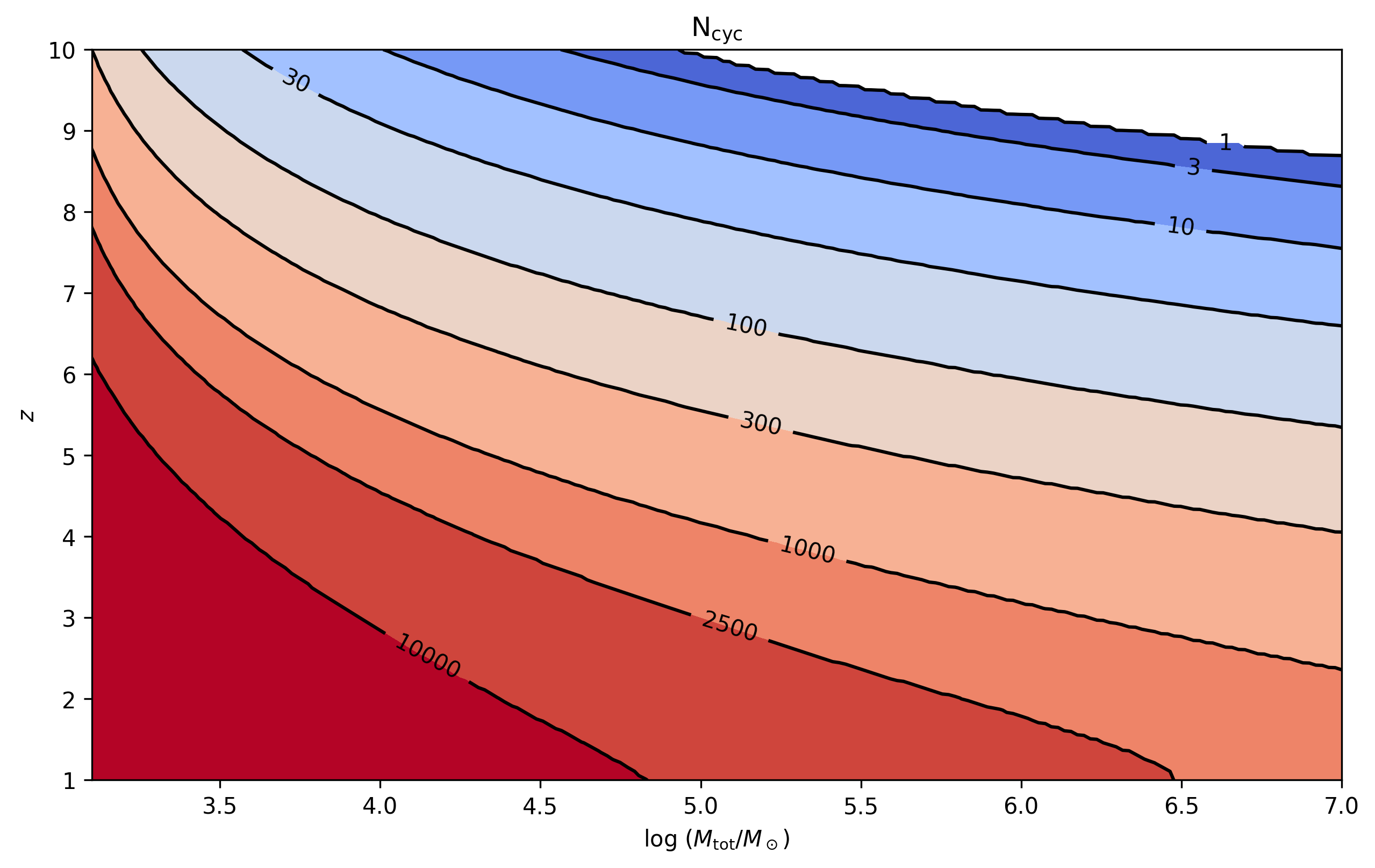}
    \caption{\small Number of cycles (neglecting post-Newtonian effects) in-band for LISA sources in the $\{M_{\rm tot}, z\}$ plane (see eq.22 of \cite{Mangiagli:2020rwz}). We count cycles from the time the system exceeds SNR=8 until the end of the inspiral phase, which we take to be at $f=2 f_{\rm ISCO}$. }
    \label{fig:cycles}
\end{figure}

In addition to this plot, the constraints on parameters for all the models using the inspiral waveform in GR, positive and negative-power polynomial cases are listed in the tables in Appendix \ref{sec:appendix_table}. A comparison with the GR case shows that the presence of $\beta_1$ and $\beta_2$ weakens the constraints on the GR parameters. We find that the constraints are controlled by the SNR and the total mass of the system. Some GR parameters tend to be better constrained when the SNR is higher, such as $\eta$, $z$ and $\Psi_c$. But the constraints on ${\cal M}_z$ and $t_c$ are tighter for systems with lower masses. This is expected since  signals from lower mass systems stay longer in the LISA band, %as shown in Figure~\ref{fig:amp_pds}, 
so that more inspiral cycles are available for constraining the parameters (see Figure \ref{fig:cycles}). 

\begin{figure}[h]
    \centering
    \includegraphics[width=0.48\textwidth]{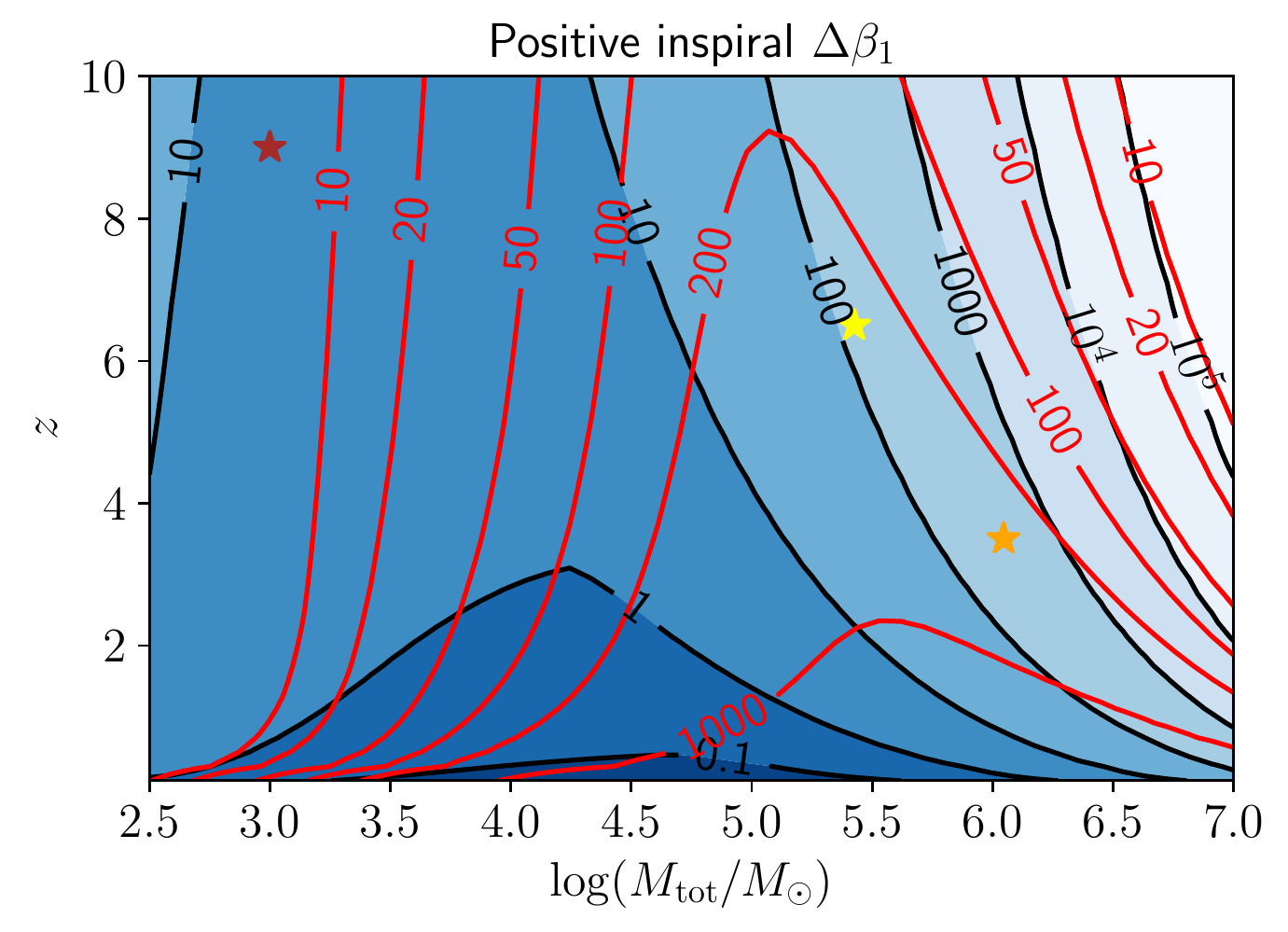}
    \includegraphics[width=0.48\textwidth]{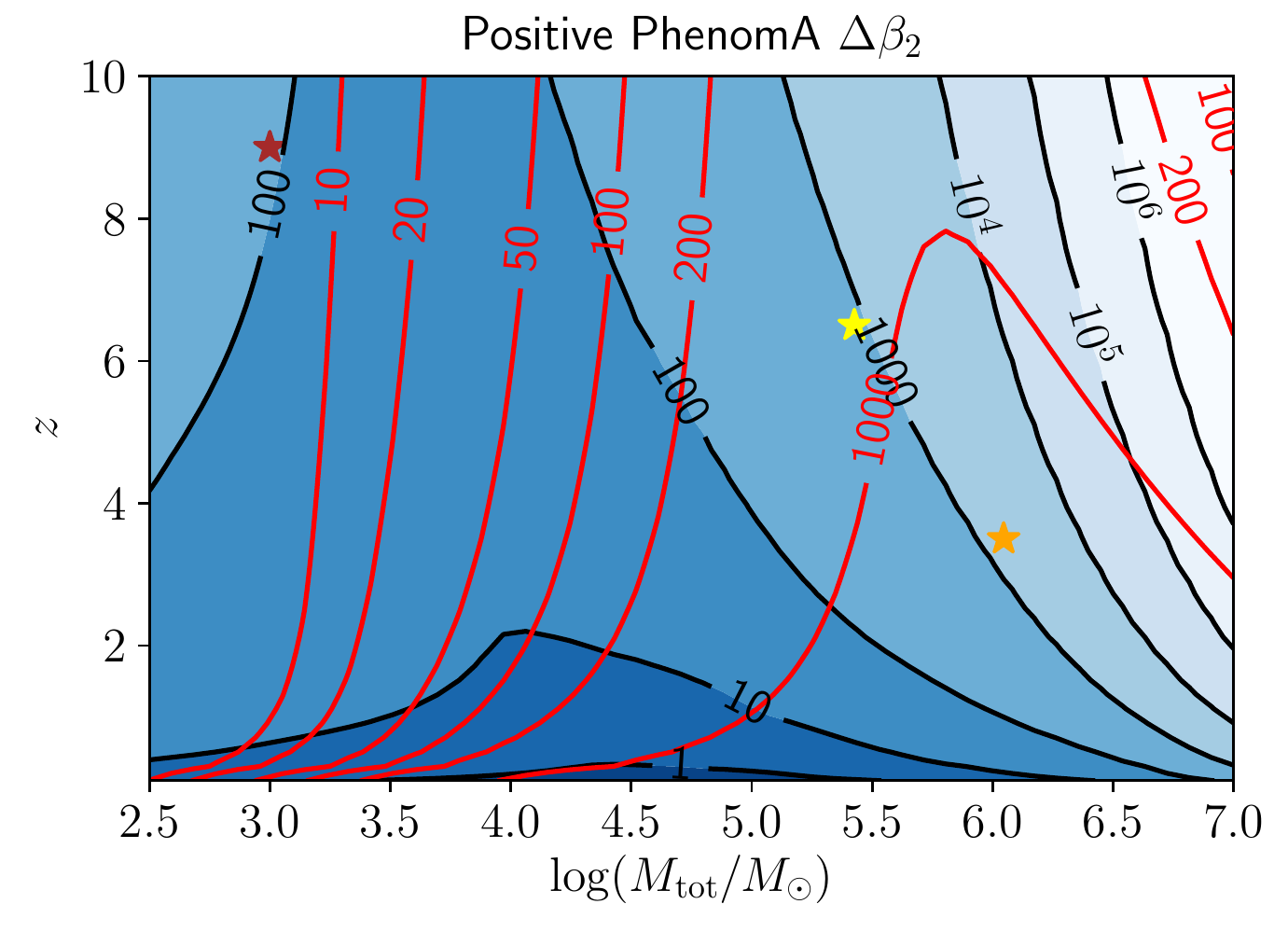}\\
    \includegraphics[width=0.48\textwidth]{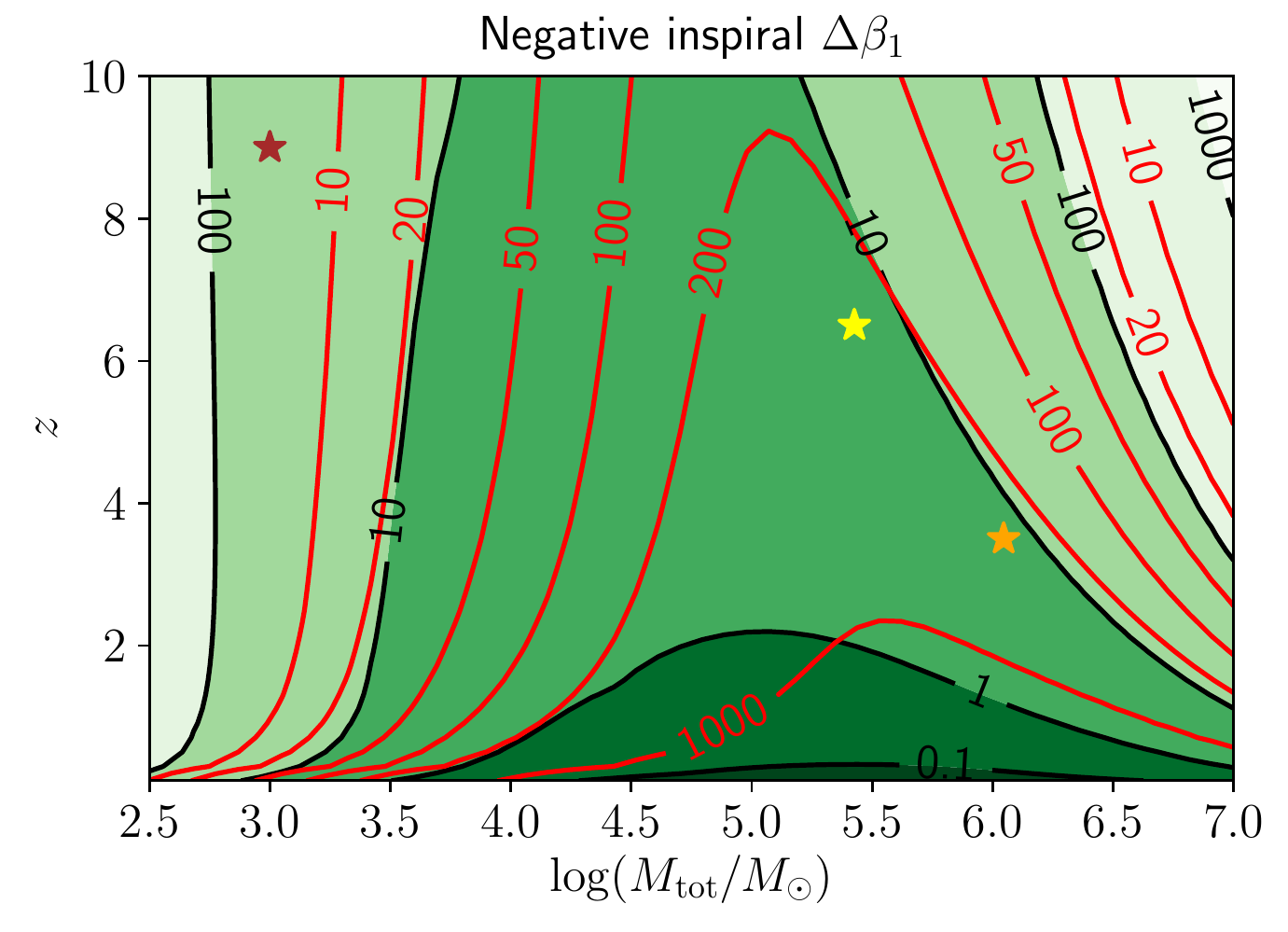}
    \includegraphics[width=0.48\textwidth]{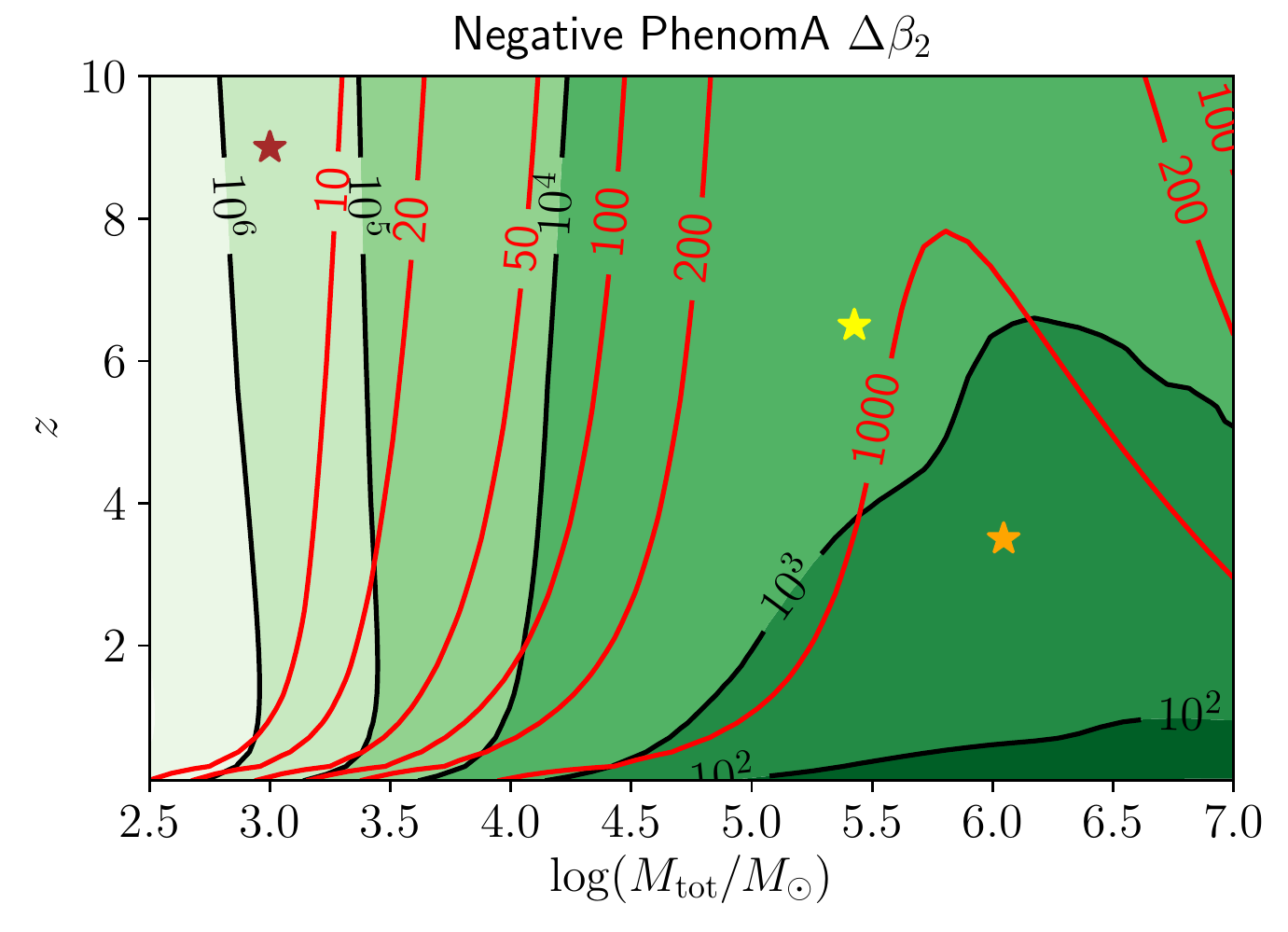}\\
    \includegraphics[width=0.4\textwidth]{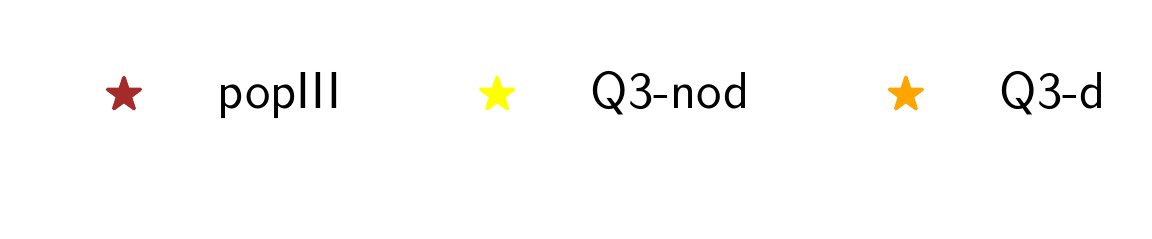}
    \caption{\small The contours of constraints on $\beta_1$ and $\beta_2$ (black) and the SNR contours (red) on the map of source total masses and redshifts for different cases. The stars show the peaks of MBH population on the map predicted by models of popIII, Q3-nod and Q3-d (\S\ref{subsec:results_IMR}). The priors on $\beta_1$ and $\beta_2$ are lifted for these plots. }
    \label{fig:beta_contour}
\end{figure}

The modified gravity parameters $\beta_1$ and $\beta_2$ are special, in the sense that they are better constrained when the signals extend to frequencies closer to $f_*$. This means that the positive-power case is best constrained by systems with $M_{\rm tot}<10^5 M_\odot$. Note also that these systems have the final stages of their inspiral -- where the modified PN terms of \S\ref{subsec:waveform_phase} are most significant -- around the peak sensitivity region of the LISA power spectral density (PSD). The left panel of Figure \ref{fig:beta_contour} shows the marginalised constraints on $\beta_1$ for both positive- and negative-power polynomial cases as a function of $\{M_{\rm tot},z\}$, with SNR overlaid in red. In particular for the positive power case, we notice that  the shape of the SNR and $\sigma_{\beta_1}$ contours are considerably different. 

For the negative-power case, the constraint contours on $\beta_1$ in Figure \ref{fig:beta_contour} (lower left panel) show greater similarity to the SNR contours. Here we expect the greatest deviation from GR to manifest in the heaviest systems, but this is compensated for by the rapidly rising PSD (and hence decreasing SNR) at low frequencies. 

In general we learn that $\beta_2$ is challenging to constrain; this is not unexpected, given it represents a second-order correction to $c_T(f)$. For the positive-power case using only the inspiral waveform (Table \ref{tab:positive_inspiral} in Appendix \ref{sec:appendix_table}), the prior on $\beta_2$ is saturated for the $M_{\rm tot}=10^6~M_\odot$ cases, indicating that we fail to obtain a meaningful constraint. Hence in right panels of Figure \ref{fig:beta_contour} we show only results that include the PhenomA merger-ringdown extension (see next subsection).

We stress that the constraints obtained are quite sensitive to the value of $f^*$. Here we have adopted a maximally conservative approach, by setting $f^*$ completely outside the LISA band for both positive- and negative-power polynomial cases. Alternatively, for a given $\{M_{\rm tot}, z\}$, $f_*$ can adopt any value provided that $(f/f_*)^n<<1$, such the expansion used in \S\ref{sec:waveforms} remains valid. An example of this can be seen in Figure \ref{fig:SNR_high}, where a lower value of $f_*=0.1$~Hz was used; an improvement of one and two orders of magnitude results for the constraints on $\beta_1$ and $\beta_2$, respectively.

Finally, the degeneracies between the parameters analysed in this section could be reduced by replacing the highly correlated parameters with new ones, for example treating the complete amplitude as a single parameter, so that parameters in the phase will not correlate to the amplitude. However, significant correlations between the phase parameters would likely remain, and we would lose the ability to estimate $\Delta z$ %Avoid \sigma as that notation hasn't been used in the paper
-- which could be significant for associating the merger to a host galaxy.
 
\subsubsection*{EFT-inspired Model}
We now move on to Fisher forecasts for the EFT case, where the beyond Einstein parameters of interest are $f_*$ and $c_0$. Recall that these parameters control the location and height of the transition seen in Figure \ref{fig:plotCS1}, with $c_0=1$ corresponding to the GR case (no transition).

\begin{figure}[ht]
    \centering
    \includegraphics[width=0.6\textwidth]{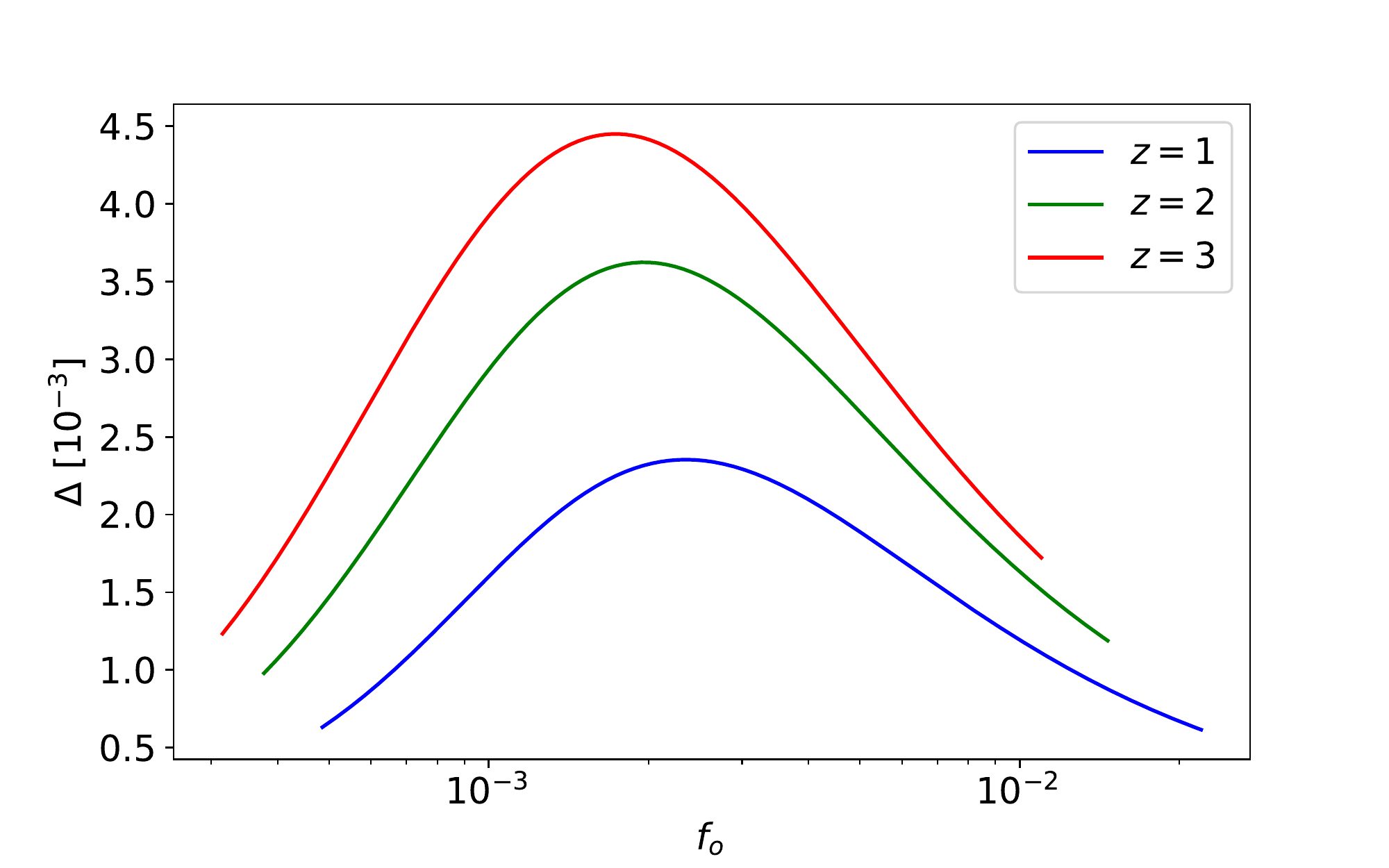}
    \caption{\small $\Delta$ as a function of observed frequency in the EFT-inspired Ansatz, at frequencies of the binary inspiral waveforms for $M_{\rm tot}=10^5~M_\odot$, $c_0=0.99$ and $f_*=3\times10^{-4}$ Hz at different redshifts.}
    \label{fig:Delta_stacked}
\end{figure}

Since the phase of the waveform in the EFT case is computed numerically, we also numerically compute the derivatives of the waveform used in the Fisher matrix. We use $c_0=0.99$ instead of $c_0=1$ for the fiducial model, because the numerical derivatives with respect to $c_0$ and $f_*$ become unstable when $c_0=1$. This is easily understood, as when there is no transition $f_*$ has no effect on the waveform, and becomes impossible to constrain. For our fiducial model we use $f_*=3\times10^{-4}$ Hz, as a typical value in the middle of the inspiral phase of LISA MBHBs. %Potential answer to minor 1?
In Figure \ref{fig:Delta_stacked} we plot the quantity $\Delta=1-\left[c_T(f_o)/c_T(f_s)\right]$ for this model; this highlights where the difference in GW speed between the source and observer is maximum. Note that the EFT-inspired model outlined in \S\ref{sec-theory-ansatze} does not have a further free parameter to control the width of the transition. Figure \ref{fig:Delta_stacked} shows the width of $\Delta(f_o)$ is quite large compared to the inspiral range of a LISA binary, meaning our detections are likely to probe only one side of the peak in $\Delta$.

\begin{figure}
    \centering
    \includegraphics[width=1.1\textwidth]{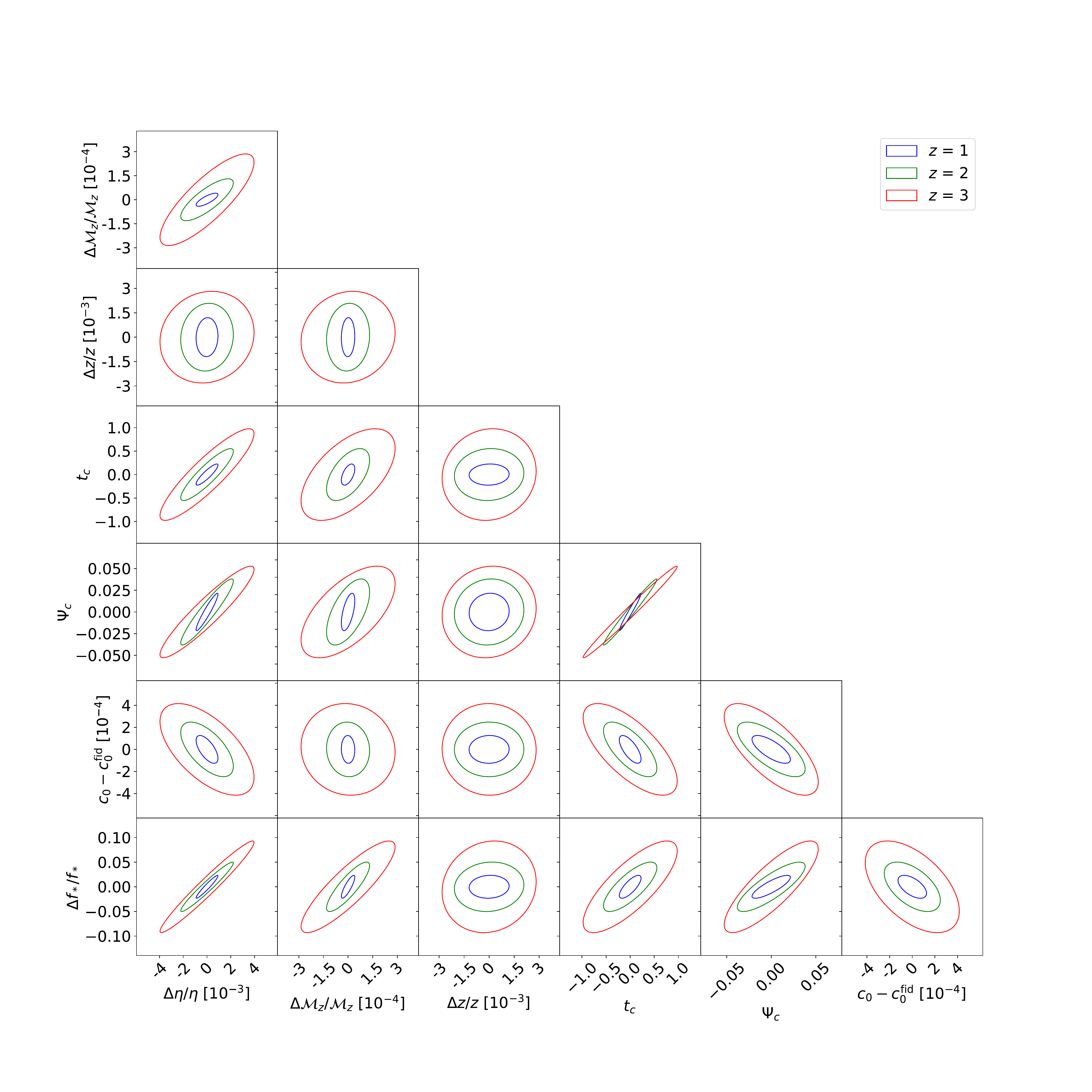}
    \caption{\small Forecast constraints for the EFT case with $M_{\rm tot}=10^5~M_\odot$, $\eta=0.25$, $t_c=0$, $\Psi_c=0$, $c_0=0.99$ and $f_*=3\times10^{-4}$ Hz at different redshifts using the inspiral waveform. We show the fractional constraints on $\eta$, ${\cal M}_z$, $z$ and $f_*$, and the constraints on $t_c$, $\Psi_c$ and $c_0$ centered at their fiducial values. }
    \label{fig:ellipse_Delta}
\end{figure}

We show the constraints for the EFT-inspired model with total mass of $10^5\,M_\odot$ at different redshifts in Figure~\ref{fig:ellipse_Delta}. We see that the constraints on GR parameters are roughly as tight as the ones in the polynomial case. However, we now obtain tight constraints on the modified gravity parameters $c_0$ and $f_*$ (compared to the fairly weak constraints on $\beta_1$ and $\beta_2$ in the polynomial models). This greater sensitivity likely comes from i) having the deviations from GR strongest in the mid-inspiral phase, where both the number of cycles and SNR accumulation are reasonable, and ii) having both parameters play comparable roles in modifying the waveform (compared to the polynomial case where $\beta_2$ is significantly subdominant to $\beta_1$ in the LISA band). 


\begin{figure}[h]
    \centering
    \includegraphics[width=0.48\textwidth]{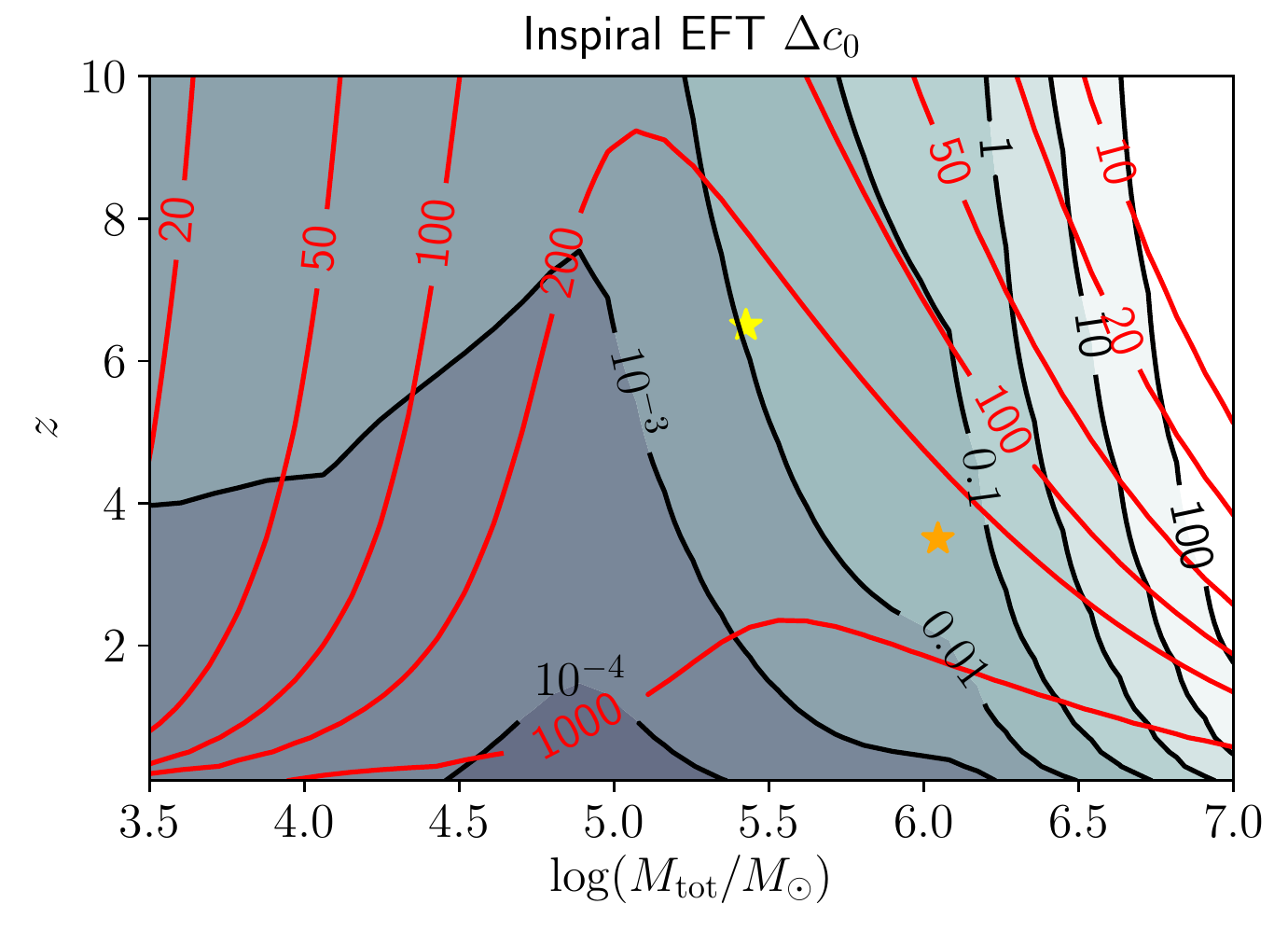}
    \includegraphics[width=0.48\textwidth]{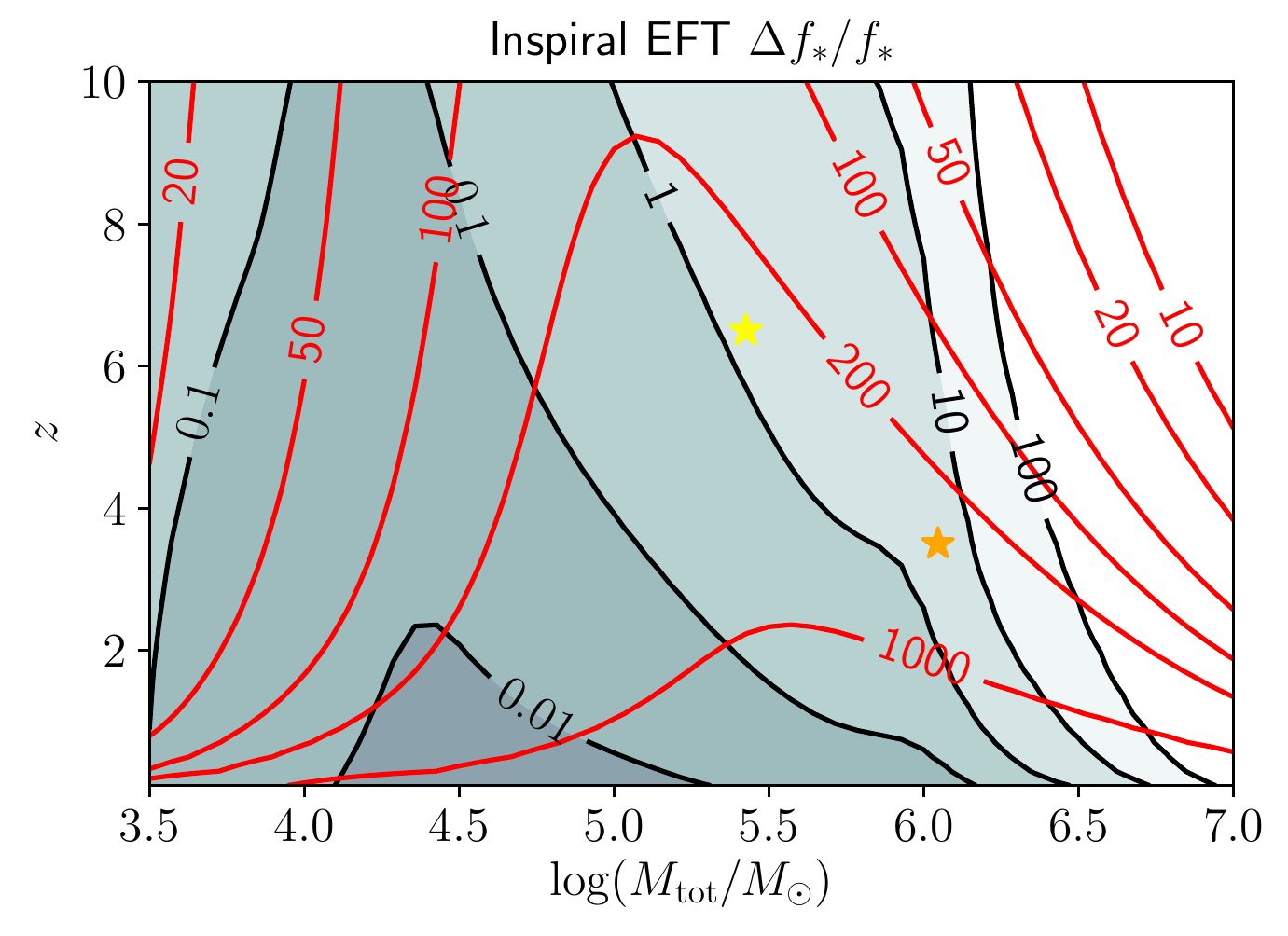}\\
    \includegraphics[width=0.3\textwidth]{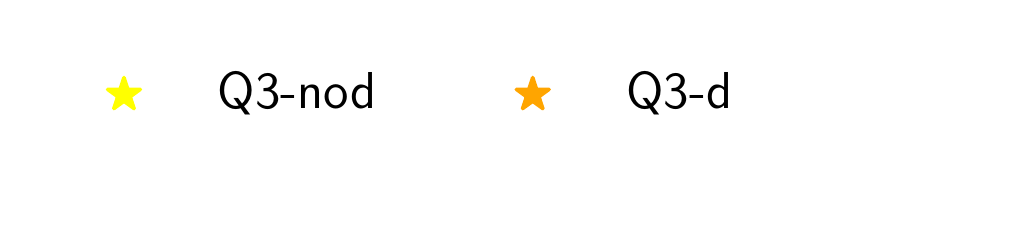}
    \caption{\small The contours of constraints on $c_0$ and fractional constraints on $f_*$ (black), along with the SNR contours (red), in the space of source total masses and redshifts, for the EFT-inspired model with inspiral signal only. The fiducial model used is $c_0 = 0.99$, $f_*=3\times10^{-4}$ Hz. The stars show the peaks of MBH population as predicted by the Q3-nodelay and Q3-delay models (\S\ref{subsec:results_IMR}). No prior bound is applied to $c_0$ and $f_*$.}
    \label{fig:EFT_contour}
\end{figure}

Figure \ref{fig:EFT_contour} shows the contours of constraints on $c_0-1$ and fractional constraints on $f_*$ in the $\{M_{\rm tot}, z\}$ plane. The lowest total mass we consider here is $10^{3.5} M_\odot$, because the numerical derivatives with respect to $c_0$ and $f_*$ become unstable for systems with lower total masses. This can be explained by the fact that lower-mass systems evolve into frequencies much higher than $f_*$, at which the effects of the transition in $c_T$ are very small -- this makes the derivatives with respect to the beyond Einstein parameters noisy. We can see from the plots that the tightest constraints on $c_0$ and $f_*$ are found by systems with $M_{\rm tot}\sim10^{4.5}-10^5 M_\odot$. This is likely because our fiducial choice of $f_*$ is located within the early inspiral stage of such systems, so that most of the transition of $c_T$ from 1 to $c_0$ takes place within the detectable signal. For lower-mass systems, only a small portion of their inspiral waveform is affected by the $c_T$ transition. 

On the other hand, for more massive systems, $f_*$ is above their inspiral frequency ranges, so that similarly most of their inspiral is also not affected by the transition of $c_T$. Similarly, varying the values of $f_*$ also has an effect on the constraints on $c_0$ and $f_*$. Taking the system with $M_{\rm tot}=10^{6} M_\odot$ and $z=1$ as an example, the fiducial value $f_*=3\times10^{-4}$ Hz sits in the middle of its inspiral waveform, as shown in Figure \ref{fig:Delta_amp_phase}. For a much larger or much smaller $f_*$, such as order of $10^{-2}$ or $10^{-6}$ Hz, the constraints are worsened by 4 orders of magnitude and 1 order of magnitude respectively. The reason is that the transition of $c_T(f)$ takes place at frequencies higher or lower than the inspiral waveform, and thus little information is obtained from the waveform to constrain $c_0$ and $f_*$. 

Given the constraints presented in \S\ref{sec:forecast}, one may ask what is the corresponding error on the propagation speed $c_T(f)$. In reality this is a source-, ansatz- and frequency-dependent statement. However, as an example, we consider an optimal source (using constraints from~\cref{fig:EFT_contour}) with $\log_{10}(M_{\rm tot}/M_{\odot}) \sim 4.75$ and $z\sim 0.5.$, under the EFT-inspired ansatz. Propagating the errors on $c_0$ and $f_*$ in the standard fashion (including their covariance), we find that deviations of $c_T(f)/c-1$ are constrained to $\lesssim 10^{-2}$ at LISA frequencies. If extrapolated to the band of ground-based detectors, this constraint becomes $\left[c_T(f)/c-1\right]\lesssim 10^{-8}$. Whilst not as tight at the bound from GW170817 and its counterpart, we stress that result uses data from an entirely distinct regime, and does not rely on the presence of any EM signals.

\subsection{Connection to other data}

We note here that if an electromagnetic counterpart for at least one LISA event is unambiguously observed, in principle this can inform the prior on $c_0$ in the EFT-style model. A time-of-flight measurement, similar that performed with GW170817, could yield constraints of order $|c_0| \lesssim 10^{-10}$. Although the electromagnetic counterpart may appear considerably delayed after the merger (e.g. a month, relative to seconds for a BNS merger), this is counteracted by the greater propagation distance of LISA sources. In this eventuality, the tight bound on $c_0$ would effectively slice through the ellipse contours in Figure \ref{fig:ellipse_Delta}. Due to the correlation with $\Psi_c$ and $t_c$, this would lead to improved constraints in all parameters. The constraint on $f_*$ would also improve, though potentially not as significantly as Figure \ref{fig:ellipse_Delta} might suggest; this is because as $c_0\rightarrow 1$ (step height in $c_T$ is decreased), the location of $f_*$ becomes harder to measure. 

A major stumbling block in this improved method is that it will likely be hard to associate electromagnetic signatures to MBH mergers with a high degree of confidence, due to both the extended delay before their appearance, and the intrinsic electromagnetic variability of galaxies themselves. Furthermore, due to the separation between the merger itself and the emitting gas, e.g. in a circumbinary disk, an EM counterpart may not be highly luminous. A thorough review of possible MBH counterpart mechanisms can be found in \cite{Bogdanovic:2021aav}. Given these uncertainties, we present constraints without assuming any prior information from a LISA EM counterpart.

\subsection{Inclusion of ringdown-merger signal}
\label{subsec:results_IMR}
For models where beyond Einstein effects grow with frequency, we naturally expect that including the merger and ringdown phases of the waveform will enhance our constraining power. For cases where the opposite  is true (like the negative-power polynomial model) one might guess that such additions will be irrelevant. In fact this turns out not to be correct: including the merger and ringdown still tightens constraints on the GR parameters, and due to the correlations between parameters, this leads to a mild improvement in $\beta_1$ and $\beta_2$. Given that the constraints on the EFT-inspired model are already strong from the inspiral alone, we will focus our attention here solely on the polynomial models.

 We perform the Fisher forecast on the same sets of systems as in \S\ref{subsec:inspiral}, but with the modified PhenomA waveform of \S\ref{sec:IMR_extension}. We present the full constraints in Appendix \ref{sec:appendix_table}, in comparison with the constraints from only the inspiral waveform that ends at $2f_{\rm ISCO}$. It is clear that the merger and ringdown phases increase the SNR values (and hence tighten constraints for all parameters), especially for intermediate and heavy mass systems which otherwise have only a short inspiral track in the LISA band. The constraint on $\beta_2$ is considerably improved in the positive power case for all systems, except for a $10^6$ $M_{\odot}$ binary at $z=2, 3$.
 
\begin{figure}
    \centering
    \includegraphics[width=1.1\textwidth]{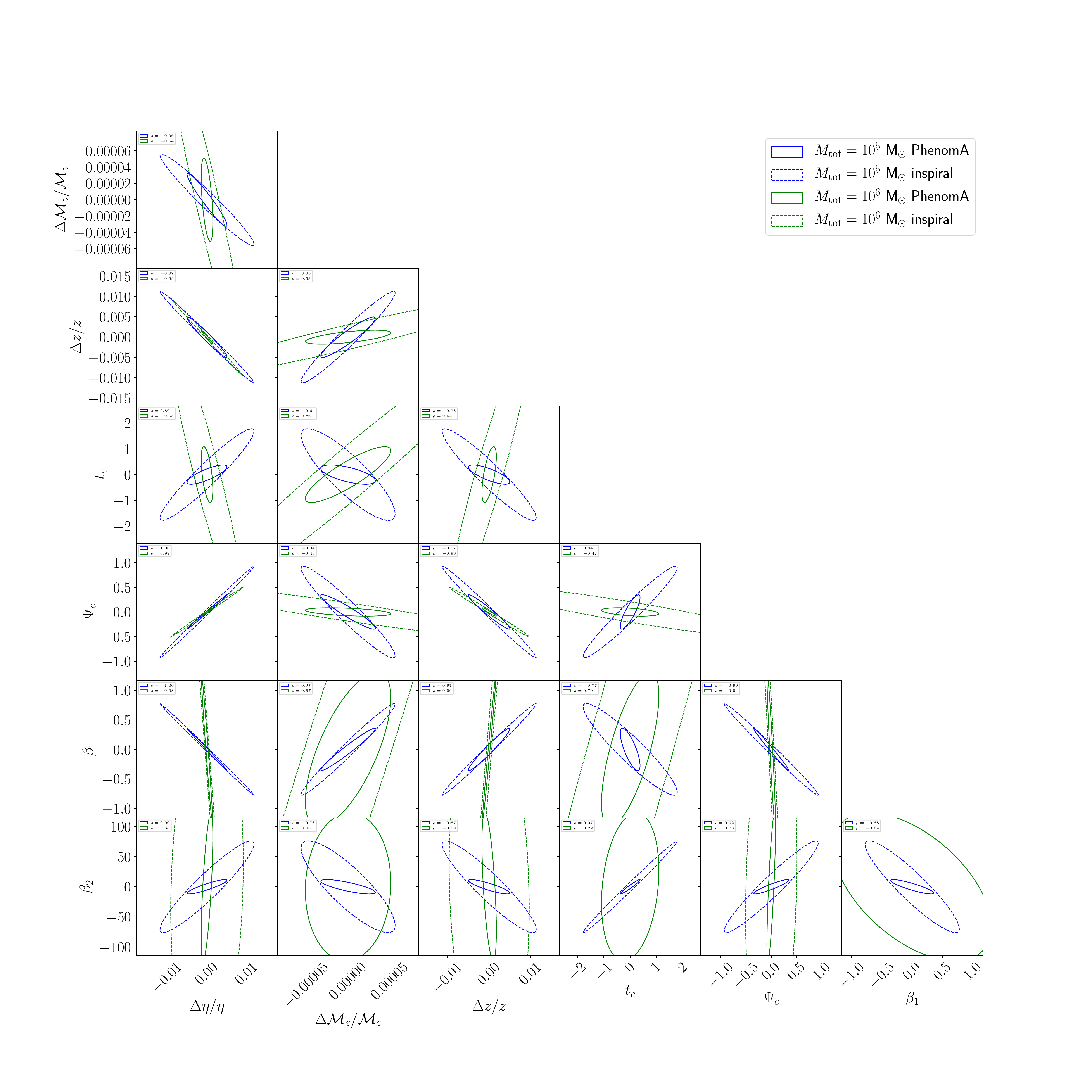}
    \caption{\small Forecast constraints with the inspiral waveform and the modified PhenomA waveform for the positive-power polynomial case, with fiducial values $M_{\rm tot}=10^5$ M$_\odot$ and $10^6$ M$_\odot$, $\eta=0.25$, $z=1$, $t_c=0$, $\Psi_c=0$, $f_*=2$ Hz, $\beta_1=0$ and $\beta_2=0$.}
    \label{fig:ellipse_phenomA}
\end{figure}

 We also place the Fisher forecast contours with the modified PhenomA waveform on top of those with the inspiral waveform for systems of $10^5$ and $10^6$ $M_\odot$ at $z=1$ in Figure \ref{fig:ellipse_phenomA}. We can see that the inclusion of merger and ringdown phases shrinks the ellipse contours, and the effect is greater for the heavier mass system. Most striking is that many ellipses change their orientation for different total masses, and some even flip their signs. In Appendix \ref{sec:integrand} we present a deeper investigation into this phenomenon. The short summary of this is that the merger and ringdown phases add dominant contributions to the integrands that yield the entries of the Fisher matrix. The parameter dependency of these new contributions can be different from the PN expansion of the inspiral, and hence the marginalised ellipses get re-oriented as per Figure \ref{fig:ellipse_phenomA}.
 
The righthand panels of Figure \ref{fig:beta_contour} show the constraints on $\beta_2$ with the PhenomA waveform as a function of $\{M_{\rm tot}, z\}$. No prior bounds on $\beta_1$ and $\beta_2$ are put in the Fisher forecast for these plots. We see that the constraints on $\beta_2$ in the negative-power case remain worse than in the positive-power case at most points, with a preference for higher-mass systems whose inspirals start closest to $f_*$ (=$2\times 10^{-7}$ Hz in this case).

Now that all four panels of Figure \ref{fig:beta_contour} have been introduced, we can consider how the areas of best constraint relate to expectations for MBHB merger rates. The three star-shaped markers in the plots mark the locations of peak merger rates predicted by three different population models from \cite{Barausse:2012fy,2016PhRvD..93b4003K} (see those works for detailed descriptions). The PopIII model predicts that the MBH merger rate peaks at low mass and high redshift, while the Q3-nodelay (Q3-nod) and Q3-delay (Q3-d) models predict it peaks at intermediate mass and low/intermediate redshifts. The best candidates for our constraints are events of $M_{\rm tot}=10^4-10^5$ M$_\odot$ for the positive-power case, since they have good constraints and relatively high SNR.  Unfortunately they are quite far from the peak of MBHB merger density for any population model. The Q3-nod and Q3-d models favour heavier MBHBs, which also lie in a region of high SNR -- hence these may favour constraints on the negative-power case. That said, there is much more information underlying these MBHB population models than can be summarised through a single peak point. In the next section we  estimate the constraints that could be obtained by combining multiple merger observations in each population model.

\begin{figure}[h!]
    \centering
    \includegraphics[width=0.6\textwidth]{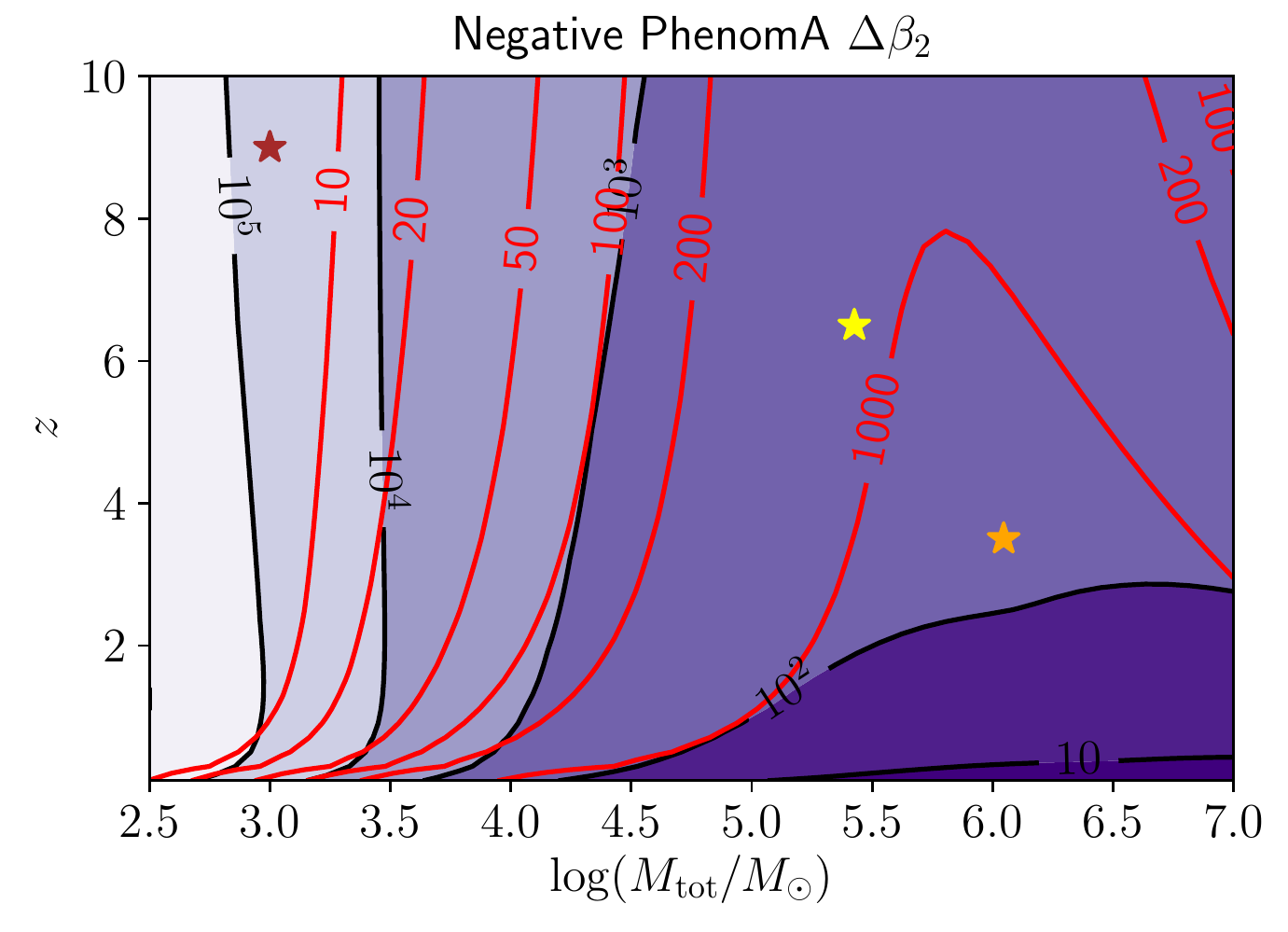}\\
    \includegraphics[width=0.4\textwidth]{plots/beta_legend.pdf}
    \caption{\small The contours of constraints on $\beta_2$ (black) and the SNR contours (red) on the map of total mass and redshift in the case with only one modified gravity parameter $\beta_2$. The stars show the peaks of MBH population on the map predicted by models of popIII, Q3-nod and Q3-d (\S\ref{subsec:results_IMR}). The prior on $\beta_2$ is lifted for this plot.}
    \label{fig:beta_contour_2}
\end{figure}

As a final exploration of the extended IMR constraints, we investigate a negative power case in which only $\beta_2\neq 0$. This case is special since it naturally fits within the constraint from GW170817, without the need to invoke further physics. We show the corresponding constraint contours of $\beta_2$ with the PhenomA waveform in Figure \ref{fig:beta_contour_2}. In general, the constraints on $\beta_2$ are tighter than the ones in which both $\beta_1$ and $\beta_2$ are varied, by around one order of magnitude. However, they still remain fairly weak, indicating that strongly-evolving deviations from GR will be challenging to detect at low frequencies, even with LISA.

\subsection{Multiple sources}
\label{sub:multiple_sources}
The constraints obtained in the previous sections are derived from single event detections; often these were picked to be relatively ideal sources, as determined by Figures \ref{fig:beta_contour} and \ref{fig:beta_contour_2}. In reality, during the full LISA mission duration we will receive a sample of events with some mass and redshift distribution, as determined by the physics of the MBH population ~\cite{Barausse:2012fy, 2016PhRvD..93b4003K, Barausse:2020mdt}. When considering a population of sources, the merger parameters are independent, whilst the beyond Einstein parameters will be common to all events\footnote{Note the introduction of redshift-dependence or environmental dependence into our beyond Einstein Ans\"atze would complicate this issue; we leave this for future work.}. As a consequence, the constraints on  $\theta_{\rm MG}$ obtained from all these events could be combined to significantly improve their determination. 

\begin{figure}[t]
    \centering
    \includegraphics[width=\textwidth]{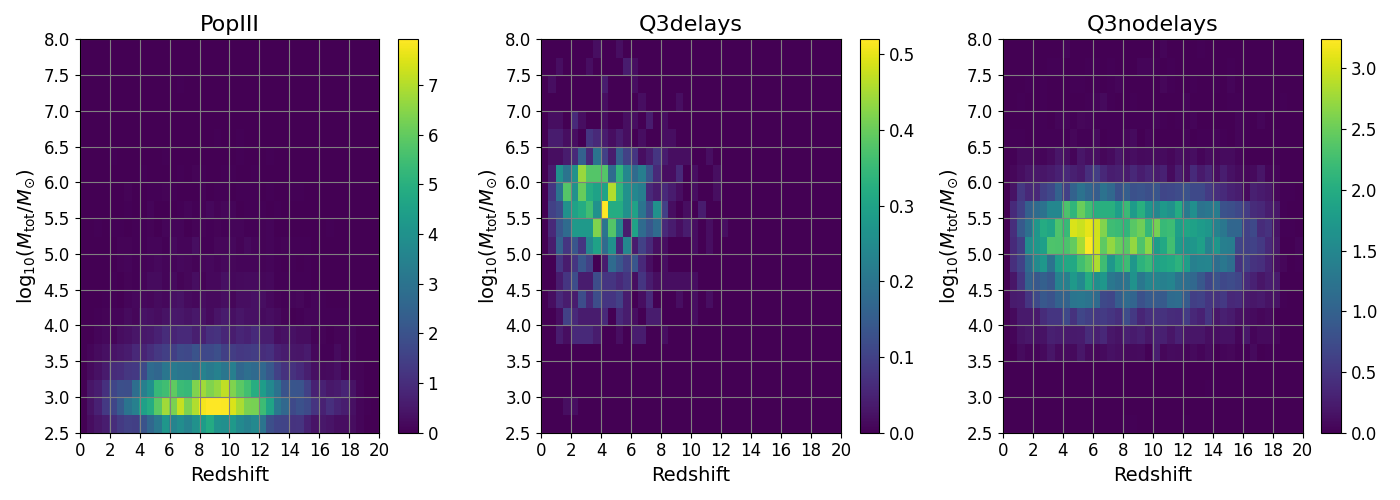}
    \caption{\small The histograms of the MBH merger populations after averaging over 50 realizations. The color scale reflects the number of events expected in each bin during a four year LISA mission (note the color scale differs between panels).}
    \label{fig:pop_hist}
\end{figure}

Let us briefly outline the procedure employed to generate realizations of the mass and redshift distribution of MBHBs for the different models (popIII, Q3-delay and Q3-nodelay), discussed in~\cite{2016PhRvD..93b4003K}, which correspond to different seeds and evolution for the MBH population. Taking the catalogues from~\cite{Catalogues_Enrico}, we obtained\footnote{For this purpose we followed the procedure described in the \texttt{readme.pdf} file in~\cite{Catalogues_Enrico}.} the mass and redshift distribution for the three population models. Given the histograms, we proceed by defining a smooth interpolation which can be used as a probability distribution function, and hence can be sampled to generate a new catalogue. The number of events in each realization of these catalogues is set by the integral of the merger rate\footnote{In reality this should be a random number drawn from a Poisson distribution with expectation value equal to the theoretical prediction for the number of events. For the scopes of our discussion we can safely ignore this point.} for the three catalogues of~\cite{Catalogues_Enrico} over the three effective years of the LISA mission (corresponding to 4 years with $75\%$ efficiency). In Figure~\ref{fig:pop_hist} we show the event distributions in the $z$-$\log_{10}(M_{\rm tot}/ M_{\odot})$ plane after averaging over 50 realizations for each population. The bin sizes for both $z$ and $\log_{10}(M_{\rm tot}/M_{\odot})$ are chosen to be $0.25$. The total number of events in the three cases are respectively $511$, $24$ and $356$ events. 

To find the systems which will contribute the most to the combined constraint, we select the bins of each population that sit above the SNR$=10$ contour, and have at least one event expected per bin. Since the Q3-delay population model cannot satisfy the last criteria (every bin has an expectation value $<1$), we discard this case. For the two remaining models, we compute the Fisher matrix for $N_{\rm bin}$ combined events in each cell of interest and invert it to obtain the covariance matrix for that cell (note this is is just the single-event covariance with errors scaled by $1/\sqrt{N_{\rm bin}}$.) We then remove all columns except for those corresponding to $\theta_{\rm MG}$ and invert again to obtain the marginalised Fisher matrix for just the beyond Einstein parameters. We sum the marginalised Fisher matrices from each cell and finally invert again to obtain the combined covariance matrix for $\theta_{\rm MG}$ from all the selected cells\footnote{We note the fully correct procedure would be to constrain the standard parameters of all systems simultaneously, along with $\theta_{\rm MG}$. Since we expect no correlation between the standard parameters of each system, the degree of approximation here is negligible.}.

\begin{table}[t]
    \centering
    \begin{tabular}{c|cc|c|cc}
    \hline
        Model & $N_{\rm tot}$ & $N_{\rm count}$ & Power & $\Delta\beta_1$ & $\Delta\beta_2$ \\
        \hline
        \multirow{2}{*}{PopIII} & \multirow{2}{*}{511} & \multirow{2}{*}{88} & Positive & $0.13$ & $9.8$ \\
         &  &  & Negative & $0.66$ & $3333.3$ \\
        \hline
        \multirow{2}{*}{Q3-nod} & \multirow{2}{*}{356} & \multirow{2}{*}{257} & Positive & $0.47$ & $49.5$ \\
         &  &  & Negative & $0.07$ & $65.7$ \\
    \hline
    \end{tabular}
    \caption{\small Total constraints on $\beta_1$ and $\beta_2$ with the inspiral waveform using events with $N_{\rm bin}>1$ and SNR $>10$ in each cell of the population histograms. $N_{\rm bin}$ gives the total number of merger events expected for each model, and $N_{\rm count}$ is the number than contribute usefully to the final combined constraint (see text).}
    \label{tab:multiple}
\end{table}

Table \ref{tab:multiple} shows the results of this process for $\beta_1$ and $\beta_2$ of the polynomial models. We see that for $\beta_1$ in both the positive- and negative-power cases, the combined constraint is approximately equal to the best-constrained region of non-vanishing size in Figure \ref{fig:beta_contour}. In essence, this means that having a population of typical GW LISA mergers is roughly equivalent to a single `golden' event in the ideal part of the $\{M_{\rm tot}, z\}$ parameter space. 

The picture for $\beta_2$ is a little more mixed, with PopIII model performing well in the positive-power case, and the Q3-nod model favouring the negative-power case. We see that the PopIII model, which generally produces lighter MBHs, yields good constraints on the positive-power model -- this is precisely in line with the discussion of \S\ref{subsec:inspiral}.

Of course, in reality we will have to work with whatever population of MBH mergers Nature gives us. If it closely resembles the Q3-delay model, for example, we will be dependent on a rare golden system to carry out the constraints forecast in this work. However, it is reassuring to see that in most cases our method has some robustness against realistic population models. Hence tests of gravity at low frequency can be carried out with LISA in (almost) any scenario.

\section{Conclusions}
\label{sec:conclude}
The development of cosmological modified gravity theories has shown that infrared departures from GR are theoretically possible. The clearest demonstration of this is screening effects, where departures from GR manifest on large scales -- a weak-field, low-density arena -- whilst being strongly suppressed in other regimes (see \cite{Joyce_2015,Burrage:2017qrf,Baker:2019gxo} for reviews). At the same time, deviations of the propagation speed of gravitational waves are a common signature of new gravitational physics. As such, it is clear that the value of $c_T$ should be probed at low energy scales, independently of existing constraints at higher frequencies. 

That said, the current tests of gravity from ground-based detectors are a force to be reckoned with. %\cite{LIGOScientific:2020tif}.
We find  it is not simple to construct a function for $c_T(f)$  which satisfies the LIGO-Virgo bounds whilst modifying the millihertz regime significantly. 
Sharp transitions for  $c_T(f)$ are needed  in the frequency band between LISA and LIGO frequencies, to ensure consistency with the results from GW170817. Future theoretical work  will be needed to explore more sophisticated models for $c_T(f)$, built from first principles,  that do not rely on this workaround.

Nevertheless, our work has established a theoretical and numerical toolkit for exploring the detectability of modified GW propagation with LISA. We implemented two {Ans\"atze} for frequency-dependent GW propagation speed, and computed the resulting modifications to the GW amplitude and (non-spinning) phase at 2.5PN order. The first {Ansatz} proposed departures of the GW propagation speed as a polynomial series in frequency for $c_T$, in which the powers can be positive or negative. The second {Ansatz} represented a smooth transition in $c_T$ from some lower value to $c_0$, taking place inside or close to the LISA band. We then performed a Fisher matrix analysis to forecast the constraints on five GR parameters and two modified gravity parameters. We compared the Fisher forecast with MCMC inference and found good agreement between them for the forecast parameter bounds, even for signals of comparatively low SNR. 

Our use of inspiral-only and a full IMR waveform represent analyses with different theoretical assumptions. If considering departures from GR, one may wish to allow for the strong-field regime itself to be modified as well; then using a (modified) PhenomA waveform, which derives from GR simulations, is not appropriate. Our inspiral-only (\S\ref{subsec:inspiral}) results represent this conservative case. However, if one is confident that the strong-field regime is identical to GR (the screened case), then our approach allows the continuation of GW propagation effects into the merger and ringdown regime. Our results using the full waveform in \S\ref{subsec:results_IMR} represent this more optimistic case. We used here a simple IMR waveform (PhenomA); this should be extended to more sophisticated, spinning waveforms for use with real data. 

Figures \ref{fig:beta_contour} and \ref{fig:EFT_contour} represent the major results of our work, showing how the constraining power of LISA for our $c_T(f)$ models is sensitive to the total mass and redshift of a MBH system. These raise the possibility that a single `golden' source may be as useful a population of less optimal systems. However, this statement clearly depends on the expected rate of MBH mergers, which is still poorly known. 
This sensitivity to the underlying MBH population increases further if redshift-dependent or cumulative corrections to GW propagation are considered. In this work we focused exclusively on the frequency dependence of $c_T$; as a result, our constraints are (unsurprisingly) always tightest from low-redshift sources. If instead the beyond Einstein effects accumulated with propagation distances -- as happens for \textit{some} modified gravity models (see e.g. \cite{Ezquiaga:2018btd,Belgacem:2019pkk} and references therein) -- then the redshift location of peaks in Figure \ref{fig:pop_hist} would also play a role in determining the constraints. For these reasons, and in view of future analyses, we developed in appendix \ref{sec:fredep} formulae that extend the discussion of the main text to include non-standard friction effects in the GW propagation.

Our method in this work has been distinctively different from that used to measure the propagation speed of GWs with event GW170817. We do not rely on the presence of an EM counterpart: for long-duration sources our analysis could be applied on-the-fly months or years before merger. This may open the possibility of multiband analyses for some sources, as considered in, e.g. \cite{Berti:2011jz,Sesana:2016ljz,Vitale:2016rfr,Carson:2019rda}.

This is not the first time modified propagation effects on the GW phase and amplitude have been computed. The ppE framework \cite{Yunes:2009ke} is a well-established formalism that shares many of the goals of this work. In fact, ppE is sufficiently general to include distinct modifications at each PN order of the phase, and can also encapsulate departures from the GR generation of GWs (not just propagation effects, as in the present work). The price paid for this powerful generality is an increased number of modified gravity parameters, such that these are usually varied and constrained one by one (see \cite{Perkins:2022fhr} for recent discussion). By focussing on a modification to $c_T$ alone, our work effectively links amplitude and various PN phase terms to vary in concert, creating a distinct signal. A mapping between our beyond Einstein parameters and those of ppE  is discussed in appendix \ref{app_ppEmap}.

The rate of ground-based GW detections will continue to rise sharply over the next decade, leading to tight constraints on gravity at the frequency of terrestrial detectors (or very exciting new results in gravitational physics). Nevertheless, LISA has a crucial role to play by opening the door to the unexplored millihertz GW regime. In this work we have developed the first tools for probing new phenomenology we may find there. Motivated by the direction of current theoretical ideas, this represents the first step of a continuing program to explore frequency-dependent effects in GW cosmology.

\acknowledgments{It is a pleasure to thank David Bacon, 
Enrico Barausse, Enis Belgacem, Emilio Bellini, Jose Maria Ezquiaga, Stefano Foffa, Noemi Frusciante, Michele Maggiore,  Nicola Tamanini, Filippo Vernizzi, and Miguel Zumalacarregui  for useful discussions. We also thank the present and past LISA Cosmology Working Group Chairs -- Robert Caldwell, Chiara Caprini, Germano Nardini, Marco Peloso, Nicola Tamanini -- for their support. We thank   Aur\'elien Hees for acting as internal referee within the LISA Consortium, as well as Nelson Chistensen for his help within the LISA PPC.
 T.B. is supported by ERC Starting Grant \textit{SHADE} (grant no. StG 949572) and a Royal Society University Research Fellowship (grant no. URF$\backslash$ R1$\backslash$180009). G.C. is supported by the I+D grant PID2020-118159GB-C41 of the Spanish Ministry of Science and Innovation. A.C. is supported by a PhD grant from the Chinese Scholarship Council (grant no.202008060014); this paper and the codes developed for it form part of his PhD thesis work.  M.F. would like to acknowledge support from the “Atracci\'{o}n de Talento” grant 2019-T1/TIC15784, his work is partially supported by the Spanish Research Agency (Agencia Estatal de Investigaci\'{o}n) through the Grant IFT Centro de Excelencia Severo Ochoa No CEX2020-001007-S, funded by MCIN/AEI/10.13039/501100011033. L.L.~is supported by a Swiss National Science Foundation Professorship grant (Nos.~170547 \& 202671). K.M. is supported by King's College London through a Postgraduate International Scholarship. M.P. was supported by STFC grants ST/P000762/1 and ST/T000791/1. M.P. acknowledges support by the European Union’s Horizon 2020 Research Council grant 724659 MassiveCosmo ERC- 2016-COG.  M.S. is supported in part by the Science and Technology Facility Council (STFC), United Kingdom, under the research grant ST/P000258/1. G.T. is partially funded by the STFC grant ST/T000813/1. D.B. acknowledges partial financial support by ASI Grant No. 2016-24-H.0. I.D.S. is supported by the Grant Agency of the Czech Republic (GAČR), under the grant number 21-16583M.
}

%\newpage

\appendix
\section{Fisher forecasts for all models}
\label{sec:appendix_table}
In this appendix we present the parameter constraints for all the models we considered in this work. The Fisher matrix analysis for the polynomial case uses flat priors on $t_c$ in $(-50,50)$, $\Psi_c$ in $(-\pi,\pi)$, $\beta_1$ in $(-20,20)$, and $\beta_2$ in $(-1000,1000)$. The positive-power cases use $f_*=2$ Hz, while the negative-power cases use $f_*=2\times10^{-7}$ Hz. The length of the signal is 30 days.

\begin{table}[H]
  \scriptsize
  \centering
  \caption{Inspiral waveform for the GR case}
  \begin{tabular}{cccccccc}        
    \hline 
    $M_{\rm tot}~[M_\odot]$ & $z$ & SNR & $\Delta {\cal M}_z / {\cal M}_z$ & $\Delta \eta / \eta$ & $\Delta z / z$ & $\Delta t_c$ & $\Delta \Psi_c$ \\
    \hline 
    $10^4$ & $1$ & $143$ & $2.97\times10^{-6}$ & $6.33\times10^{-4}$ & $5.63\times10^{-3}$ & $0.34$ & $1.38\times10^{-1}$\\
    $10^4$ & $2$ & $88$ & $6.73\times10^{-6}$ & $1.04\times10^{-3}$ & $9.26\times10^{-3}$ & $0.57$ & $2.05\times10^{-1}$\\
    $10^4$ & $3$ & $69$ & $1.11\times10^{-5}$ & $1.37\times10^{-3}$ & $1.20\times10^{-2}$ & $0.77$ & $2.53\times10^{-1}$\\
    $10^5$ & $1$ & $1021$ & $4.51\times10^{-6}$ & $1.54\times10^{-4}$ & $7.92\times10^{-4}$ & $0.11$ & $1.98\times10^{-2}$\\
    $10^5$ & $2$ & $595$ & $1.34\times10^{-5}$ & $3.42\times10^{-4}$ & $1.37\times10^{-3}$ & $0.28$ & $4.09\times10^{-2}$ \\
    $10^5$ & $3$ & $449$ & $2.65\times10^{-5}$ & $5.57\times10^{-4}$ & $1.83\times10^{-3}$ & $0.51$ & $6.37\times10^{-2}$ \\
    $10^6$ & $1$ & $2561$ & $3.39\times10^{-5}$ & $3.19\times10^{-4}$ & $3.17\times10^{-4}$ & $0.81$ & $3.12\times10^{-2}$ \\
    $10^6$ & $2$ & $882$ & $1.36\times10^{-4}$ & $1.14\times10^{-3}$ & $9.31\times10^{-4}$ & $4.06$ & $1.10\times10^{-1}$ \\
    $10^6$ & $3$ & $448$ & $3.27\times10^{-4}$ & $2.53\times10^{-3}$ & $1.87\times10^{-3}$ & $11.5$ & $2.41\times10^{-1}$ \\
    $10^7$ & $1$ & $350$ & $6.25\times10^{-4}$ & $2.64\times10^{-3}$ & $2.35\times10^{-3}$ & $46.6$ & $2.33\times10^{-1}$\\
    $10^7$ & $2$ & $85$ & $2.33\times10^{-3}$ & $5.40\times10^{-3}$ & $9.66\times10^{-3}$ & $49.9$ & $3.90\times10^{-1}$ \\
    $10^7$ & $3$ & $34$ & $6.00\times10^{-3}$ & $1.29\times10^{-2}$ & $2.42\times10^{-2}$ & $50.0$ & $9.05\times10^{-1}$ \\
    \hline 
  \end{tabular}
\end{table}

\begin{table}[H]
  \scriptsize
  \centering
  \caption{Full PhenomA waveform for the GR case}
  \begin{tabular}{cccccccc}        
    \hline 
    $M_{\rm tot}~[M_\odot]$ & $z$ & SNR & $\Delta {\cal M}_z / {\cal M}_z$ & $\Delta \eta / \eta$ & $\Delta z / z$ & $\Delta t_c$ & $\Delta \Psi_c$ \\
    \hline 
    $10^4$ & $1$ & $143$ & $2.88\times10^{-6}$ & $6.02\times10^{-4}$ & $5.66\times10^{-3}$ & $3.09\times10^{-1}$ & $1.30\times10^{-1}$\\
    $10^4$ & $2$ & $88$ & $6.37\times10^{-6}$ & $9.44\times10^{-4}$ & $9.30\times10^{-3}$ & $4.82\times10^{-1}$ & $1.84\times10^{-1}$\\
    $10^4$ & $3$ & $69$ & $1.03\times10^{-5}$ & $1.19\times10^{-3}$ & $1.20\times10^{-2}$ & $5.59\times10^{-1}$ & $2.15\times10^{-1}$\\
    $10^5$ & $1$ & $1037$ & $3.74\times10^{-6}$ & $1.04\times10^{-4}$ & $7.85\times10^{-4}$ & $4.81\times10^{-2}$ & $1.24\times10^{-2}$\\
    $10^5$ & $2$ & $619$ & $1.04\times10^{-5}$ & $1.97\times10^{-4}$ & $1.33\times10^{-3}$ & $8.85\times10^{-2}$ & $2.12\times10^{-2}$ \\
    $10^5$ & $3$ & $482$ & $1.97\times10^{-5}$ & $2.80\times10^{-4}$ & $1.74\times10^{-3}$ & $1.24\times10^{-1}$ & $2.80\times10^{-2}$ \\
    $10^6$ & $1$ & $7075$ & $1.85\times10^{-5}$ & $7.46\times10^{-5}$ & $1.23\times10^{-4}$ & $6.89\times10^{-2}$ & $5.84\times10^{-3}$ \\
    $10^6$ & $2$ & $4042$ & $6.67\times10^{-5}$ & $2.17\times10^{-4}$ & $2.34\times10^{-4}$ & $2.75\times10^{-1}$ & $1.67\times10^{-2}$ \\
    $10^6$ & $3$ & $2878$ & $1.48\times10^{-4}$ & $4.24\times10^{-4}$ & $3.61\times10^{-4}$ & $6.85\times10^{-1}$ & $3.23\times10^{-2}$ \\
    $10^7$ & $1$ & $6660$ & $2.08\times10^{-4}$ & $3.39\times10^{-4}$ & $2.90\times10^{-4}$ & $2.54$ & $2.55\times10^{-2}$\\
    $10^7$ & $2$ & $2062$ & $7.91\times10^{-4}$ & $1.10\times10^{-3}$ & $1.16\times10^{-3}$ & $12.4$ & $8.24\times10^{-2}$ \\
    $10^7$ & $3$ & $966$ & $1.85\times10^{-3}$ & $1.99\times10^{-3}$ & $2.83\times10^{-3}$ & $29.7$ & $1.50\times10^{-1}$ \\
    \hline 
  \end{tabular}
\end{table}

\begin{table}[H]
  \scriptsize
  \centering
  \caption{Inspiral waveform for the positive-power case}
  \begin{tabular}{cccccccccc}        
    \hline 
    $M_{\rm tot}~[M_\odot]$ & $z$ & SNR & $\Delta {\cal M}_z / {\cal M}_z$ & $\Delta \eta / \eta$ & $\Delta z / z$ & $\Delta t_c$ & $\Delta \Psi_c$ & $\Delta \beta_1$ & $\Delta \beta_2$  \\ 
    \hline 
    $10^4$ & $1$ & $143$ & $3.07\times10^{-5}$ & $2.12\times10^{-2}$ & $5.63\times10^{-3}$ & $1.36$ & $3.03$ & $1.01\times10^{-1}$ & $16.8$ \\
    $10^4$ & $2$ & $88$ & $4.88\times10^{-5}$ & $2.51\times10^{-2}$ & $9.27\times10^{-3}$ & $1.69$ & $3.04$ & $1.98\times10^{-1}$ & $31.1$ \\
    $10^4$ & $3$ & $69$ & $6.57\times10^{-5}$ & $2.77\times10^{-2}$ & $1.20\times10^{-2}$ & $2.05$ & $3.03$ & $3.08\times10^{-1}$ & $45.5$ \\
    $10^5$ & $1$ & $1021$ & $3.77\times10^{-5}$ & $7.94\times10^{-3}$ & $8.06\times10^{-4}$ & $1.21$ & $0.63$ & $5.23\times10^{-1}$ & $51.5$ \\
    $10^5$ & $2$ & $595$ & $9.98\times10^{-5}$ & $1.71\times10^{-2}$ & $1.45\times10^{-3}$ & $3.06$ & $1.29$ & $1.76$ & $194$ \\
    $10^5$ & $3$ & $449$ & $1.64\times10^{-4}$ & $2.38\times10^{-2}$ & $2.04\times10^{-3}$ & $4.75$ & $1.73$ & $3.37$ & $404$ \\
    $10^6$ & $1$ & $2561$ & $1.27\times10^{-4}$ & $6.94\times10^{-3}$ & $7.59\times10^{-4}$ & $3.83$ & $0.41$ & $5.93$ & $956$ \\
    $10^6$ & $2$ & $882$ & $2.93\times10^{-4}$ & $1.17\times10^{-2}$ & $1.65\times10^{-3}$ & $11.5$ & $0.65$ & $15.9$ & $996$ \\
    $10^6$ & $3$ & $448$ & $4.00\times10^{-4}$ & $1.06\times10^{-2}$ & $2.24\times10^{-3}$ & $17.0$ & $0.61$ & $18.6$ & $999$ \\
    \hline 
  \end{tabular}
  \label{tab:positive_inspiral}
\end{table}

\begin{table}[H]
  \scriptsize
  \centering
  \caption{Inspiral waveform for the positive-power case with $c_0=0.8$}
  \begin{tabular}{ccccccccccc}        
    \hline 
    $M_{\rm tot}~[M_\odot]$ & $z$ & SNR & $\Delta {\cal M}_z / {\cal M}_z$ & $\Delta \eta / \eta$ & $\Delta z / z$ & $\Delta t_c$ & $\Delta \Psi_c$ & $\Delta \beta_1$ & $\Delta \beta_2$ & $c_0$ \\ 
    \hline 
    $10^4$ & $1$ & $138$ & $4.24\times10^{-5}$ & $2.38\times10^{-2}$ & $5.85\times10^{-3}$ & $1.33$ & $3.03$ & $1.21\times10^{-1}$ & $17.4$ & $3.47\times10^{-2}$ \\
    $10^4$ & $2$ & $84$ & $6.74\times10^{-5}$ & $2.78\times10^{-2}$ & $9.62\times10^{-3}$ & $1.87$ & $3.04$ & $2.32\times10^{-1}$ & $31.6$ & $5.65\times10^{-2}$ \\
    $10^4$ & $3$ & $66$ & $9.08\times10^{-5}$ & $3.05\times10^{-2}$ & $1.24\times10^{-2}$ & $2.45$ & $3.03$ & $3.58\times10^{-1}$ & $45.1$ & $7.19\times10^{-2}$ \\
    $10^5$ & $1$ & $984$ & $5.31\times10^{-5}$ & $8.62\times10^{-3}$ & $8.30\times10^{-4}$ & $1.14$ & $0.63$ & $6.00\times10^{-1}$ & $45.5$ & $5.07\times10^{-3}$ \\
    $10^5$ & $2$ & $573$ & $1.40\times10^{-4}$ & $1.85\times10^{-2}$ & $1.45\times10^{-3}$ & $2.71$ & $1.30$ & $2.00$ & $171$ & $9.39\times10^{-3}$ \\
    $10^5$ & $3$ & $433$ & $2.31\times10^{-4}$ & $2.57\times10^{-2}$ & $1.99\times10^{-3}$ & $4.14$ & $1.75$ & $3.83$ & $358$ & $1.34\times10^{-2}$ \\
    $10^6$ & $1$ & $2467$ & $1.67\times10^{-4}$ & $7.03\times10^{-3}$ & $4.50\times10^{-4}$ & $5.51$ & $0.40$ & $6.24$ & $927$ & $5.78\times10^{-3}$ \\
    $10^6$ & $2$ & $849$ & $3.76\times10^{-4}$ & $1.10\times10^{-2}$ & $1.36\times10^{-3}$ & $16.2$ & $0.59$ & $15.5$ & $995$ & $1.15\times10^{-2}$ \\
    $10^6$ & $3$ & $432$ & $5.54\times10^{-4}$ & $9.95\times10^{-3}$ & $2.75\times10^{-3}$ & $21.5$ & $0.57$ & $18.0$ & $999$ & $1.45\times10^{-2}$ \\
    \hline 
  \end{tabular}
  \label{tab:positive_inspiral}
\end{table}

\begin{table}[H]
  \scriptsize
  \centering
  \caption{Full PhenomA waveform for the positive-power case}
  \begin{tabular}{cccccccccc}        
    \hline 
    $M_{\rm tot}~[M_\odot]$ & $z$ & SNR & $\Delta {\cal M}_z / {\cal M}_z$ & $\Delta \eta / \eta$ & $\Delta z / z$ & $\Delta t_c$ & $\Delta \Psi_c$ & $\Delta \beta_1$ & $\Delta \beta_2$  \\ 
    \hline 
    $10^4$ & $1$ & $143$ & $2.62\times10^{-5}$ & $1.56\times10^{-2}$ & $1.56\times10^{-2}$ & $0.80$ & $1.75$ & $8.05\times10^{-2}$ & $4.30$ \\
    $10^4$ & $2$ & $88$ & $4.44\times10^{-5}$ & $2.06\times10^{-2}$ & $2.15\times10^{-2}$ & $1.26$ & $2.01$ & $1.71\times10^{-1}$ & $7.41$ \\
    $10^4$ & $3$ & $69$ & $6.19\times10^{-5}$ & $2.41\times10^{-2}$ & $2.62\times10^{-2}$ & $1.63$ & $2.21$ & $2.80\times10^{-1}$ & $11.4$ \\
    $10^5$ & $1$ & $1037$ & $2.14\times10^{-5}$ & $3.29\times10^{-3}$ & $3.27\times10^{-3}$ & $0.25$ & $2.29\times10^{-1}$ & $2.35\times10^{-1}$ & $7.85$ \\
    $10^5$ & $2$ & $619$ & $5.75\times10^{-5}$ & $6.91\times10^{-3}$ & $6.98\times10^{-3}$ & $0.49$ & $4.58\times10^{-1}$ & $7.78\times10^{-1}$ & $26.8$ \\
    $10^5$ & $3$ & $482$ & $1.04\times10^{-4}$ & $1.04\times10^{-2}$ & $1.08\times10^{-2}$ & $0.69$ & $6.67\times10^{-1}$ & $1.62$ & $56.9$ \\
    $10^6$ & $1$ & $7075$ & $3.35\times10^{-5}$ & $1.00\times10^{-3}$ & $1.17\times10^{-3}$ & $0.72$ & $5.85\times10^{-2}$ & $8.82\times10^{-1}$ & $88.1$ \\
    $10^6$ & $2$ & $4042$ & $9.79\times10^{-5}$ & $1.47\times10^{-3}$ & $1.88\times10^{-3}$ & $2.39$ & $8.37\times10^{-2}$ & $2.11$ & $325$ \\
    $10^6$ & $3$ & $2878$ & $1.96\times10^{-4}$ & $1.78\times10^{-3}$ & $2.55\times10^{-3}$ & $4.76$ & $9.47\times10^{-2}$ & $3.79$ & $649$ \\
    \hline 
  \end{tabular}
\end{table}

\begin{table}[H]
  \scriptsize
  \centering
  \caption{Inspiral waveform for the negative-power case}
  \begin{tabular}{cccccccccc}        
    \hline 
    $M_{\rm tot}~[M_\odot]$ & $z$ & SNR & $\Delta {\cal M}_z / {\cal M}_z$ & $\Delta \eta / \eta$ & $\Delta z / z$ & $\Delta t_c$ & $\Delta \Psi_c$ & $\Delta \beta_1$ & $\Delta \beta_2$  \\ 
    \hline 
    $10^5$ & $1$ & $1021$ & $5.04\times10^{-5}$ & $5.09\times10^{-4}$ & $7.92\times10^{-4}$ & $0.20$ & $5.52\times10^{-2}$ & $3.21\times10^{-1}$ & $606$ \\
    $10^5$ & $2$ & $595$ & $1.11\times10^{-4}$ & $9.49\times10^{-4}$ & $1.37\times10^{-3}$ & $0.49$ & $9.92\times10^{-2}$ & $5.43\times10^{-1}$ & $805$ \\
    $10^5$ & $3$ & $449$ & $1.80\times10^{-4}$ & $1.40\times10^{-3}$ & $1.85\times10^{-3}$ & $0.87$ & $1.43\times10^{-1}$ & $7.14\times10^{-1}$ & $882$ \\
    $10^6$ & $1$ & $2561$ & $4.04\times10^{-4}$ & $1.35\times10^{-3}$ & $4.24\times10^{-4}$ & $2.25$ & $1.20\times10^{-1}$ & $6.59\times10^{-1}$ & $315$ \\
    $10^6$ & $2$ & $882$ & $1.43\times10^{-3}$ & $4.41\times10^{-3}$ & $1.37\times10^{-3}$ & $10.5$ & $3.88\times10^{-1}$ & $1.71$ & $620$ \\
    $10^6$ & $3$ & $448$ & $2.61\times10^{-3}$ & $7.74\times10^{-3}$ & $2.64\times10^{-3}$ & $24.3$ & $6.78\times10^{-1}$ & $2.52$ & $757$ \\
    $10^7$ & $1$ & $350$ & $6.62\times10^{-3}$ & $6.93\times10^{-3}$ & $5.34\times10^{-3}$ & $49.5$ & $5.31\times10^{-1}$ & $2.69$ & $310$ \\
    $10^7$ & $2$ & $85$ & $2.29\times10^{-2}$ & $1.99\times10^{-2}$ & $1.94\times10^{-2}$ & $49.9$ & $1.45$ & $6.84$ & $593$ \\
    $10^7$ & $3$ & $34$ & $3.58\times10^{-2}$ & $3.15\times10^{-2}$ & $3.60\times10^{-2}$ & $50.0$ & $2.27$ & $8.64$ & $652$ \\
    \hline 
  \end{tabular}
\end{table}

\begin{table}[H]
  \scriptsize
  \centering
  \caption{Full PhenomA waveform for the negative-power case}
  \begin{tabular}{cccccccccc}        
    \hline 
    $M_{\rm tot}~[M_\odot]$ & $z$ & SNR & $\Delta {\cal M}_z / {\cal M}_z$ & $\Delta \eta / \eta$ & $\Delta z / z$ & $\Delta t_c$ & $\Delta \Psi_c$ & $\Delta \beta_1$ & $\Delta \beta_2$  \\ 
    \hline 
    $10^5$ & $1$ & $1037$ & $3.57\times10^{-5}$ & $2.83\times10^{-4}$ & $8.07\times10^{-4}$ & $7.17\times10^{-2}$ & $2.83\times10^{-2}$ & $2.56\times10^{-1}$ & $518$ \\
    $10^5$ & $2$ & $619$ & $7.56\times10^{-5}$ & $4.60\times10^{-4}$ & $1.36\times10^{-3}$ & $1.30\times10^{-1}$ & $4.32\times10^{-2}$ & $4.36\times10^{-1}$ & $714$ \\
    $10^5$ & $3$ & $482$ & $1.17\times10^{-4}$ & $5.93\times10^{-4}$ & $1.77\times10^{-3}$ & $1.80\times10^{-1}$ & $5.32\times10^{-2}$ & $5.74\times10^{-1}$ & $802$ \\
    $10^6$ & $1$ & $7075$ & $1.24\times10^{-4}$ & $2.04\times10^{-4}$ & $1.29\times10^{-4}$ & $1.49\times10^{-1}$ & $1.59\times10^{-2}$ & $2.90\times10^{-1}$ & $172$ \\
    $10^6$ & $2$ & $4042$ & $4.07\times10^{-4}$ & $5.80\times10^{-4}$ & $3.06\times10^{-4}$ & $5.95\times10^{-1}$ & $4.50\times10^{-2}$ & $7.46\times10^{-1}$ & $349$ \\
    $10^6$ & $3$ & $2878$ & $7.70\times10^{-4}$ & $1.01\times10^{-3}$ & $6.05\times10^{-4}$ & $1.34$ & $7.82\times10^{-2}$ & $1.19$ & $474$ \\
    $10^7$ & $1$ & $6660$ & $7.67\times10^{-4}$ & $4.35\times10^{-4}$ & $1.10\times10^{-3}$ & $2.94$ & $3.45\times10^{-2}$ & $6.20\times10^{-1}$ & $107$ \\
    $10^7$ & $2$ & $2062$ & $3.14\times10^{-3}$ & $1.33\times10^{-3}$ & $4.79\times10^{-3}$ & $14.9$ & $1.07\times10^{-1}$ & $1.98$ & $261$ \\
    $10^7$ & $3$ & $966$ & $7.32\times10^{-3}$ & $2.55\times10^{-3}$ & $1.13\times10^{-2}$ & $34.5$ & $2.11\times10^{-1}$ & $3.64$ & $394$ \\
    \hline 
  \end{tabular}
\end{table}

The constraints for the EFT-inspired case are shown below. We apply flat priors on $t_c$ in $(-50,50)$ and $\Psi_c$ in $(-\pi,\pi)$. We do not apply prior bounds on $c_0$ and $f_*$, since the constraints are already strong.

\begin{table}[H]
  \scriptsize
  \centering
  \caption{Inspiral waveform for the EFT-induced case}
  \begin{tabular}{cccccccccc}        
    \hline 
    $M_{\rm tot}~[M_\odot]$ & $z$ & SNR & $\Delta {\cal M}_z / {\cal M}_z$ & $\Delta \eta / \eta$ & $\Delta z / z$ & $\Delta t_c$ & $\Delta \Psi_c$ & $\Delta c_0$ & $\Delta f_*/f_*$  \\ 
    \hline 
    $10^4$ & $1$ & $144$ & $1.08\times10^{-4}$ & $3.79\times10^{-3}$ & $5.60\times10^{-3}$ & $0.54$ & $2.46\times10^{-1}$ & $3.91\times10^{-4}$ & $2.06\times10^{-2}$ \\
    $10^4$ & $2$ & $88$ & $2.28\times10^{-4}$ & $5.98\times10^{-3}$ & $9.23\times10^{-3}$ & $0.91$ & $3.22\times10^{-1}$ & $5.40\times10^{-4}$ & $2.70\times10^{-2}$ \\
    $10^4$ & $3$ & $69$ & $3.67\times10^{-4}$ & $7.81\times10^{-3}$ & $1.20\times10^{-2}$ & $1.24$ & $3.66\times10^{-1}$ & $7.08\times10^{-4}$ & $3.39\times10^{-2}$ \\
    $10^5$ & $1$ & $1024$ & $2.69\times10^{-5}$ & $6.06\times10^{-4}$ & $7.89\times10^{-4}$ & $0.15$ & $1.42\times10^{-2}$ & $8.28\times10^{-5}$ & $1.53\times10^{-2}$ \\
    $10^5$ & $2$ & $596$ & $8.62\times10^{-5}$ & $1.46\times10^{-3}$ & $1.37\times10^{-3}$ & $0.36$ & $2.50\times10^{-2}$ & $1.63\times10^{-4}$ & $3.30\times10^{-2}$ \\
    $10^5$ & $3$ & $450$ & $1.88\times10^{-4}$ & $2.60\times10^{-3}$ & $1.85\times10^{-3}$ & $0.64$ & $3.47\times10^{-2}$ & $2.73\times10^{-4}$ & $6.12\times10^{-2}$ \\
    $10^6$ & $1$ & $2576$ & $9.71\times10^{-5}$ & $2.57\times10^{-3}$ & $1.33\times10^{-3}$ & $0.72$ & $9.12\times10^{-3}$ & $2.63\times10^{-3}$ & $1.87\times10^{-1}$ \\
    $10^6$ & $2$ & $886$ & $4.41\times10^{-4}$ & $1.07\times10^{-2}$ & $3.86\times10^{-3}$ & $3.45$ & $2.92\times10^{-2}$ & $7.98\times10^{-3}$ & $5.97\times10^{-1}$ \\
    $10^6$ & $3$ & $451$ & $1.11\times10^{-3}$ & $2.58\times10^{-2}$ & $8.18\times10^{-3}$ & $9.20$ & $5.87\times10^{-2}$ & $1.68\times10^{-2}$ & $1.33$ \\
    \hline 
  \end{tabular}
\end{table}

\section{Behaviour of Fisher integrands}
\label{sec:integrand}
\begin{figure}
    \centering
    \includegraphics[width=1.05\textwidth]{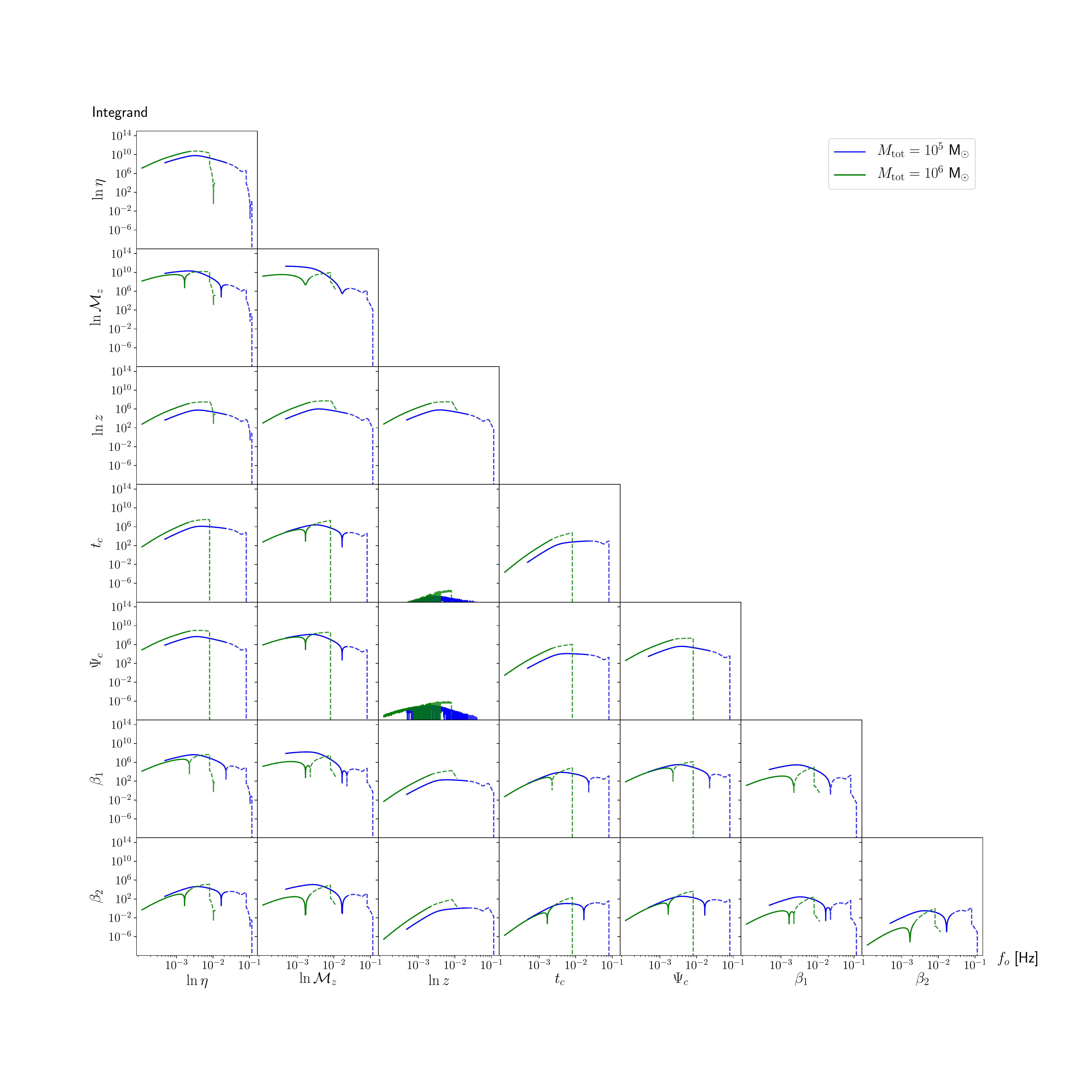}
    \caption{\small The Fisher matrix integrands of each pair of the parameters for the positive-power case with $M_{\rm tot}=10^5$ (blue) and $10^6$ (green) M$_{\odot}$. The x-axis is the frequency, and the y-axis is the integrand. The two parameters used for the integrand in each cell are indicated next to the axis ticks. The solid lines show the inspiral phase, while the dashed lines show the merger and ringdown extensions.}
    \label{fig:integrand}
\end{figure}

In this appendix we plot the Fisher matrix integrands for the positive-power case with $M_{\rm tot}=10^5$ and $10^6$ M$_{\odot}$ in Figure \ref{fig:integrand}. These assist with analysing the features seen in Fisher matrix ellipse contours, for example, the rotations of different cases. The integrand is defined as
\begin{equation}
    I(\theta_i,\theta_j) = \frac{f}{S_n(f)} \left[\left(\frac{\partial h}{\partial \theta_i}\right)\left(\frac{\partial h}{\partial \theta_j}\right)^* + \left(\frac{\partial h}{\partial \theta_i}\right)^* \left(\frac{\partial h}{\partial \theta_j}\right) \right]
\end{equation}
where $\theta_i$ is the parameter of interest. The overall factor of $f$ is included so that the Fisher matrix element is computed by integrating the integrand in $\log f$ space. Figure \ref{fig:integrand} shows that the integrands for different pairs of parameters.% could have a large difference.
%The integrands for $\ln \eta$ and $\ln {\cal M}_z$ are the largest;% since the derivatives of waveforms with respect to them have the largest magnitudes. On the other hand, the integrands for
For the parameter pairs $(\ln z,t_c)$ and $(\ln z,\Psi_c)$ the integrand is effectively zero, with only a small contribution arising from numerical noise and imperfect numerical derivatives. This is because $z$ only features in the amplitude of the waveform, while $t_c$ and $\Psi_c$ are only contained in the phase. Therefore we have
\begin{align}
    I(\ln z,t_c) = & \frac{f}{S_n(f)} \left[\left(\frac{\partial h}{\partial \ln z}\right)\left(\frac{\partial h}{\partial t_c}\right)^* + \left(\frac{\partial h}{\partial \ln z}\right)^* \left(\frac{\partial h}{\partial t_c}\right) \right]\nonumber\\
    = & \frac{f}{S_n(f)} \left[\left(\frac{\partial A}{\partial \ln z} e^{i\Psi}\right)\left(-iAe^{-i\Psi}\frac{\partial \Psi}{\partial t_c}\right) + \left(\frac{\partial A}{\partial \ln z}e^{-i\Psi}\right) \left(iAe^{i\Psi}\frac{\partial \Psi}{\partial t_c}\right) \right] \nonumber\\
    = & 0.
\end{align}
and similarly for $(\ln z,\Psi_c)$. \\
\\
In Figure \ref{fig:integrand} we observe that the shapes of the integrand can change substantially with the mass of the system. We further see that including the merger-ringdown part of the signal (dashed lines) can add a significant extra contribution, and in some cases may dominate the final integral that enters the Fisher matrix. The oscillatory features seen at high frequencies are due to resonances in the LISA PSD, and the sharp endpoints come from the merger-ringdown model visible in Figure \ref{fig:amp_pds}.

These substantial changes are responsible for the contour variations in seen in Figure \ref{fig:ellipse_phenomA}. Extending the integrand in the IMR case results in larger Fisher matrix elements, and hence correspondingly tighter forecast ellipses relative to the inspiral-only cases.

In general the areas
under the green curves ($M_{\rm tot}=10^6 M_{\odot}$) are less than the areas under the blue curves ($M_{\rm tot}=10^5 M_{\odot}$ case; this results in the correlation coefficient $|\rho|$ being smaller for the more massive system, and hence generally less diagonal Fisher ellipses. This occurs for the parameter pairs of $(\ln \eta,\ln {\cal M}_z)$, $(\ln \eta,\beta_2)$, $(t_c,\beta_1)$, $(t_c,\beta_2)$, $(\Psi_c,\beta_1)$, $(\Psi_c,\beta_2)$ and $(\beta_1,\beta_2)$. However, because the Fisher matrix has sizeable off-diagonal elements, it's inverse (the covariance) is not always trivial to predict; this could be the reason the behaviour described above is not universal.

\section{Theoretical motivations for the EFT Ansatz}
\label{theorymotEFT}

\smallskip
We are motivated by the arguments of \cite{deRham:2018red}: suppose there exists a scalar  theory valid up to a strong coupling scale $\Lambda$, with new physics (e.g extra degrees of freedom) entering at the scale $M\,\le\,\Lambda$. Let us assume  a homogeneous scalar background 
$\phi_0(t)$ that spontaneously breaks
Lorentz invariance, $\phi_0(t)\,=\,\alpha\,\Lambda\,t$,
 parameterised with a constant parameter $\alpha$ (although it may be mildly time-dependent, with $|\dot \alpha/\alpha|\,\le\, H$). The spontaneous breaking of Lorentz invariance typically leads to a scalar speed different to that of light. We consider for example the partial UV completion
of Eq. (6) in \cite{deRham:2018red}. It leads to a dispersion relation,
\be
\omega^2\,=\,k^2-\alpha^2\,\frac{\omega^2 M^2}{M^2-\omega^2+k^2}.
\label{scalDR}
\ee
%%%%%%%%%%%%%%%%%%%%
%\begin{figure}[h!]
%\centering
 % \includegraphics[width = 0.5 \textwidth]{plotCS}
 %\caption{\it Plot of $c_s$ as a function of $k/k_\star$ for    $c_0\,=\,0.1$ (dot dashed green), $c_0\,=\,0.3$ (dashed blue),  $c_0\,=\,0.8$ (black).}
 %\label{fig:plot1}
%\end{figure}
%%%%%%%%%%%%%%%%%%%
The propagation speed is defined through the dispersion relation
\be
\omega^2\,=\,c^2(t, \,k)\,k^2.
\ee
Therefore \eqref{scalDR} leads to a  scalar  speed given by %(we understand the time-dependence from now on)
\be
\label{resCS}
c_s^2(k)\,=\,1+\frac{k_\star^2}{k^2}-\frac{k_\star^2}{k^2}\,\sqrt{1+2 \left(1-c_0^2 \right) \,\frac{k^2}{k_\star^2}},.
\ee
Although motivated by scalar theories, we adopt this expression in the tensor case for simplicity.
Here
\bea
k_\star&=&\frac{M }{\sqrt{2}\,c_0}\hskip1cm;\hskip1cm c_0^2\,=\,\frac{1}{1+
\alpha^2}
\eea
Note that the function \eqref{resCS} has the properties 
\bea
c_s(k\ll k_\star)&=&c_0,
\label{asym1}
\\
c_s(k \gg k_\star)&=&1,
\label{asym2}
\eea
showing consistency with GR at large cases.
%See Fig \ref{fig:plot1}.
\smallskip
Rewriting tensor speed in (\ref{resCS}) in terms of frequency ($f\,\equiv\,2\pi\,k$), one obtains
\be\label{cTexp}
c_T(f)\,=\,\left[ 
1+\frac{f_\star^2}{f^2}-\frac{f_\star^2}{f^2}\,\sqrt{1+2 \left(1-c_0^2 \right) \,\frac{f^2}{f_\star^2} }
\right]^{1/2},
\ee
as presented in (\ref{cTexpA}).
We can analytically compute the slope of the speed
\bea
n_T(f)&\equiv&\frac{d\, \ln c_T}{d \,\ln f}\,=\,
\frac{1+ \left(1-c_0^2\right) f^2/f^2_\star-\sqrt{1+2 \left(1-c_0^2\right) f^2/f^2_\star }}{\sqrt{1+2 \left(1-c_0^2\right) f^2/f^2_\star }\left( 1+  f^2/f^2_\star-\sqrt{1+2 \left(1-c_0^2\right) f^2/f^2_\star }\right)}
\eea
\smallskip
There exists an inflection point at
\be
\label{posIP}
\frac{f_{\rm in}}{f_\star}\,=\,\sqrt{\frac{c_0 \left(\sqrt 2+c_0\right)}{1-c_0^2} }
\ee
which is an increasing function of $c_0$. 
 At the inflection point the  slope of $c_T(f)$ is maximal, resulting
\be
n^{\rm{max}}_T (f_{\rm in})\,=\,\frac{ (1-c_0^2)}{\left(1+\sqrt{2} \,c_0\right)^2}
\ee
which is a decreasing function of $c_0$. To investigate deviations from GR with the above speed profile, we compute the dimensionless quantity $\Delta$ defined in (\ref{eq:Delta_def}). For a given $c_0$, the maximum GR deviation depends on redshift
\be
\Delta_{\rm max}(c_0,z)=A(c_0) \left(1-\frac{1}{(1+z)^{n(c_0)}}\right).
\ee
We find that parameters $A$ and $n$ both decrease linearly with $c_0$. We perform least squares polynomial fits to obtain the expression in (\ref{eq:deltamax}). These results hint at possible independent redshift mapping of GW sources, provided we can accurately estimate $\Delta_{\rm max}$.

\section{Recovering a luminal \texorpdfstring{$c_T$}{cT} at high frequencies}
\label{sec:luminalrecovery}

As discussed in \S\ref{sec-theory-ansatze}, for any viable deviation in the tensor sound speed $c_T$ from the luminal speed $c$ to be observable in the LISA frequency band, the more complete gravitational theory, valid beyond its EFT description, must efficiently suppress this deviation to within a relative deviation of $\mathcal{O}(10^{-15})$ in the LIGO band. For a simple quantitative comparison of the constraints, let us consider the LIGO bound $|1-c_T(f\sim10~{\rm Hz})/c| \lesssim 10^{-15}$, which for the power-law parametrisation~\eqref{eq:cT_power} approximately implies that
\begin{equation}
 |\beta_n| \lesssim 10^{-15-n} (f_*/{\rm Hz})^n \,.
 \label{eq:GW170817}
\end{equation}
In comparison, in \S\ref{sec:forecast}, we found for the positive and negative powers that
\begin{align}
    & |\beta_1| \lesssim 0.065 (f_*/{\rm Hz}) \,,  & & |\beta_2| \lesssim 2.5 (f_*/{\rm Hz})^{2} \,, \\
    & |\beta_1| \lesssim 1.4\times10^{-8} (f_*/{\rm Hz})^{-1} \,, & & |\beta_2| \lesssim 2.6\times10^{-12} (f_*/{\rm Hz})^{-2} \,,
\end{align}
for the PopIII and Q3-nod cases, respectively.
Hence, if the functional forms are maintained to LIGO scales, these constraints are weaker than that of GW170817.
In the case of the EFT-inspired $c_T(f)$ function~\eqref{cTexpA}, we note that for $f_*\ll10$~Hz, $c_T(f)$ reduces in the LIGO band to a negative power law with $n=-1$ and $\beta_1 = \sqrt{2(1-c_0^2)}/2\approx\sqrt{1-c_0}$ for $c_0\approx1$.
Thus, in the LIGO band, our constraint from \S\ref{sec:forecast} can roughly be interpreted as
\begin{equation}
    |\beta_1| \lesssim 3\times10^{-6} (f_*/{\rm Hz})^{-1} \,,
\end{equation}
which is also weaker than eq.~\eqref{eq:GW170817}.

\begin{figure}
    \centering
    \resizebox{0.55\textwidth}{!}{\includegraphics{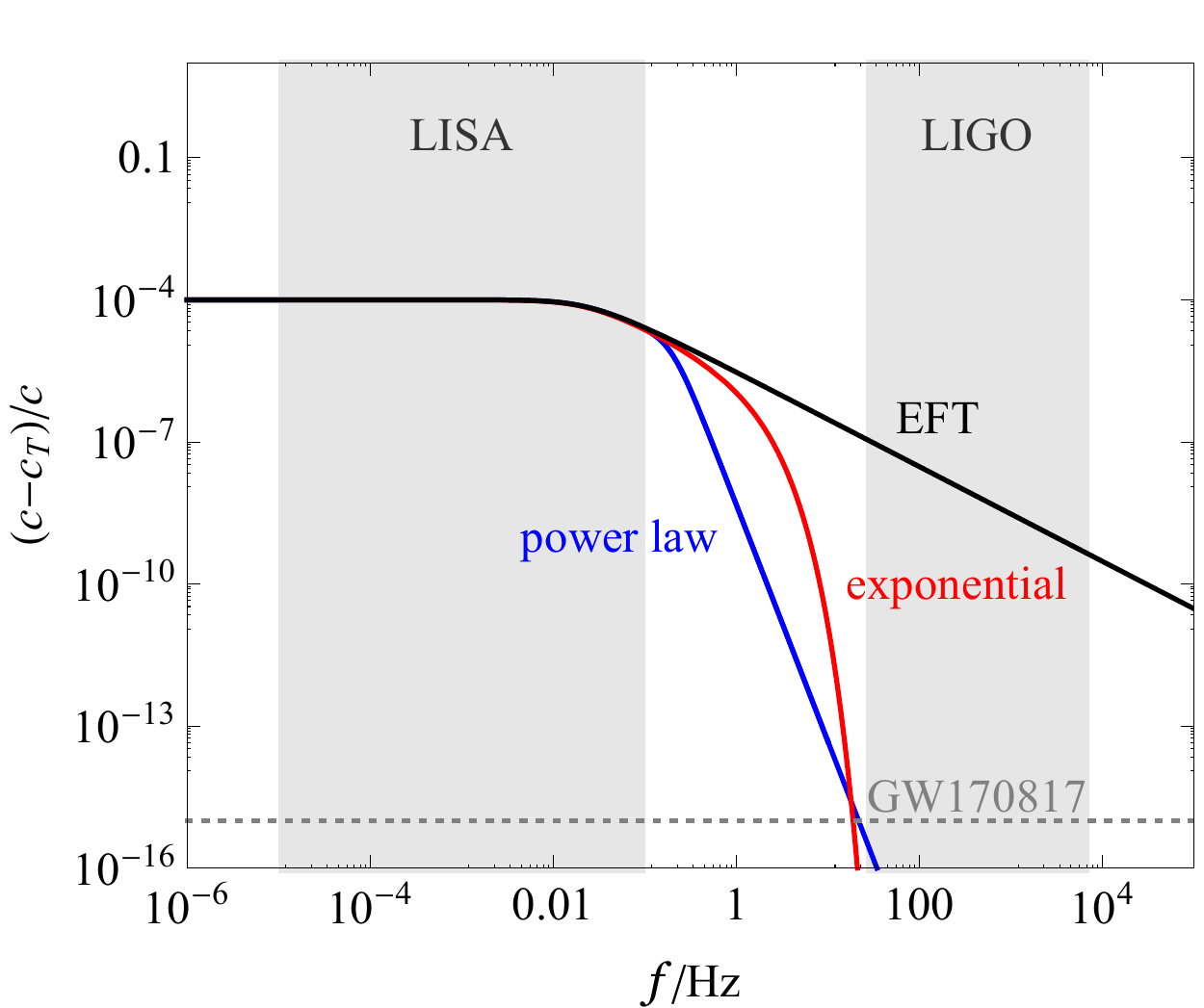}}
    \caption{
     A deviation of $(c-c_0)/c = 10^{-4}$ in the EFT-inspired Ansatz~\eqref{cTexpA} (black curve) with $f_*=3\times10^{-4}$~Hz, motivated by our forecasts in \S\ref{sec:forecast} and observable with LISA, can be efficiently suppressed with a power-law (blue) or exponential (red) suppression, eqs.~\eqref{eq:cTinterpol1} and \eqref{eq:cTinterpol2}, from higher-order corrections to satisfy the GW170817 bound (gray dotted) while remaining an accurate description in the LISA band. The parameters are chosen as $\tilde{f}_*=0.2$~Hz, $p=2$ and $\tilde{f}_*=1$~Hz and $p=\nicefrac{1}{2}$ for the power-law and exponential suppressions, respectively.
    }
    \label{fig:cT_interpolation_efficiency}
\end{figure}

Observable modifications introduced with the functional forms of $c_T(f)$
%introduced with the leading-order corrections
in eqs.~\eqref{eq:cT_power} and \eqref{cTexpA} can thus not be suppressed efficiently enough to satisfy the GW170817 bound.
However, higher-order corrections may in principle kick in to suppress the remaining deviations in the LIGO band.
We shall briefly inspect here some requirements on the functional forms of $c_T(f)$ that a more complete UV description of
%an observationally compatible theory should satisfy.
a theory should satisfy to remain observable in the LISA band while remaining compatible with the LIGO constraint.
For this purpose we shall consider a power-law and exponential suppression of the tensor sound speed of the forms
\begin{eqnarray}
    \tilde{c}_T(f) & = & \frac{c_T(f) + (f/\tilde{f}_*)^{2p}}{1 + (f/\tilde{f}_*)^{2p}} \label{eq:cTinterpol1} \,, \\
    \tilde{c}_T(f) & = & 1 - \left[ 1 - c_T(f) \right] e^{-(f/\tilde{f}_*)^{2p}} \label{eq:cTinterpol2} \,,
\end{eqnarray}
respectively.
The parameters $\tilde{f}_*$ and $p$ shall be chosen such that $c_T(f)$, given by eqs.~\eqref{eq:cT_power} or \eqref{cTexpA}, is valid in the LISA band and $(c-c_T)/c < 10^{-15}$ for LIGO.
For simplicity, we shall focus only on the EFT Ansatz, which in the high-frequency limit can however also be interpreted in terms of a $n=-1$ power-law Ansatz.
%
%In Fig.~\ref{fig:cT_interpolation_efficiency} we show some viable ranges for $\tilde{f}_*$ and $p$ for each scenario.
%
%Particularly, we find that for an exponential or power-law suppression with $p\gtrsim1$ or $p\gtrsim4$, our analysis remains fully viable.
Figure~\ref{fig:cT_interpolation_efficiency} shows how a deviation of $(c-c_0)/c = 10^{-4}$ in eq.~\eqref{cTexpA} with $f_*=3\times10^{-4}$~Hz, motivated by our forecasts in \S\ref{sec:forecast}, can be efficiently suppressed with eqs.~\eqref{eq:cTinterpol1} or \eqref{eq:cTinterpol2} in the LIGO band.
Particularly, we find that for an exponential or power-law suppression with $p\gtrsim\nicefrac{1}{2}$ or $p\gtrsim2$, our forecasts remain valid for a potential signature detectable in the LISA band that is hidden to LIGO.

%%%%%%%%%%%%%%%%%%%%%%%%%%%%%%%%%%%%%%%%%%%%%%%%%%%%%%%

\section{Future directions: general parametrization of GW propagation} \label{sec:fredep}

In \S\ref{sec-theory-ansatze}, we motivated a frequency-dependent group velocity $c_T(f)$ from the fact that, in many models of modified gravity (including quantum gravity), the modification of the dispersion relation can be written as a modified dispersion relation $\omega^2-k^2\to F(\omega,k)=0$. There, we assumed that all the time- or redshift-dependence of $c_T$ was implicit in the frequency $f$. Now we relax that assumption and consider a non-trivial function $c_T(z,f)$ of the redshift and the frequency, also including a non-trivial modification to the cosmological friction term. In fact, modified theories of gravity usually predict modifications not only the propagation speed, but also the cosmological friction experienced by GWs as well as their luminosity distance. While redshift-dependent deformations of the friction and of the luminosity distance have already been considered in the literature for all these aspects of GW propagation (see \cite{Belgacem:2019pkk} and references therein), the case of frequency-dependent of mixed redshift-frequency modifications is almost virgin territory (see \cite{Creminelli:2018xsv,Creminelli:2019nok} for scenarios motivating this possibility). Here, we will take some steps in this direction, generalizing the parametrization of \cite{Nishizawa:2017nef} to a polynomial parametrization including all these corrections.

\subsection{Equation of motion and amplitude}
\label{subsec:eom_amp}
Theories of modified gravity can deform the evolution equation 
of tensor modes in a variety of ways. Here we are especially interested to new contributions that depend on the GW momentum. A possible structure for the modified GW evolution equation in Fourier space is (omitting the GW polarization index) 
\bea\label{gen_ev_eq}
h_{ij}''(\eta, {\bf k})+2\,{\cal H}\,\left[ 1+ \Gamma_\alpha (\eta, k) \right]\, h_{ij}'(\eta, {\bf k} )
+ k^2\,\left[1+ \Gamma_\beta (\eta, k)\right]^2\, h_{ij}(\eta, {\bf k})
=0\,,
\eea
where  primes denote derivatives of conformal time $d\eta=dt/a$ and ${\cal H}\,=\,a'/a$. Eq. (\ref{gen_ev_eq}) describes propagating massless modes in scenarios with modifications of both the cosmological friction term and of the graviton dispersion relation.

The momentum-dependent modifications of the graviton dispersion relations are controlled by the quantity $\Gamma_\beta(\eta, k)$ in eq.\ \eqref{gen_ev_eq}. Such function contributes to the GW speed as
\be\label{cTdef}
c_T(\eta, k)\,=\,1+ \Gamma_\beta (\eta, k)\,,
\ee
which generalizes eq.\ (\ref{eq:cT_power}).

The momentum-dependent modifications of the friction term is controlled by the quantity $\Gamma_\alpha(\eta, k)$ and generalizes the Mukhanov--Sasaki equation stemming from \eqref{act1}. Its homogeneous part, often denoted as $-\delta(\eta)$ \cite{Belgacem:2019pkk}, is well studied and is expected, for example, in scenarios with non-minimal couplings between scalars and gravity. The momentum-dependent part of $\Gamma_\alpha$ has received much less attention, especially for what respects its consequences for modified GW waveforms. A time- and momentum-dependent $\Gamma_\alpha(\eta,k)$ could be motivated by graviton decay into dark energy, Lorentz-violation models and quantum-gravity scenarios such as non-commutative spacetimes. However, while it is not especially difficult to build models with infrared deformations from classical modified gravity scenarios, it is more challenging to obtain them from theories of quantum gravity, where the main effects are expected to happen at very short, UV scales \cite{Calcagni:2019kzo,Calcagni:2019ngc}. In particular, non-commutative spacetimes do modify the propagation of GWs at the non-commutativity scale characterizing time-space uncertainty. In fact, it is possible to re-interpret the corrections to the GW equation of motion (\ref{gen_ev_eq}) in terms of an effective, scale-dependent scale factor as it appears in non-commutative settings (Appendix \ref{app-eveqGC}). However, the non-commutative scale is $O(l_{\rm Pl})$ and the effect on the propagation of GWs at cosmological scales is totally negligible. For instance, the deviation from the GR luminosity distance is $\sim O(10^{-120})$ \cite{Calcagni:2019ngc}. A possibility could be to devise a non-commutative model where the uncertainty scale is the Hubble radius $H^{-1}$, so that corrections could take place at infrared distances. We will not make an attempt to build a viable and robust model along this line, since our approach here will be mainly phenomenological.

It is not difficult to formally solve the evolution equation at short wavelengths much smaller than the horizon size $\lambda\ll H^{-1}$, using, e.g., the techniques of \cite{Nishizawa:2017nef,Belgacem:2019pkk}. One finds that the mode function  $h(\eta, {\bf k} )$ reads
\bea\label{solLF1}
h(\eta, {\bf k} )&=&h(\eta_{e}, {\bf k} )\,
\frac{a(\eta_{e})\,\sqrt{c_T(\eta_{e}, k)}}{a(\eta)\,\sqrt{c_T(\eta, k)}}\,\exp{\left[
-\int_{\eta_{e}}^\eta\,d \hat \eta\,{\cal H}(\hat \eta)\,\Gamma_{\alpha} (\hat \eta, k) \right]}
\nonumber\\
&&
\,
\times \,
\exp{ \left[i\,k\,
\int_{\eta_{e}}^\eta\,d \hat \eta\,c_T(\hat \eta,k) \right]}
\,,
\eea
where the integration constants are chosen such to match with the GW mode at emission when $\eta=\eta_{e}$. The terms in the first line control how the GW amplitude is modified with respect to the one at emission by the GW expansion (notice the scale factor at the denominator) and by modifications of gravity. The term in the second line controls the evolution of the GW phase, which is sensitive to $c_T$. Importantly, to derive this solution we did not have to make any assumption on the momentum dependence of the functions $\Gamma_{\alpha}$, $\Gamma_{\beta}$: these functions could have rich momentum-dependent profiles with sudden changes, and transitorily large first derivatives as a function of the GW frequency.
 
We do not make any explicit assumption on screening mechanisms and the value of $c_T$ nearby the source but, in order to satisfy the stringent GW170817 constraints, we assume that the speed of GWs is equal to one at the frequencies $f_{\rm gb} \sim 10$ Hz of ground-based detectors such as LIGO-Virgo:
 \be\label{ct1a}
 c_T(\eta,\,f_{\rm gb})\,=\,1\,,
 \ee
while at much lower frequencies it can be different from one.

\subsection{Parametrizations for \texorpdfstring{$\Gamma_\alpha$}{Gammaalpha} and \texorpdfstring{$\Gamma_\beta$}{Gammabeta}}\label{sec-para}

The momentum dependence in $\Gamma_\alpha$ and $\Gamma_\beta$ eventually translates to corrections in physical observables and a choice of parametrization is needed. One can decide to parametrize directly the observables, but from a theoretical as well as a model-building point of view it may be more convenient to start with a parametrization for $\Gamma_\alpha$ and $\Gamma_\beta$ and then work one's way through the observables. We will follow this route.

Nishizawa \cite{Nishizawa:2017nef} proposed a momentum-dependent parametrization given by constant coefficients and positive powers of momentum. In our notation and extending this parametrization to, redshift- or time-dependent coefficients $\alpha_n$, 
\beq\label{nishi}
\Gamma_\alpha(z,k) = \sum_{n=0}^\infty \alpha_n(z)\,\left(\frac{k}{k_*}\right)^n\,,
\eeq
where $k_*$ is a reference scale that we can take to be the typical frequency $f_*=k_*/(2\pi)$ to which the detector is sensitive (in analogy with CMB parametrizations via a pivot scale). In a model-independent approach, one typically truncates the series up to a finite order $N>0$. We took $N=2$ in all the above sections. The $n=0$ term corresponds to the homogeneous correction
\beq
\alpha_0(z)=-\delta(z)\,.
\eeq
We can apply the same parametrization also to $\Gamma_\beta$, with different coefficients:
\beq\label{nishi2}
\Gamma_\beta(z,k) = \sum_{n=0}^\infty \beta_n(z)\,\left(\frac{k}{k_*}\right)^n= \sum_{n=0}^\infty \beta_n(z)\,\left(\frac{f}{f_*}\right)^n\,,
\eeq
eventually truncated to a finite order $N$. The coefficient $\beta_0(0)$ is the frequency-independent correction to the propagation speed measured locally today ($z=0$), so that in the absence of frequency dependence $c_T(z=0)=1+\beta_0(0)$. However, 
\eqref{ct1a} implies that one would have to fine tune the $\beta_{\pm n}$ so that $\Gamma_\beta=0$ exactly at LIGO-Virgo frequency, unless 
\beq\label{beta0}
\beta_0(0)=0\,,
\eeq
a choice made also in eq. (\ref{eq:cT_power}). In this way, the local propagation speed is exactly equal to the speed of light.

Seen as perturbative expansions, eqs. \eqref{nishi} and \eqref{nishi2} encode effects appearing at $k/k_*\lesssim 1$ (small frequencies or large scales). However, {\it a priori} one may conceive corrections that become important at $k/k_*\gtrsim 1$ (large frequencies, small scales), in which case we can take negative powers of momenta:
\beq\label{nishi3}
\Gamma_\alpha(z,k) = \sum_{n=0}^\infty \alpha_{-n}(z)\,\left(\frac{k_*}{k}\right)^n\,,\qquad
\Gamma_\beta(z,k) = \sum_{n=0}^\infty \beta_{-n}(z)\,\left(\frac{k_*}{k}\right)^n\,.
\eeq
One could combine small- and large-frequency corrections in a single truncated Laurent series, but in the lack of a theoretical justification for it (i.e., if $\Gamma_{\alpha,\beta}$ were special functions of $k$ with a non-trivial Laurent representation, such as $e^k/k+e^{1/k}$) it is more practical to study each series separately in their convergence region (small or large $k$). We will represent both cases at once with the notation
\beq\label{nishi4}
\Gamma_\alpha(z,f) = \sum_{n=0}^\infty \alpha_{\pm n}(z)\,\left(\frac{f}{f_*}\right)^{\pm n},\qquad
\Gamma_\beta(z,f) = \sum_{n=0}^\infty \beta_{\pm n}(z)\,\left(\frac{f}{f_*}\right)^{\pm n}.
\eeq

This parametrization assumes that time and momentum dependence are factored out, at least term by term. In general, this may not be the case at an exact level. There exist models (not necessarily important for GW astronomy, such as non-commutative spacetimes \cite{Brandenberger:2002nq}) where the time-momentum dependence is mixed, for instance of the form $\Gamma_{\alpha,\beta}(\eta+\ell^2 k,\eta-\ell^2 k)$, where $\ell$ is a characteristic or fundamental length scale of the system. However, asymptotically any such expression will in general admit a small- or large-$k$ expansion of the above form, where the time dependence is factored out.

In the rest of this section, we use this model-independent, parametrization-dependent approach to calculate the general $(z,f)$-dependent modifications of the luminosity distance and the waveform phase. We will need the general expansion of $f_s(f_o)$ stemming from formula \eqref{mairel1}:
\be\label{diffreG}
\frac{1+\Gamma_\beta(\ze,f_s)}{f_s}=\frac{1}{1+\ze} \frac{1+\Gamma_\beta(0,f_o)}{f_o}\,,
\ee
where we used $z_o=0$. Plugging \eqref{nishi4}, one has
\be\label{fexp}
f_s = f_o\sum_{l=0}^\infty \gamma_{\pm l}(\ze)\left(\frac{f_o}{f_*}\right)^{\pm l},\qquad
\gamma_0(\ze) = \frac{1+\beta_0(\ze)}{1+\beta_0(0)}(1+\ze)\,,
\ee
where higher-order coefficients can be calculated explicitly. For our purposes, it will be sufficient to expand up to the next-to-leading term in the observed frequency. For the series of positive powers,
\be
\gamma_1(\ze) =\frac{[1+\beta_0(\ze)][(1+\ze)\beta_1(\ze)-\beta_1(0)]}{[1+\beta_0(0)]^2}(1+\ze)\,,
\ee
while for the series of negative powers one has
\be
\gamma_{-1}(\ze) =
-\frac{(1+\ze)\beta_1(0)[1+2\beta_0(\ze)+\beta_0^2(\ze)]-[1+\beta_0(0)]^2\beta_1(\ze)}{[1+\beta_0(0)]^2[1+\beta_0(\ze)]}\,.
\ee
These coefficients simplify if we assume \eqref{beta0}:
\bea
\gamma_0(\ze) &=& [1+\beta_0(\ze)](1+\ze)\,,\label{ga0be0}\\
\gamma_1(\ze) &=& [1+\beta_0(\ze)][(1+\ze)\beta_1(\ze)-\beta_1(0)](1+\ze)\,,\\
\gamma_{-1}(\ze) &=&
-\frac{(1+\ze)\beta_1(0)[1+2\beta_0(\ze)+\beta_0^2(\ze)]-\beta_1(\ze)}{1+\beta_0(\ze)}\,.
\eea
From now on, we ignore PN corrections but the reader should keep in mind that the full expressions to be compared with simulated or data should include these terms and may display degeneracies.
 
\subsection{Luminosity distance}

In terms of $\Gamma_\alpha$ and $\Gamma_\beta$, the GW luminosity distance reads (\S\ref{sec_pre}):
\bea
\dl^{\rm GW}(\ze,f)
&=& \left(1+\ze \right) \,r_{\rm com}^{\rm GW}\,\sqrt{\frac{c_T(t_s, f_s)}{
c_T(t_o, f)}}
 \, \exp{\left[\int_0^{\ze}\,\frac{\Gamma_{\alpha} (\tilde z, f) }{1+\tilde z}\,d \tilde z 
\right]}\,,\label{ludi1}
\eea
where we set $a(t_o)=1$ and $f=f_o$ is the frequency as measured by the observer. When $\Gamma_\alpha(z,f)=-\delta(z)$ and $\Gamma_\beta=0$, we get the very same result of \cite{Belgacem:2019pkk}, while when $\Gamma_\alpha=0$ we obtain eq.\ (\ref{eq:distance}).

The parametrization \eqref{ludi1} in terms of $\Gamma_{\alpha,\beta}$ can be further manipulated without loss of generality under the assumption that corrections to general relativity are small. To make notation more compact and connect with the observable studied in \cite{Belgacem:2019pkk} for standard sirens, we also assume that the propagation of photons is unmodified, so that the optical luminosity distance is the usual
\be
\dl^{\rm EM}(\ze)=\left(1+\ze \right)\,r_{\rm com}\,,\qquad r_{\rm com}:=\int^{t_{\rm em}}\frac{d \tilde t}{a(\tilde t)}\,.
\ee
In particular, the ratio between the comoving distance of a GW source and its optical counterpart is
\be
\hspace{-.8cm}\frac{r_{\rm com}^{\rm GW}}{r_{\rm com}}=1+\langle\langle\Gamma_\beta\rangle\rangle=1+\sum_{l=0}^\infty\langle\langle\beta_{\pm l}\rangle\rangle\left(\frac{f}{f_*}\right)^{\pm l}\,,\qquad \langle\langle\beta\rangle\rangle:=\frac{1}{r_{\rm com}}\int^{t_{\rm em}}\frac{d\tilde t}{a(\tilde t)}\,\beta[z(\tilde t)]\,.
\ee
When $\Gamma_{\alpha,\beta}\ll 1$ for any redshift and frequency, assuming $\beta_0(0)=0$ and from eqs.\ \eqref{ludi1}, \eqref{diffreG} and \eqref{ga0be0} one has
\bea
\frac{\dl^{\rm GW}}{\dl^{\rm EM}}(\ze,f)
\!\!&=&\!\!\frac{1}{\sqrt{1+\ze}}\sqrt{\frac{f_s}{f}}
 \, \exp{\left[\int_0^{\ze}\,\frac{\Gamma_{\alpha} (\tilde z, f) }{1+\tilde z}\,d \tilde z 
\right]}\frac{r_{\rm com}^{\rm gw}}{r_{\rm com}}\nonumber\\
\!\!&=& \!\!\sqrt{\frac{\gamma_0(\ze)}{1+\ze}}\sqrt{1+\sum_{l=1}^\infty \frac{\gamma_{\pm l}(\ze)}{\gamma_0(\ze)}\left(\frac{f}{f_*}\right)^{\pm l}} \sum_{m=0}^\infty\frac{1}{m!}\left[\int_0^{\ze}\,\frac{\Gamma_{\alpha} (\tilde z, f) }{1+\tilde z}\,d \tilde z \right]^m\frac{r_{\rm com}^{\rm gw}}{r_{\rm com}}\nonumber\\
\!\!&=& \!\!\sum_{q=0}^\infty\frac{\sqrt{\pi[1+\beta_0(\ze)]}}{2\Gamma(1+q) \Gamma(3/2-q)}\left[\sum_{l=1}^\infty \frac{\gamma_{\pm l}(\ze)}{\gamma_0(\ze)}\left(\frac{f}{f_*}\right)^{\pm l}\right]^q\sum_{m=0}^\infty\frac{1}{m!}\left[\int_0^{\ze}\,\frac{\Gamma_{\alpha} (\tilde z, f) }{1+\tilde z}\,d \tilde z \right]^m\frac{r_{\rm com}^{\rm gw}}{r_{\rm com}}\nonumber\\
\!\!&=&\!\! 1+\sum_{n=0}^\infty b_{\pm n}(\ze)\left(\frac{f}{f_*}\right)^{\pm n},\label{ludi3}
\eea
where the first two coefficients are
\bea
b_0(\ze) &=&\sqrt{1+\beta_0(\ze)}\,\exp\left[\int_0^{\ze}\,\frac{\alpha_{0}(\tilde z)}{1+\tilde z}\,d \tilde z\right]-1+\langle\langle\beta_0(\ze)\rangle\rangle\,,\label{b1}\\
b_{\pm 1}(\ze) &=& \exp\left[\int_0^{\ze}\frac{\alpha_{0}(\tilde z)}{1+\tilde z}\,d \tilde z\right]\left[1+\langle\langle\beta_0(\ze)\rangle\rangle\right]\nonumber\\
&&\times\left[\frac{\gamma_{\pm 1}(\ze)}{2(1+\ze)\sqrt{1+\beta_0(\ze)}}+\sqrt{1+\beta_0(\ze)}\int_0^{\ze}\,\frac{\alpha_{\pm 1}(\tilde z)}{1+\tilde z}\,d \tilde z\right]\nonumber\\
&&+\exp\left[\int_0^{\ze}\,\frac{\alpha_{0}(\tilde z)}{1+\tilde z}\,d \tilde z\right]\sqrt{1+\beta_0(\ze)}\langle\langle\beta_{\pm 1}(\ze)\rangle\rangle\,.\label{b2}
\eea
Let us calculate the corrections \eqref{b1} and \eqref{b2} to the GW luminosity distance for some analytic examples.
\begin{enumerate}
\item For the kink profile
\beq\label{binomkin}
c_T(f)=\frac{c_0+(f/f_{*})^{2p}}{1+(f/f_{*})^{2p}}\,,
\eeq 
where $p>0$, we must assume that $c_0=c_0(z)$ is a function of redshift, otherwise all corrections vanish. If $p>1/2$ and assuming no correction to the friction term ($\Gamma_\alpha=0$), we have an expansion in positive powers of the frequency:
\bea
&&\alpha_{n}=0\,,\\
&&\beta_0=c_0-1\,,\qquad\beta_1=0\,,\qquad \beta_2=1-c_0\label{binomkin2a}\\
&&\gamma_1=0\,,\label{binomkin2b}%\qquad \,,
\eea
so that
{ \be
b_0(\ze) =\sqrt{c_0(\ze)}+\langle\langle c_0(\ze)\rangle\rangle-2\,,\qquad 
b_{1}(\ze) =0\,,\label{b2a}
\ee}
with the condition that $c_0(0)=1$. In this case, the correction to the luminosity distance only depends on the redshift. A concrete example falling into this analysis is the $p=1$ profile
\begin{equation}\label{cTtoy}
c_T(k)=\frac{d\omega}{dk}\,=\,\frac{c_0+k^2/\Lambda^2}{1+k^2/\Lambda^2}\,,
\end{equation}
which stems from the modified dispersion relation
\begin{equation}
\omega = k+(c_0-1)\Lambda\,{\rm arctan}\frac{k}{\Lambda}\,.
\end{equation}

The case $p=1/2$ is less trivial:
\bea
&&\alpha_{n}=0\,,\\
&&\beta_0=c_0-1\,,\qquad\beta_1=1-c_0\,,\qquad \beta_2=c_0-1\\
&&\gamma_1=0\,,%\qquad \,,
\eea
so that there is a frequency dependence at the next-to-leading order:
\bea
b_0(\ze) &=&\sqrt{c_0(\ze)}+\langle\langle c_0(\ze)\rangle\rangle-2\,,\\
b_{1}(\ze) &=&\sqrt{c_0(\ze)}\left\{\frac{1}{2}[1-c_0(\ze)]\langle\langle c_0(\ze)\rangle\rangle(1+\ze)+1-\langle\langle c_0(\ze)\rangle\rangle\right\}\,.\label{b2b}
\eea
\item For the example of \S\ref{secgmnotes}, from (\ref{cTexp})
%\be
%c_T=\left[1+\left(\frac{f_*}{f}\right)^2-\left(\frac{f_*}{f}\right)^2\sqrt{1+2(1-c_0^2)\left(\frac{f}{f_*}\right)^2}\right]^{1/2},
%\ee
one has
\bea
&&\alpha_{n}=0\,,\\
&&\beta_0=c_0-1\,,\qquad\beta_1=0\,,\qquad \beta_2=\frac{(1-c_0^2)^2}{4c_0}\label{binomkin4a}\\
&&\gamma_1=0\,,\label{binomkin4b}%\qquad \,,
\eea
and one recovers the case \eqref{b2a}. In order to remove this degeneracy between models at the theoretical level, one has to calculate the higher-order coefficients $\gamma_2$ and $b_2$ but, in practice, at the experimental level the two models are indistinguishable.
\end{enumerate}

\subsection{Phase}

Let us see how the perturbative approach can be used to evaluate the phase in eq.\ \eqref{fasemod}.

For example, we assume that the time-dependence $c_T$ at present times is very mild, and negligible. Eq. \eqref{eq:df_dt_delta} at lowest order in the PN expansion is
\bea\label{frecon3}
 \frac{ d f}{d t_o}&=& 
\frac{c^2_T(t_o, f)}{c^2_T(t_s, f_s )}
  \frac{ 1-{\partial \ln c_T(t_s, f_s)}/{\partial \ln f_s}} { 1-{\partial \ln c_T(t_o, f)}/{\partial \ln f_o}} 
  \,\frac{96\,( \pi\,{\cal M}_s\,f_s )^{\frac{11}{3}}}{5 \pi\,{\cal M}_s^2\,(1+z)^2}\,.
\eea
Inverting this expression and integrating yields the formal relation for the frequency dependence of the time as measured by the observer:
 \bea\label{timet1}
 t_o(f)-t_c&=&\frac{5}{96\,\pi^{\frac83} \left(1+\ze\right)^{\frac53}\,{\cal M}_s^{\frac53}}
 \,\int_{f_c}^{f}\,
 \left[ \frac{ 1-\partial \ln c_T(t_o, \tilde f)/\partial \ln \tilde f}{1-\partial \ln c_T(t_s, \tilde f_s)/\partial \ln \tilde f_s}\right]\,\frac{c_T^{\frac53} (\tilde f)}{c_T^{\frac53}(\tilde f_s)}
 \frac{d \tilde f}{\tilde f^\frac{11}{3}}\,,
 \eea
 where $\tilde f_s=f_s(\tilde f)$ has the functional dependence of \eqref{mairel1} with $f=f_o$ replaced by $\tilde f$. Equation \eqref{timet1} can be substituted in eq.\ \eqref{fasemod}, and the integral performed numerically (depending on $c_T(f))$. The effect of modified  gravity is contained in the frequency dependence of $c_T(f)$.
 
Using eqs.\ \eqref{cTdef} and \eqref{diffreG} and the expansions \eqref{nishi4} and \eqref{fexp}, we get 
\be
1-\frac{\partial \ln c_T(f)}{\partial \ln f}=1-\frac{\sum_{n=1}^\infty (\pm n)\beta_{\pm n} (f/f_*)^{\pm n}}{1+\sum_{m=0}^\infty \beta_{\pm m} (f/f_*)^{\pm m}}=1\mp\frac{\beta_{\pm1}}{1+\beta_0}\left(\frac{f}{f_*}\right)^{\pm 1}+O(f^{\pm 2}),
\ee
so that, if $\beta_0(0)=0$,
\bea
 t_o(f)-t_c&=&\frac{5}{96\,\pi^{\frac83} {\cal M}_s^{\frac53}}\,\int_{\bar f}^{f}
 \left[ \frac{ 1-\partial \ln c_T(t_o, \tilde f)/\partial \ln \tilde f}{1-\partial \ln c_T(t_s, \tilde f_s)/\partial \ln \tilde f_s}\right]\left(\frac{\tilde f}{\tilde f_s}\right)^{\frac53} \frac{d \tilde f}{\tilde f^{\frac{11}{3}}}\nonumber\\
 &\simeq&\frac{5}{96\,\pi^{\frac83} {\cal M}_s^{\frac53}}
 \int_{\bar f}^{f}\,\frac{1\mp\beta_{\pm1}(0)\left(\frac{\tilde f}{f_*}\right)^{\pm 1}}{1\mp\frac{\beta_{\pm1}(\ze)}{1+\beta_0(\ze)}\left(\frac{\tilde f_s}{f_*}\right)^{\pm 1}}\left(\frac{\tilde f}{\tilde f_s}\right)^{\frac53}
 \frac{d \tilde f}{\tilde f^\frac{11}{3}}\nonumber\\
 &\simeq&\frac{5}{96\,\pi^{\frac83}{\cal M}_s^{\frac53}}
 \int_{\bar f}^{f}\,
 \left[1\mp\beta_{\pm1}(0)\left(\frac{\tilde f}{f_*}\right)^{\pm 1}\pm\frac{\beta_{\pm1}(\ze)}{1+\beta_0(\ze)}\left(\frac{\tilde f_s}{f_*}\right)^{\pm 1}\right]\!\left(\frac{\tilde f}{\tilde f_s}\right)^{\frac53}
 \frac{d \tilde f}{\tilde f^\frac{11}{3}}\nonumber\\
  &\simeq&\frac{5}{96\,\pi^{\frac83}{\cal M}_s^{\frac53}\gamma_0^{\frac53}(\ze)}
 \int_{\bar f}^{f}\,
 \left\{1\pm\frac{\beta_{\pm1}(\ze)\gamma_0^{\pm 1}(\ze)-\beta_{\pm1}(0)[1+\beta_0(\ze)]}{1+\beta_0(\ze)}\left(\frac{\tilde f}{f_*}\right)^{\pm 1}\right\}
 \frac{d \tilde f}{\tilde f^\frac{11}{3}}\nonumber\\
 &=&-\frac{5}{256\pi^{\frac83}{\cal M}_s^{\frac53}\gamma_0^{\frac53}(\ze)}\left[1\pm\frac{8}{8\mp 3}\frac{\beta_{\pm1}(\ze)\gamma_0^{\pm 1}(\ze)-\beta_{\pm1}(0)[1+\beta_0(\ze)]}{1+\beta_0(\ze)}\left(\frac{f}{f_*}\right)^{\pm 1}\right]\frac{1}{f^\frac83}\nonumber\\
 &&-\left\{\vphantom{\frac12}f\leftrightarrow  f_c\right\}\,,\label{timet2}
 \eea
 and from eq.\ \eqref{fasemod} we get the final expressions with $+$ sign to leading order when $f/f_*\ll1$,
 \bea
 \Psi^{\rm pos}(f)&\simeq&
  \frac{3}{128[\pi{\cal M}_s\gamma_0(\ze)]^{\frac53}}\left\{1+4\frac{\beta_{1}(\ze)\gamma_0(\ze)-\beta_{1}(0)[1+\beta_0(\ze)]}{1+\beta_0(\ze)}\frac{f}{f_*}\right\}\frac{1}{f^\frac53}
  \nonumber\\
 && -\left\{\vphantom{\frac12}f\leftrightarrow f_c\right\}- \Psi_c-\frac{\pi}{4}\,,\label{phaseplus}
 \eea
 and the one with $-$ sign to leading order when $f/f_*\gg1$:
 \bea
 \Psi^{\rm neg}(f)&\simeq&
  \frac{3}{128[\pi{\cal M}_s\gamma_0(\ze)]^{\frac53}}\left\{1-\frac{5}{11}\frac{\beta_{-1}(\ze)\gamma_0^{-1}(\ze)-\beta_{-1}(0)[1+\beta_0(\ze)]}{1+\beta_0(\ze)}\frac{f_*}{f}\right\}\frac{1}{f^\frac53}  \nonumber\\
 && -\left\{\vphantom{\frac12}f\leftrightarrow f_c\right\}- \Psi_c-\frac{\pi}{4} \,.\label{phaseminus}
 \eea
 Using eqs.\ (\ref{MuMz}) and (\ref{fexp}),
 \begin{equation}
 \frac{1}{{\cal M}_s}=\frac{\gamma_0}{{\cal M}_o}\left[1+\sum_{l=1}^{+\infty}\gamma_{\pm l}\left(\frac{f}{f_*}\right)^{\pm l}\right],
 \end{equation}
 and the phases (\ref{phaseplus}) and (\ref{phaseminus}) in terms of the variable $u$ are, respectively,
  \bea
 \Psi^{\rm pos}(u)&\simeq&
  \frac{3}{128}\left(1+\left\{4\frac{\beta_{1}(\ze)\gamma_0(\ze)-\beta_{1}(0)[1+\beta_0(\ze)]}{1+\beta_0(\ze)}+\frac53\frac{\gamma_1(z)}{\gamma_0(z)}\right\}\frac{u}{u_*}\right)\frac{1}{u^\frac53}
  \nonumber\\
 && -\left\{\vphantom{\frac12}u\leftrightarrow u_c\right\}- \Psi_c-\frac{\pi}{4}\,,\label{phaseplusu}
 \eea
 and
 \bea
 \Psi^{\rm neg}(u)&\simeq&
  \frac{3}{128}\left(1-\left\{\frac{5}{11}\frac{\beta_{-1}(\ze)\gamma_0^{-1}(\ze)-\beta_{-1}(0)[1+\beta_0(\ze)]}{1+\beta_0(\ze)}-\frac53\frac{\gamma_{-1}(z)}{\gamma_0(z)}\right\}\frac{u_*}{u}\right)\frac{1}{u^\frac53}  \nonumber\\
 && -\left\{\vphantom{\frac12}u\leftrightarrow u_c\right\}- \Psi_c-\frac{\pi}{4} \,.\label{phaseminusu}
 \eea
 Notice that the coefficients of the leading correction in $u^{\pm1}$ vanish identically in the approximation of redshift independence, where $\gamma_0=1$, $\gamma_{\pm 1}=0$, $\beta_0=0$ and $\beta_1={\rm const}$. However, this does not imply that the $z$-dependent case introduces extra terms in the $u$-expansion with respect to the simplest case analyzed in this paper. In fact, it is easy to convince oneself that switching on redshift dependence only changes the coefficients of the expansion, but not the expansion itself. Comparing, for instance, eqs.\ (\ref{to_MG}) and (\ref{timet2}) for the positive-power case at zero order in the PN expansion and using eqs.\ (\ref{defuf0}), (\ref{eq:cT_u}) and (\ref{cTappr}), one sees that the structure of the time difference is $t_o-t_c\sim u^{-8/3}(1+A u+B u^2)$ in both cases, the only difference being in the coefficients $A$ and $B$. Including PN terms does not create any mismatch in the powers of $u$. In particular, the phase in the positive-power case has the structure of eq.\ (\ref{posPsi}) (coefficients omitted),
 \begin{equation}\nonumber
 \Psi^{\rm pos}(u)\sim u^{-\frac53}\left(1+u^\frac23+u+u^\frac43+u^\frac53\ln u+\dots\right)+\dots\,,
 \end{equation}
 while in the negative-power case it has the structure of eq.\ (\ref{negPsi}),
 \begin{align}
 \Psi^{\rm neg}(u)\sim& u^{-\frac53}\left(\dots+u^{-\frac53}+u^{-\frac43}+u^{-1}+u^{-\frac23}+u^{-\frac13}\right.\nonumber\\
 &\left.\qquad+1+u^\frac13+u^\frac23+u+u^\frac43+u^\frac53\ln u+\dots\right)+\dots\,.\nonumber
 \end{align}
 Introducing a set of coefficients $\phi_n^{\pm}$,
 \begin{align}
 &\Psi^{\rm pos}(u)= \frac{3}{128}\frac{1}{u^\frac53}\left[\sum_{n=0}^{4}\phi_n^+(z)\, u^{\frac{n}{3}}+\phi^+_5(z) u^{\frac{5}{3}}\ln u\right]\nonumber\\
  &\quad\qquad\qquad-\left\{\vphantom{\frac12}u\leftrightarrow u_c\right\}- \Psi_c-\frac{\pi}{4},\qquad \phi^+_1=0\,,\label{psiphipos}\\
  &\Psi^{\rm neg}(u)= \frac{3}{128}\frac{1}{u^\frac53}\left[\phi^-_{-5}(z) u^{-\frac{5}{3}}+\sum_{n=-4}^{4}\phi_n^-(z)\, u^{\frac{n}{3}}+\phi^-_5(z) u^{\frac{5}{3}}\ln u\right]\nonumber\\
  &\quad\qquad\qquad-\left\{\vphantom{\frac12}u\leftrightarrow u_c\right\}- \Psi_c-\frac{\pi}{4}.\label{psiphineg}
 \end{align}

 \subsection{Comparison with the ppE framework}\label{app_ppEmap}
 
 We conclude comparing our polynomial expansion with the coefficients of the Parametrized Post-Einsteinian (ppE) framework \cite{Yunes:2009ke,Yunes:2016jcc}. In the latter case, the waveform is written as
\begin{equation}\label{hppE}
h_{\rm ppE}(u) = A_{\rm ppE}(u)\,e^{i\Psi_{\rm ppE}(u)},  
 \end{equation}
where the phase is parametrized as
 \begin{equation}\label{ppEphase}
\Psi_{\rm ppE}(u)=\frac{3}{128}\frac{1}{u^\frac53}\sum_{n=0}^5\phi_n^{\rm ppE}u^\frac{n}{3}-\left\{\vphantom{\frac12}u\leftrightarrow u_c\right\}- \Psi_c-\frac{\pi}{4},
 \end{equation}
where we stopped the series at the 2.5 PN level and the parameters of the expansion are the coefficients $\phi_i^{\rm ppE}$. Although the amplitude can also be expanded in terms of the frequency \cite{Cornish:2011ys,Nishizawa:2017nef}, we focus here on the comparison of the phase, as for our analysis this holds information on the most parameters.
%
%, in general this is less used in the ppE framework since those corrections are deemed to be sub-dominant with respect to the terms coming from the phase \cite{}.
%
%This is a first difference with respect to our formalism, where amplitude and phase corrections mix together on equal grounds in observables such as the luminosity distance (\ref{ludi3})--(\ref{b2}). In the following, we concentrate on the phase.
%
A mapping of the amplitude coefficients between the positive-power case and the ppE formalism can be found in \cite{Nishizawa:2017nef} (where $u\propto f^{1/3}$).

Comparing eqs.\ (\ref{psiphipos}) and (\ref{psiphineg}) with (\ref{ppEphase}), we can map our expansion into the ppE one. The coefficients $\phi_i^\pm(z)$ can be read off from eqs.\ (\ref{phaseplusu}) and (\ref{phaseminusu}) and their higher-order generalization.

A difference with respect to the ppE framework is in the expansion itself, which is carried out in multiples of $u^{(n-5)/3}$ in the ppE case at any order. This expansion is the same as in the positive-power case but only up to the $n=4$ term. The $n=5$ term has, in our case, an extra $\ln u$ factor. Therefore, the degeneracy between the two expansions is broken at the 2.5 PN approximation. In the negative-power case, the difference lies in all the negative-power terms of the tower.

Third, our coefficients are redshift-dependent while usually the $\phi_i^{\rm ppE}$ are not, although in this paper we focused on $z$-independent cases for simplicity.

Fourth, it is clear that, modulo the discrepancy in the powers, our expansion and the ppE one are physically equivalent because they both encode beyond-Einstein-gravity effects as a perturbative frequency expansion. However, while the starting point of the ppE formalism is the waveform (\ref{hppE}), ours is the Mukhanov--Sasaki equation (\ref{gen_ev_eq}), where modifications to GR appear in the friction term and in the propagation speed of GWs. Therefore, while the end result in terms of waveform-related, perturbatively expanded observables is essentially the same, from our perspective we are able to connect these corrections more directly to physical models, where the input from the theory appears at the level of the GW action and propagation equation. This is also the reason why the basic expansion for us is in terms of $f$ and is later recast in terms of the less fundamental but much more convenient variable $u$, while the ppE framework is already cast in terms of $u$. As we have seen, however, this extra step on our side is not particularly difficult because the $f$ and $u$ expansions differ by negligible terms.

Along the same lines, we note that in our positive-power expansion the $n=1$ term is always zero, while in the ppE framework this term is as free as the others. This is an important characterization of our expansion because it is a consequence of implementing a simple \emph{Ansatz} (a polynomial of positive powers) to a basic object such as the propagation equation (\ref{gen_ev_eq}) of GWs. In this respect, our expansion could be thought as unveiling the fundamental structure of the theory more directly than the ppE framework. However, in the negative-power case the $n=1$ term is non-vanishing, a reminder that the ``prediction'' $\phi^+_1=0$ strongly depends on how the frequency expansion of the propagation speed is defined.

Fifth, it may be worth mentioning a departure point not intrinsic to our formalism but, rather, in the practical way we handled it. Typically, ppE parameters are constrained one at a time, since varying all at once produces in very weak constraints. In contrast, here we varied multiple parameters together without marginalizing.

\subsection{Alternative perspective on eq.\ (\ref{gen_ev_eq})}\label{app-eveqGC}

An alternative implicit parametrization of the effective Mukhanov--Sasaki equation (\ref{gen_ev_eq}) makes use of a momentum-dependent friction term
 \begin{equation}
 {\cal H} (1+\Gamma_\alpha) = {\cal H}_k\equiv \frac{\partial_\eta a_k}{a_k}\,,
 \end{equation}
 where $a_k=a_k(\eta)$ is an effective scale factor that depends both on time and on momentum. Then, eq.\ (\ref{solLF1}) simplifies to
\bea\label{solLF1ak}
h(\eta, {\bf k} )&=&h(\eta_{e}, {\bf k} )\,
\frac{a_k(\eta_{e})\,\sqrt{c_T(\eta_{e}, k)}}{a_k(\eta)\,\sqrt{c_T(\eta, k)}}\,
\exp{ \left[i\,k\,
\int_{\eta_{e}}^\eta\,d \hat \eta\,c_T(\hat \eta,k) \right]}.
\eea
The advantage of using $a_k$ would be about interpretation. On one hand, it can be seen as an effective time- and momentum-dependent effective mass $M^2_k(\eta)=M^2_{\rm Pl} a_k^2$ in the action for the perturbation mode. In quantum gravity as well as in particular in non-commutative models where time and space coordinates obey an uncertainty relation \cite{Brandenberger:2002nq}, one is usually able to write down a four-dimensional action of the following form, where $\tau$ is a time parameter not necessarily equal to conformal time $\eta$:
\begin{equation}
S=\frac12\int d\eta\,d^3\bm{k}\,M^2_k(\tau)\left(\frac{d h_{-k}}{d\tau}\frac{d h_k}{d\tau}-c_T^2k^2 h_{-k} h_k\right). \label{S2nc2}
\end{equation}
The equation of motion for the Mukhanov--Sasaki variable $u_k=a_k h_k$ is
\beq\label{msenc}
\frac{d^2 u_k}{d\tau^2}+\left(c_T^2k^2-\frac{1}{a_k}\frac{d^2 a_k}{d\tau^2}\right)u_k=0\,,
\ee
where the function $a_k(\tau)$ depends on the theory or model.

On the other hand, a momentum-dependent scale factor admits a neat physical interpretation if we extend the separate-universe approach \cite{Wands:2000dp} to the case of a propagating gravitational wave. In this case, at the linear order in perturbation theory one regards the wave as a local deformation of a FRW spacetime, so that $1-1/a_k$ is the redshift measured along the trajectory of the wave.

\section{Constraints on modified GW dispersion relations, following \cite{Yunes:2016jcc} }
\label{app_rough}

Yunes et al  \cite{Yunes:2016jcc}
 develop an argument for obtaining a rough bound on modified GW dispersion relations, focussing on  the GW
 frequencies of   ground-based GW detectors. The same argument
 can be generalised, and applied to space-based detectors as LISA. In  this 
 brief Appendix, we make use 
 of the argument of  \cite{Yunes:2016jcc}  to derive rough
 bounds on some of the parameters controlling deviations from $c_T=1$, as
 they appear in our polynomial
 Ansatz  \ref{eq:cT_power}. 
 
 \smallskip
 
 We  start briefly discussing the general idea behind the method of  \cite{Yunes:2016jcc}.  Then,
 we  apply the argument
  to the case of LISA,  and  translate it to  special cases of  our polynomial Ansatz of eq \eqref{eq:cT_power}. 
   The starting point of \cite{Yunes:2016jcc} is to introduce  an Ansatz of for the  modified dispersion relation
   of GW:   
   \begin{equation} \label{yunes_disp_rel}
   E^2= (p c)^2+
   {\mathbb{A}} \,(p c)^{\alpha}\,,
   \end{equation}
   which depends on two free parameters $  {\mathbb{A}}$ and $\alpha$.  
   A modified dispersion as \eqref{yunes_disp_rel}  affects the speed of GW: its deviation
   from the speed of light can be  parameterised in  terms of the quantity  $\delta_g$ as
   \begin{equation}
   \label{defdelg}
\delta_g\,\equiv\,\Big| 1-\frac{c_T}{c} \Big|\,=\,\frac{|1-\alpha|}{2}\,{\mathbb{A}}\, h^{\alpha-2}\,f^{\alpha-2} \,,
  \end{equation}
  where, as in the main text, $c_T$ is the GW speed.
  
  We used the fact that
the GW energy $E$ appearing in eq \eqref{yunes_disp_rel}  is related to its frequency by
\begin{equation}
 E\,=\,h \,f\,,
\end{equation}
with $h$  the Planck constant, expressed as
\begin{equation}
\label{planck_constant}
h\,=\,\left( \frac{4.1\times 10^{-13}\,{\rm eV}}{100 {\rm Hz}}\right)\,.
\end{equation}

The idea of  \cite{Yunes:2016jcc} is to make use~\footnote{Notice that, instead, the LIGO-Virgo collaboration, in \cite{LIGOScientific:2019fpa,LIGOScientific:2020tif,LIGOScientific:2021sio}, investigate the consequences of Ansatz
 \eqref{yunes_disp_rel} for the GW waveforms, and make use of information from measured GW waveforms
 to derive bounds on ${\mathbb{A}}$ as a function of $\alpha$.   } of existing bounds on $\delta_g$
for deriving bounds on ${\mathbb{A}}$ as a function of $\alpha$. 
   \cite{Yunes:2016jcc} was published before the 
   GW170817 event, and 
   makes use of existing bounds on the graviton mass \cite{Yunes:2016jcc} as starting point.
    Here,  we make use of the GW170817  
  bound on $\delta_g$ from the delay between EM and GW signals
\begin{equation}  \label{boudg2}
\delta_g^{\rm delay}\le
10^{-15}\,.
\end{equation}
This bound applies to  frequencies of ground-based detectors, around $ 100$ Hz, and corresponds
to the choice   $\alpha=2$ in the expression \eqref{defdelg}.

Inverting eq \eqref{defdelg}, we can write the equality
\begin{equation}
{\mathbb{A}}\,=\,\frac{2\,\delta_g}{|1-\alpha|}\,h^{-\alpha}\,\left(\frac{f}{100 {\rm Hz}}\right)^{-\alpha}\,
\,h^{2}\,\left(\frac{f}{100 {\rm Hz}}\right)^{2}\,
\left(100 {\rm Hz} \right)^{2-\alpha}\,.
\end{equation}
Substituting the value of $h$ from eq \eqref{planck_constant}, and the bound on $\delta_g$
 in eq \eqref{boudg2}, we obtain by extrapolation 
 the following  upper bound on the parameter $\mathbb{A}$:
\begin{equation}
{\mathbb{ A}}\,\le\,\frac{3\times10^{-40}}{|1-\alpha|}\,
\,\left(\frac{f}{100 {\rm Hz}}\right)^{2-\alpha}\,\left(\frac{10^{13}}{4.1}\right)^{\alpha}\,{\rm eV}^{2-\alpha}\,.
\end{equation}
For the case of LISA, choosing say a pivot scale $f_\star\,=\,10^{-4}$ Hz, we obtain the bound 
\begin{equation}
\label{bound2a}
{\mathbb A}_{\rm LISA}\,\le\,\frac{3\times10^{-52}}{|1-\alpha|}\,
\,\left(\frac{10^{19}}{4.1}\right)^{\alpha}\,{\rm eV}^{2-\alpha}\,,
%\left(\frac{10^{13}}{4.1}\right)^{-2}
\end{equation}
for the quantity  ${\mathbb A}$ as a function of $\alpha$.

\smallskip

While the arguments of \cite{Yunes:2016jcc} discussed so far are  designed for the 
parameterisation \eqref{yunes_disp_rel} of the modified GW dispersion relations,  they can be translated with some assumptions into bounds on the dimensionless parameters $\beta_i$ of the parametrization
in eq \eqref{eq:cT_power} of the main text. For example, let us focus on  the case of positive power $n=2$. Hence, we switch on the parameter $\beta_2$ in eq \eqref{eq:cT_power}, 
and assume $c_0=1$ and $\beta_i=0$ for $i\neq2$. We expect very stringent bounds on $\beta_2$. In fact, the quantity $\delta_g$ now  reduces to 
\begin{equation}
\label{dg_red}
\delta_g\,=\,\Big| 1-c_T \Big|\,=\,\big| \beta_2 \big|\,\left( \frac{f}{f_\star}\right)^2\,.
\end{equation}
Comparing eqs \eqref{dg_red} and \eqref{defdelg}, we find they coincide when choosing
$\alpha=4$, when identifying 
$\big| \beta_2 \big|\,\equiv\,
({3}/{2})\,{\mathbb{A}}\, f_\star^2\,h^{2}
$, with $f_\star$ a pivot scale higher than the LISA band, say $f_\star\,=\,1$ Hz. We can then use eq \eqref{bound2a} to derive an upper bound on the dimensionless quantity $\big| \beta_2 \big|$
as
\begin{eqnarray}
\big| \beta_2 \big|&\le&\frac94\,10^{-52}\,\left( \frac{10^{19}}{4.1}\right)^4
\,{\rm eV}^{-2}\,\left( 1\,{\rm Hz}\right)^2\left(\frac{4.1\times 10^{-13} {\rm eV}}{100 {\rm Hz}} \right)^2
\\
&\le&1.3 \times 10^{-7}\,,
\end{eqnarray}
so a very stringent bound indeed.

%\bibliographystyle{apsrev4-1}
%\bibliography{FreqTest_refs}


\end{document}